\documentclass{lmcs}
\pdfoutput=1

\usepackage{lastpage}
\lmcsdoi{21}{2}{2}
\lmcsheading{}{\pageref{LastPage}}{}{}%
{Feb.~29,~2024}{Apr.~08,~2025}{}


\usepackage{amssymb}
\usepackage{amsmath}
\usepackage{multicol}
\usepackage{xspace}
\usepackage{hyperref}
\usepackage{cleveref}
\usepackage[utf8]{inputenc}
\usepackage{booktabs}

\Crefname{exa}{Example}{Examples}\crefname{exa}{Example}{Examples}
\Crefname{thm}{Theorem}{Theorems}\crefname{thm}{Theorem}{Theorems}
\Crefname{defi}{Definition}{Definitions}\crefname{defi}{Definition}{Definitions}
\Crefname{prop}{Proposition}{Propositions}\crefname{prop}{Proposition}{Propositions}
\Crefname{lem}{Lemma}{Lemmas}\crefname{lem}{Lemma}{Lemmas}
\Crefname{cor}{Corollary}{Corollaries}\crefname{cor}{Corollary}{Corollaries}
\Crefname{rem}{Remark}{Remarks}\crefname{rem}{Remark}{Remarks}
\Crefname{figure}{Figure}{Figures}\crefname{figure}{Figure}{Figures}
\Crefname{table}{Table}{Tables}\crefname{table}{Table}{Tables}
\Crefname{section}{Section}{Sections}\crefname{section}{Section}{Sections}

\usepackage{tikz}
\usetikzlibrary{backgrounds, positioning, arrows, automata, calc}
\tikzstyle{state}=[thick,minimum size=18pt, circle,draw]
\tikzstyle{transition}=[->,thick,>=stealth,shorten >=1pt,shorten <=1pt]
\tikzstyle{final}=[after node path={ node[state, scale=.8] at (\tikzlastnode) {} }]
\tikzstyle{initial}=[after node path={%
	[to path={[transition] (\tikztostart) -- (\tikztotarget)}]%
	(\tikzlastnode.180)++(180:13pt) edge (\tikzlastnode)%
}]
\tikzset{
	bg/.default={},
	bg/.style={execute at end picture={
			\begin{scope}[on background layer]
				\node[xshift=-1mm, yshift=-1mm] (sw) at (current bounding box.south west) {};
				\node[xshift=1mm, yshift=1mm] (ne) at (current bounding box.north east) {};
				\node[xshift=1mm, yshift=-1mm] (nw) at (current bounding box.north west) {};
				\fill[fill=black!10,rounded corners] (sw) rectangle (ne);
				
				\ifx&#1&\else
				\node[anchor=north east, xshift=2pt] at (nw) {#1};
				\fi
			\end{scope}
	}},
}

\newcommand{\BB}{\ensuremath{\mathbb{B}}\xspace}
\newcommand{\DD}{\ensuremath{\mathbb{D}}\xspace}
\newcommand{\NN}{\ensuremath{\mathbb{N}}\xspace}
\newcommand{\QQ}{\ensuremath{\mathbb{Q}}\xspace}
\newcommand{\RR}{\ensuremath{\mathbb{R}}\xspace}
\newcommand{\ZZ}{\ensuremath{\mathbb{Z}}\xspace}

\newcommand{\A}{\mathcal{A}}
\newcommand{\B}{\mathcal{B}}
\newcommand{\C}{\mathcal{C}}
\newcommand{\D}{\mathcal{D}}
\newcommand{\trans}[3]{#1\xrightarrow[]{#2}#3}

\newcommand{\Func}[1]{{\mathsf{#1}}}
\newcommand{\Val}{\Func{Val}}
\newcommand{\Inf}{\Func{Inf}}
\newcommand{\Sup}{\Func{Sup}}
\newcommand{\DSum}{\Func{DSum}}
\newcommand{\LimInf}{\Func{LimInf}}
\newcommand{\LimSup}{\Func{LimSup}}
\newcommand{\LimInfAvg}{\Func{LimInfAvg}}
\newcommand{\LimSupAvg}{\Func{LimSupAvg}}
\newcommand{\MPL}{\underline{\Func{MeanPayoff}}}
\newcommand{\MPH}{\overline{\Func{MeanPayoff}}}

\newcommand{\err}{\texttt{err}}
\newcommand{\req}{\texttt{rq}}

\newcommand{\gra}{\texttt{gr}}
\newcommand{\tick}{\texttt{tk}}
\newcommand{\other}{\texttt{oo}}

\newcommand{\TopolClosure}{{\it TopolCl}}
\newcommand{\TopolInterior}{{\it TopolInt}}
\newcommand{\safe}[1]{{\it SafetyCl}(#1)}
\newcommand{\cosafe}[1]{{\it CoSafetyCl}(#1)}

\newcommand{\prefixeq}{\preceq}
\newcommand{\prefix}{\prec}
\newcommand{\st}{\;\ifnum\currentgrouptype=16 \middle\fi|\;}
\newcommand{\Subject}[1]{\paragraph*{#1.}}

\newcommand{\CompClass}[1]{{\textsc{#1}}\xspace}
\newcommand{\PTime}{\CompClass{PTime}}
\newcommand{\ExpTime}{\CompClass{ExpTime}}
\newcommand{\PSpace}{\CompClass{PSpace}}
\newcommand{\PSpaceC}{\CompClass{PSpace}-complete\xspace}
\newcommand{\PSpaceH}{\CompClass{PSpace}-hard\xspace}
\newcommand{\ExpSpace}{\CompClass{ExpSpace}}

\newcommand{\Entry}[1]{\begin{tabular}[c]{@{}c@{}} #1\end{tabular}}
\newcommand{\RefEntry}[2]{\Entry{#1\\#2}}

\newcommand{\avg}{\text{\normalfont avg}}

\begin{document}
	\title[Safety and Liveness of Quantitative Properties and Automata]{Safety and Liveness of Quantitative Properties and Automata}
	\titlecomment{The present article combines and extends~\cite{DBLP:conf/fossacs/HenzingerMS23} and~\cite{DBLP:conf/concur/BokerHMS23}.}
	\thanks{
		This work was supported in part by the ERC-2020-AdG 101020093 and the Israel Science Foundation grant 2410/22.
		N. Mazzocchi was affiliated with ISTA when this work was submitted for publication.
	}
	
	\author[U.~Boker]{Udi {Boker}\lmcsorcid{0000-0003-4322-8892}}[a]
	\author[T.~A.~Henzinger]{Thomas A. {Henzinger}\lmcsorcid{0000-0002-2985-7724}}[b]
	\author[N.~Mazzocchi]{Nicolas {Mazzocchi}\lmcsorcid{0000-0001-6425-5369}}[b,c]
	\author[N.~E.~Sara\c{c}]{N. Ege {Sara\c{c}}\lmcsorcid{0009-0000-2866-8078}}[b]
	
	\address{Reichman University, Herzliya, Israel}
	\email{udiboker@runi.ac.il}
	
	\address{Institute of Science and Technology Austria (ISTA), Klosterneuburg, Austria}
	\email{tah@ista.ac.at, esarac@ista.ac.at}
	
	\address{Slovak University of Technology in Bratislava, Slovak Republic}
	\email{nicolas.mazzocchi@stuba.sk}
	
	
	
	
		
	\begin{abstract}
		Safety and liveness stand as fundamental concepts in formal languages, playing a key role in verification. 
		The safety-liveness classification of boolean properties characterizes whether a given property can be falsified by observing a finite prefix of an infinite computation trace (always for safety, never for liveness).
		In the quantitative setting, properties are arbitrary functions from infinite words to partially-ordered domains.
		Extending this paradigm to the quantitative domain, where properties are arbitrary functions mapping infinite words to partially-ordered domains, we introduce and study the notions of quantitative safety and liveness.
		
		First, we formally define quantitative safety and liveness, and prove that our definitions induce conservative quantitative generalizations of both the safety-progress hierarchy and the safety-liveness decomposition of boolean properties.
		Consequently, like their boolean counterparts, quantitative properties can be $\min$-decomposed into safety and liveness parts, or alternatively, $\max$-decomposed into co-safety and co-liveness parts.
		We further establish a connection between quantitative safety and topological continuity and provide alternative characterizations of quantitative safety and liveness in terms of their boolean analogs.
		
		Second, we instantiate our framework with the specific classes of quantitative properties expressed by automata.
		These quantitative automata contain finitely many states and rational-valued transition weights, and their common value functions $\Inf$, $\Sup$, $\LimInf$, $\LimSup$, $\LimInfAvg$, $\LimSupAvg$, and $\DSum$ map infinite words into the totally-ordered domain of real numbers.
		For all common value functions, we provide a procedure for deciding whether a given automaton is safe or live, we show how to construct its safety closure, and we present a $\min$-decomposition into safe and live automata.
	\end{abstract}

	\maketitle
	\newpage
	
	\tableofcontents

	\section{Introduction}	
	
	\Subject{Boolean safety and liveness}
	Safety and liveness are elementary concepts in the semantics of computation~\cite{DBLP:journals/tse/Lamport77}.
	They can be explained through the thought experiment of a \emph{ghost monitor}---an imaginary device that watches an infinite computation trace (word) at runtime, one observation (letter) at a time, and always maintains the set of \emph{possible prediction values} to reflect the satisfaction of a given property.
	Let $\varPhi$ be a boolean property, meaning that $\varPhi$ divides all infinite traces into those that satisfy~$\varPhi$, and those that violate~$\varPhi$.
	After any finite number of observations,~\texttt{True} is a possible prediction value for $\varPhi$ if the observations seen so far are consistent with an infinite trace that satisfies~$\varPhi$, and~\texttt{False} is a possible prediction value for $\varPhi$ if the observations seen so far are consistent with an infinite trace that violates~$\varPhi$.
	When~\texttt{True} is no possible prediction value, the ghost monitor can reject the hypothesis that $\varPhi$ is satisfied.
	The property $\varPhi$ is \emph{safe} if and only if the ghost monitor can always reject a violating hypothesis $\varPhi$ after a finite number of observations.
	Orthogonally, the property $\varPhi$ is \emph{live} if and only if the ghost monitor can never reject a hypothesis $\varPhi$ after a finite number of observations:
	for all infinite traces, after every finite number of observations,~\texttt{True} remains a possible prediction value for~$\varPhi$.
	
	The safety-liveness classification of properties is fundamental in verification.
	In the natural topology on infinite traces---the ``Cantor topology''---the safety properties are the closed sets, and the liveness properties are the dense sets~\cite{DBLP:journals/ipl/AlpernS85}.
	For every property~$\varPhi$, the location of $\varPhi$ within the Borel hierarchy that is induced by the Cantor topology---the so-called ``safety-progress hierarchy''~\cite{ChangMP93}---indicates the level of difficulty encountered when verifying~$\varPhi$.
	On the first level, we find the safety and co-safety properties, the latter being the complements of safety properties, i.e., the properties whose falsehood (rather than truth) can always be rejected after a finite number of observations by the ghost monitor.
	More sophisticated verification techniques are needed for second-level properties, which are the countable boolean combinations of first-level properties---the so-called ``response'' and ``persistence'' properties~\cite{ChangMP93}.
	Moreover, the orthogonality of safety and liveness leads to the following celebrated fact:
	\emph{every} property can be written as the intersection of a safety property and a liveness property~\cite{DBLP:journals/ipl/AlpernS85}.
	This means that every property $\varPhi$ can be decomposed into two parts:
	a safety part---which is amenable to simple verification techniques, such as invariants---and a liveness part---which requires heavier verification paradigms, such as ranking functions.
	Dually, there is always a disjunctive decomposition of $\varPhi$ into co-safety and co-liveness.
	
	\Subject{Quantitative safety and liveness}
	So far, we have retold the well-known story of safety and liveness for \emph{boolean} properties.
	A boolean property $\varPhi$ is formalized mathematically as the \emph{set} of infinite computation traces that satisfy~$\varPhi$, or equivalently, the characteristic \emph{function} that maps each infinite trace to a truth value.
	Quantitative generalizations of the boolean setting allow us to capture not only correctness properties, but also performance properties~\cite{DBLP:conf/concur/HenzingerO13}.
	In this paper we reveal the story of safety and liveness for such \emph{quantitative} properties, which are functions from infinite traces to an arbitrary set $\DD$ of \emph{values}.
	In order to compare values, we equip the value domain $\DD$ with a partial order~$<$, and we require $(\DD,<)$ to be a complete lattice.
	The membership problem~\cite{DBLP:journals/tocl/ChatterjeeDH10} for an infinite trace $w$ and a quantitative property $\varPhi$ asks whether $\varPhi(w)\geq v$ for a given threshold value $v\in\DD$.
	Correspondingly, in our thought experiment, the ghost monitor attempts to reject hypotheses of the form $\varPhi(w)\geq v$, which cannot be rejected as long as all observations seen so far are consistent with an infinite trace $w$ with $\varPhi(w)\geq v$.
	We will define $\varPhi$ to be a \emph{quantitative safety} property if and only if every wrong hypothesis of the form $\varPhi(w)\geq v$ can always be rejected by the ghost monitor after a finite number of observations, and we will define $\varPhi$ to be a \emph{quantitative liveness} property if and only if some wrong hypothesis of the form $\varPhi(w)\geq v$ can never be rejected by the ghost monitor after any finite number of observations.
	We note that in the quantitative case, after every finite number of observations, the set of possible prediction values for $\varPhi$ maintained by the ghost monitor may be finite or infinite, and in the latter case, it may not contain a minimal or maximal element.

	\Subject{Examples}
	Suppose we have four observations:
	observation $\req$ for ``request a resource,''  $\gra$ for ``grant the resource,'' $\tick$ for ``clock tick,'' and $\other$ for ``other.''
	The boolean property \textit{Resp} requires that every occurrence of~$\req$ in an infinite trace is followed eventually by an occurrence of~$\gra$.
	The boolean property \textit{NoDoubleReq} requires that no occurrence of~$\req$ is followed by another~$\req$ without some~$\gra$ in between.
	The quantitative property \textit{MinRespTime} maps every infinite trace to the largest number $k$ such that there are at least $k$ occurrences of~$\tick$ between each~$\req$ and the closest subsequent~$\gra$.
	The quantitative property \textit{MaxRespTime} maps every infinite trace to the smallest number $k$ such that there are at most $k$ occurrences of~$\tick$ between each~$\req$ and the closest subsequent~$\gra$.
	The quantitative property \textit{AvgRespTime} maps every infinite trace to the lower limit value $\liminf$ of the infinite sequence $(v_i)_{i \geq 1}$, where $v_i$ is, for the first $i$ occurrences of~$\tick$, the average number of occurrences of~$\tick$ between~$\req$ and the closest subsequent~$\gra$.
	Note that the values of \textit{AvgRespTime} can be $\infty$ for some computations, including those for which the value of \textit{Resp} is~\texttt{True}.
	This highlights that boolean properties are not embedded in the limit behavior of quantitative properties.
		
	The boolean property \textit{Resp} is live because every finite observation sequence can be extended with an occurrence of~$\gra$.
	In fact, \textit{Resp} is a second-level liveness property (namely, a response property), because it can be written as a countable intersection of co-safety properties. 
	The boolean property \textit{NoDoubleReq} is safe because if it is violated, it will be rejected by the ghost monitor after a finite number of observations, namely, as soon as the ghost monitor sees a~$\req$ followed by another occurrence of~$\req$ without an intervening~$\gra$.
	According to our quantitative generalization of safety, \textit{MinRespTime} is a safety property.
	The ghost monitor always maintains the minimal number $k$ of occurrences of~$\tick$ between any past~$\req$ and the closest subsequent~$\gra$ seen so far; the set of possible prediction values for \textit{MinRespTime} is then $\{0,1,\ldots,k\}$.
	Every hypothesis of the form ``the \textit{MinRespTime}-value is at least~$v$'' is rejected by the ghost monitor as soon as $k<v$; if such a hypothesis is violated, this will happen after some finite number of observations.
	Symmetrically, the quantitative property \textit{MaxRespTime} is co-safe, because every wrong hypothesis of the form ``the \textit{MaxRespTime}-value is at most~$v$'' will be rejected by the ghost monitor as soon as
	the smallest possible prediction value for \textit{MaxRespTime}, which is the maximal number of occurrences of~$\tick$ between any past~$\req$ and the closest subsequent~$\gra$ seen so far, goes above~$v$.
	By contrast, the quantitative property \textit{AvgRespTime} is both live and co-live because no hypothesis of the form ``the \textit{AvgRespTime}-value is at least~$v$,'' nor of the form ``the \textit{AvgRespTime}-value is at most~$v$,'' can ever be rejected by the ghost monitor after a finite number of observations.
	All nonnegative real numbers and $\infty$ always remain possible prediction values for \textit{AvgRespTime}.
	Note that a ghost monitor that attempts to reject hypotheses of the form $\varPhi(w) \geq v$ does not need to maintain the entire set of possible prediction values, but only the $\sup$ of the set of possible prediction values, and whether or not the $\sup$ is contained in the set.
	Dually, updating the $\inf$ (and whether it is contained) suffices to reject hypotheses of the form $\varPhi(w) \leq v$.	
	
	\Subject{Quantitative safety and liveness in automata}
	The notions of safety and liveness consider system properties in full generality: every set of system executions---even the uncomputable ones---can be seen through the lens of the safety-liveness dichotomy.
	To bring these notions more in line with practical requirements, their projections onto formalisms with desirable closure and decidability properties, such as $\omega$-regular languages, have been studied thoroughly in the boolean setting.
	For example,~\cite{DBLP:journals/dc/AlpernS87} gives a construction for the safety closure of a B\"uchi automaton and shows that B\"uchi automata are closed under the safety-liveness decomposition.
	In turn,~\cite{DBLP:journals/fmsd/KupfermanV01} describes an efficient model-checking algorithm for B\"uchi automata that define safety properties.
	
	Similarly to how boolean automata (e.g., regular and $\omega$-regular automata) define classes of boolean properties amenable to boolean verification, quantitative automata (e.g., limit-average and discounted-sum automata) define classes of quantitative properties amenable to quantitative verification.
	Quantitative automata generalize standard boolean automata with weighted transitions and a value function that accumulates an infinite sequence of rational-valued weights into a single real number, a generalization of acceptance conditions of $\omega$-regular automata.
	
	We study the projection of the quantitative safety-liveness dichotomy onto the properties definable by common quantitative automata.
	First, we show how certain attributes of quantitative automata simplify the notions of safety and liveness.
	Then, we use these simplifications to study safety and liveness of the classes of quantitative automata with the value functions $\Inf$, $\Sup$, $\LimInf$, $\LimSup$, $\LimInfAvg$, $\LimSupAvg$, and $\DSum$~\cite{DBLP:journals/tocl/ChatterjeeDH10}.
	In \cref{fig:intro}a, we describe a quantitative automaton using the value function $\LimSup$ to express the long-term maximal power consumption of a device, which is neither safe nor live.
	
	\begin{figure}[t]\centering
		\hspace*{-20pt}
			\noindent\begin{minipage}[c]{.4\linewidth}
				\scalebox{0.9}{
					\begin{tikzpicture}[bg={(a)}, node distance =2cm]
						\node[state, label=center:$p_0$] (0) {};
						\node[yshift=-1cm] (i) at (0) {};
						\node[state, xshift=.5cm, right of = 0, label=center:$p_1$] (1) {};
						\node[state, below of = 1, label=center:$p_2$] (2) {};
						\node[state, below of = 0, label=center:$p_3$] (3) {};
						
						\path[transition]
						(i) edge (0)
						(0) edge[loop left] node[left] {\small on:2} (0)
						(1) edge[bend right=15] node[above] {\small on:2} (0)
						(2) edge[bend right=8, pos=.3] node[right] {\small on:2} (0)
						(0) edge[bend right=0] node[below] {\small off:0} (1)
						(1) edge[loop right] node[right] {\small off:0} (1)
						(2.60) edge[bend right=55, pos=.6] node[left] {\small off:0} (1.300)
						(0) edge[bend right=8,pos=.85] node[yshift=-2pt, left] {\small eco:1} (2)
						(1) edge[bend left=55] node[right] {\small eco:1} (2)
						(2) edge[loop right] node[right] {\small eco:1} (2)
						(0) edge[bend right=55] node[left] {\small err:0} (3)
						(1) edge[bend left=0, pos=.8] node[above left] {\small err:0} (3)
						(2) edge[bend left=15] node[below] {\small err:0} (3)
						(3) edge[loop left] node[left] {\small $\Sigma$:0} (3)
						;	
					\end{tikzpicture}
				}
			\end{minipage}
			\hspace*{-7pt}
			\noindent\begin{minipage}[c]{.21\linewidth}
				\scalebox{0.9}{
					\begin{tikzpicture}[bg={(b)}, node distance =1.8cm]
						\node[state, label=center:$p_0$](0) {};
						\node[state, below of = 0, label=center:$p_3$] (1) {};
						\node[xshift=-1cm] (i) at (0) {};
						
						\path[transition]
						(i) edge (0)
						(0) edge[loop above] node {\small on,eco,off:2} (0)
						(0) edge node[right] {\small err:0} (1)
						(1) edge[loop right] node {\small $\Sigma$:0} (1)
						;
					\end{tikzpicture}
				}
			\end{minipage}
			\hspace*{-7pt}
			\noindent\begin{minipage}[c]{.37\linewidth}\centering
				\scalebox{0.9}{
					\begin{tikzpicture}[bg={(c)}, node distance =2cm]
						\node (center1) {};
						\node[state, label=center:$p_0$, shift=(180:1.5cm)] (0) at (center1) {};
						\node[state, label=center:$p_1$, shift=(300:1.5cm)] (1) at (center1) {};
						\node[state, label=center:$p_2$, shift=(60:1.5cm)] (2) at (center1) {};
						\node[yshift=1cm] (i) at (0) {};
						
						\path[transition]
						(i) edge (0)
						(0) edge[loop below] node {\small \renewcommand*{\arraystretch}{.5}\begin{tabular}{c}on:2\\err:0\end{tabular}} (0)
						(1) edge[bend left=15, pos=.4] node[xshift=-.2cm, below] {\small on:2} (0)
						(2) edge[] node[right] {\small on:2} (0)
						(0) edge[pos=.6] node[xshift=.2cm, above] {\small off:0} (1)
						(1) edge[loop right] node[right] {\small off:0, err:0} (1)
						(2) edge[bend left=15] node[right] {\small off:0} (1)
						(0) edge[bend left=15, pos=.7] node[yshift=.1cm, left] {\small eco:1} (2)
						(1) edge[] node[left] {\small eco:1} (2)
						(2) edge[loop right] node[right] {\small eco:1, err:0} (2)
						;	
					\end{tikzpicture}
				}
			\end{minipage}
		\caption{\label{fig:intro}
			\textbf{(a)} A $\LimSup$-automaton $\A$ modeling the long-term maximal power consumption of a device.
			\textbf{(b)} An $\Inf$-automaton (or a $\LimSup$-automaton) expressing the safety closure of $\A$.
			\textbf{(c)} A $\LimSup$-automaton expressing the liveness component of the decomposition of $\A$.}
	\end{figure}
	
	\Subject{Contributions and overview}
	First, we focus on quantitative properties in their entire generality (\cref{sec:preliminaries1,sec:QuantitativeSafety,sec:Hierarchy,sec:QuantitativeLiveness}).
	We formally introduce quantitative safety as well as safety closure, namely the property that increases the value of each trace as little as possible to achieve safety.
	Then, we prove that our generalization of the boolean setting preserves classical desired behaviors.
	In particular, we show that a quantitative property $\varPhi$ is safe if and only if $\varPhi$ equals its safety closure.
	Moreover, for totally-ordered value domains, a quantitative property is safe if and only if for every value $v$, the set of executions whose value is at least $v$ is safe in the boolean sense.
	We demonstrate a close relation between safety properties and continuous functions with respect to the dual Scott topology of their value domain.
	Pushing further, we define discounting properties on metrizable totally-ordered value domains, characterize them through uniform continuity, and show that they coincide with the conjunction of safety and co-safety.

	We then generalize the safety-progress hierarchy to quantitative properties.
	We first define limit properties.
	For $\Val \in\{\Inf, \Sup, \LimInf, \LimSup\}$, the class of $\Val$-properties captures those for which the value of each infinite trace can be derived by applying the limit function $\Val$ to the infinite sequence of values of finite prefixes.
	We prove that $\Inf$-properties coincide with safety, $\Sup$-properties with co-safety, $\LimInf$-properties are suprema of countably many safety properties, and $\LimSup$-properties infima of countably many co-safety properties.
	The $\LimInf$-properties generalize the boolean persistence properties of~\cite{ChangMP93}; the $\LimSup$-properties generalize their response properties.
	For example, \textit{AvgRespTime} is a $\LimInf$-property.
	
	We continue with introducing quantitative liveness and co-liveness, and prove that their relations with quantitative safety and co-safety further preserve the classical boolean facts.
	In particular, we show that in every value domain there is a unique property which is both safe and live, and then as a central result, we provide a safety-liveness decomposition that holds for every quantitative property, i.e., every quantitative property is the pointwise minimum of a safety and a liveness property.
	We also prove that, like for boolean properties, there exists a liveness-liveness decomposition for every nonunary quantitative property.	
	Moreover, we provide alternative characterizations of liveness for quantitative properties that have the ability to express the least upper bound over their values, namely, supremum-closed.
	For such properties, we show that a property is live iff for every value $v$, the set of executions whose value is at least $v$ is live in the boolean sense.

	Second, we focus on quantitative automata (\cref{sec:preliminaries2,sec:ConstFuncProb,sec:safety,sec:liveness}).
	In contrast to general quantitative properties, these automata use functions on the totally-ordered domain of the real numbers (as opposed to a more general partially-ordered domain).
	Quantitative automata also have the restriction that only finitely many weights (those on the automaton transitions) can contribute to the value of an execution.
	In this setting, we carry the notion of safety (resp. co-safety, discounting) from properties to value functions, and show that a value function is safe (resp. co-safe, discounting) iff every quantitative automaton equipped with this value function expresses a safety (resp. co-safety, discounting) property.
	For example, $\Inf$ is a safe value function, and $\DSum$ is a discounting value function, therefore both safe and co-safe thanks to our characterization in the general setting.
		
	We prove that the considered classes of quantitative automata are supremum-closed.
	Together with the total-order constraint, this helps us simplify the study of their safety and liveness thanks to our alternative characterizations from the first part.
	These simplified characterizations prove useful for checking safety and liveness of quantitative automata, constructing their safety closure, and decomposing them into safety and liveness components.
	
	For example, let us recall the quantitative automaton in \cref{fig:intro}a.
	Since it is supremum-closed, we can construct its safety closure in \PTime by computing the maximal value it can achieve from each state.
	The safety closure of this automaton is shown in \cref{fig:intro}b.
	For the value functions $\Inf$, $\Sup$, $\LimInf$, $\LimSup$, $\LimInfAvg$, and $\LimSupAvg$, the safety closure of a given automaton is an $\Inf$-automaton, while for $\DSum$, it is a $\DSum$-automaton.
	
	Evidently, one can check if a quantitative automaton $\A$ is safe by checking if it is equivalent to its safety closure, i.e., if $\A(w) = \safe{\A}(w)$ for every execution $w$.
	This allows for a \PSpace procedure for checking the safety of $\Sup$-, $\LimInf$-, and $\LimSup$-automata~\cite{DBLP:journals/tocl/ChatterjeeDH10}, but not for $\LimInfAvg$- and $\LimSupAvg$-automata, whose equivalence check is undecidable~\cite{DBLP:conf/csl/DegorreDGRT10,DBLP:conf/concur/ChatterjeeDEHR10,DBLP:journals/tcs/HunterPPR18}.
	For these cases, we use the special structure of the safety-closure automaton for reducing safety checking to the problem of whether an automaton expresses a constant function.
	We show that the latter problem is \PSpaceC for $\LimInfAvg$- and $\LimSupAvg$-automata, by a somewhat involved reduction to the limitedness problem of distance automata, and obtain an \ExpSpace decision procedure for their safety check.
	
	Thanks to our alternative characterization of liveness, one can check if a quantitative automaton $\A$ is live by checking if its safety closure is universal with respect to its maximal value, i.e., if $\safe{\A}(w) \geq \top$ for every execution $w$, where $\top$ is the supremum over the values of $\A$.
	For all value functions we consider except $\DSum$, the safety closure is an $\Inf$-automaton, which allows for a \PSpace solution to liveness checking~\cite{DBLP:conf/vmcai/KupfermanL07,DBLP:journals/tocl/ChatterjeeDH10}, which we show to be optimal.
	Yet, it is not applicable for $\DSum$-automata, as the decidability of their universality check is an open problem. 
	Nonetheless, as we consider only universality with respect to the maximal value of the automaton, we can reduce the problem again to checking whether an automaton expresses a constant function, which we show to be in \PSpace for $\DSum$-automata.
	This yields a \PSpaceC solution to the liveness check of $\DSum$-automata.
	
	Finally, we investigate the safety-liveness decomposition for quantitative automata.
	Recall the automaton from \cref{fig:intro}a and its safety closure from \cref{fig:intro}b.
	The liveness component of the corresponding decomposition is shown in \cref{fig:intro}c.
	Intuitively, it ignores $\err$ and provides information on the power consumption as if the device never fails.
	Then, for every execution $w$, the value of the original automaton on $w$ is the minimum of the values of its safety closure and the liveness component on $w$.
	Since we identified the value functions $\Inf$ and $\DSum$ as safe, their safety-liveness decomposition is trivial.
	For the classes of automata we study, we provide \PTime safety-liveness decompositions.
	Moreover, for deterministic $\Sup$-, $\LimInf$-, and $\LimSup$-automata, we give alternative \PTime decompositions that preserve determinism.
	
	We note that our alternative characterizations of safety and liveness of quantitative properties extend to co-safety and co-liveness.
	Our results for the specific automata classes are summarized in \cref{tbl:Complexity} and most are already implemented~\cite{DBLP:conf/isola/ChalupaHMS24,chalupa2025automatinganalysisquantitativeautomata}.
	While we focus on automata that resolve nondeterminism by $\sup$, their duals hold for quantitative co-safety and co-liveness of automata that resolve nondeterminism by $\inf$, as well as for deterministic automata.
	We leave the questions of co-safety and co-liveness for automata that resolve nondeterminism by $\sup$ open.
	
	\Subject{Related Work}
	To the best of our knowledge, previous definitions of safety and liveness in nonboolean domains make implicit assumptions about the specification language or implicitly use boolean safety and liveness~\cite{DBLP:conf/csl/KatoenSZ14,DBLP:journals/acta/FaranK18,DBLP:journals/tr/QianSCP22,DBLP:conf/cav/BansalV19}.
	We identify three notable exceptions -- \cite{DBLP:conf/atva/WeinerHKPS13,DBLP:journals/isci/LiDL17,DBLP:conf/nfm/GorostiagaS22}.
	
	In 	~\cite{DBLP:conf/atva/WeinerHKPS13}, the authors study a notion of safety for the rational-valued min-plus weighted automata on finite words.
	They take a weighted property as $v$-safe for a given rational $v$ when for every execution $w$, if the hypothesis that the value of $w$ is strictly less than $v$ is wrong (i.e., its value is at least $v$), then there is a finite prefix of $w$ to witness it.
	Then, a weighted property is safe when it is $v$-safe for \emph{some} value $v$.
	Given a nondeterministic weighted automaton $\A$ and an integer $v$, they show that it is undecidable to check whether $\A$ is $v$-safe.
	By contrast, our definition quantifies over \emph{all} values and nonstrict lower-bound hypotheses.
	Moreover, for this definition, we show that checking safety of all common classes of quantitative automata is decidable, even in the presence of nondeterminism.
	
	In~\cite{DBLP:journals/isci/LiDL17}, the authors present a safety-liveness decomposition on multi-valued truth domains, which are bounded distributive lattices.
	Their motivation is to provide algorithms for model-checking properties on multi-valued truth domains.
	While their definitions admit a safety-liveness decomposition, our definition of liveness captures strictly fewer properties, leading to a stronger safety-liveness decomposition theorem. In addition, our definitions also fit naturally with the definitions of emptiness, equivalence, and inclusion for quantitative languages~\cite{DBLP:journals/tocl/ChatterjeeDH10}.

	In~\cite{DBLP:conf/nfm/GorostiagaS22}, the authors generalize the framework of~\cite{DBLP:conf/birthday/PeledH18} to nonboolean value domains.
	Their definitions do not allow for a safety-liveness decomposition since their notion of safety is too permissive and their liveness too restrictive.
	They also do not have a fine-grained classification of nonsafety properties.
	We further elaborate on the relationships between the definitions of~\cite{DBLP:journals/isci/LiDL17,DBLP:conf/nfm/GorostiagaS22} and ours in the relevant sections below.
	
	Our study shows that determining whether a given quantitative automaton expresses a constant function is key to deciding safety and liveness, in particular for automata classes in which equivalence or universality checks are undecidable or open.
	To the best of our knowledge, this problem has not been studied before.

	\begin{table}[t]\centering
		\bgroup \renewcommand*{\arraystretch}{1.2}
		\footnotesize \begin{tabular}{ccccc}
				\toprule
				\phantom{$\big($}& $\Inf$ & $\Sup$, $\LimInf, \LimSup$ & $\LimInfAvg, \LimSupAvg$ & $\DSum$ \\
				\midrule
				\Entry{Safety closure\\construction} &
				$O(1)$ & \multicolumn{2}{c}{\RefEntry{\PTime}{\cref{cl:SafetyClosure,cl:SafetyClosureValDet}} }& $O(1)$ \\
				\midrule
				\Entry{Constant-function\\check}&
				\multicolumn{4}{c}{\RefEntry{\PSpaceC}{\cref{cl:ConstantCheckBasic,cl:ConstantCheckLimAvg,cl:ConstantCheckForDSum}}}\\
				\midrule
				Safety check& 
				$O(1)$ & \multicolumn{1}{c}{\RefEntry{\PSpaceC}{\cref{cl:SafetyCheckBasic}}}
				& \RefEntry{\ExpSpace; \PSpaceH}{\cref{cl:SafetyCheck,cl:SafetyCheckLowerBound}}  & $O(1)$ \\
				\midrule
				Liveness check& \multicolumn{4}{c}{\RefEntry{\PSpaceC}{\cref{cl:LivenessCheck}}}\\
				\midrule
				\Entry{Safety-liveness\\decomposition}& 
				$O(1)$ & \multicolumn{2}{c}{\RefEntry{\PTime}{\cref{cl:SafetyLivenessDecompositionLiminf,cl:SafetyLivenessDecompositionSup,cl:SafetyLivenessDecompositionLimAvg}}}
				& $O(1)$ \\
				\bottomrule

		\end{tabular}
		\egroup
		\caption{The complexity of performing the operations on the left column with respect to nondeterministic automata with the value function specified on the top row.}\label{tbl:Complexity}
	\end{table}

	\section{Quantitative Properties} \label{sec:preliminaries1}
	
	Let $\Sigma = \{a,b,\ldots\}$ be a finite alphabet of letters (observations).
	An infinite (resp. finite) word (trace) is an infinite (resp. finite) sequence of letters $w \in \Sigma^\omega$ (resp. $u \in \Sigma^*$).
	For $n \in \NN$, we denote by $\Sigma^n$ the set of finite words of length $n$.
	Given $u \in \Sigma^*$ and $w \in \Sigma^* \cup \Sigma^\omega$, we write $u \prefix w$ (resp.\ $u \prefixeq w$) when $u$ is a strict (resp.\ nonstrict) prefix of~$w$.
	We denote by $|w|$ the length of $w \in \Sigma^* \cup \Sigma^\omega$ and, given $a \in \Sigma$, by $|w|_a$ the number of occurrences of $a$ in $w$.
	For $w \in \Sigma^* \cup \Sigma^\omega$ and $0 \leq i < |w|$, we denote by $w[i]$ the $i$th letter of $w$.
	
	A \emph{value domain} $\DD$ is a poset.
	We assume that $\DD$ is a nontrivial (i.e., $\bot \neq \top$) complete lattice.
	Whenever appropriate, we write $0$ or $-\infty$ instead of $\bot$ for the least element $\inf \DD$, and $1$ or $\infty$ instead of $\top$ for the greatest element $\sup \DD$.
	We respectively use the terms minimum and maximum for the greatest lower bound and the least upper bound of finitely many elements.
	
	A \emph{quantitative property} is a total function $\varPhi : \Sigma^\omega \to \DD$ from the set of infinite words to a value domain.
	A boolean property $P \subseteq \Sigma^\omega$ is a set of infinite words.
	We use the boolean domain $\BB = \{0,1\}$ with $0 < 1$ and, in place of $P$, its \emph{characteristic property} $\varPhi_P : \Sigma^\omega \to \BB$, which is defined by $\varPhi_P(w) = 1$ if $w \in P$, and $\varPhi_P(w) = 0$ if $w \notin P$. 
	When we say just \emph{property}, we mean a quantitative one.
	
	Given a property $\varPhi$ and a finite word $u \in \Sigma^*$, let $P_{\varPhi, u} = \{\varPhi(uw) \st w \in \Sigma^\omega\}$.
	A property $\varPhi$ is \emph{$\sup$-closed} (resp. \emph{$\inf$-closed}) when for every finite word $u \in \Sigma^*$ we have that $\sup P_{\varPhi, u} \in P_{\varPhi, u}$  (resp. $\inf P_{\varPhi, u} \in P_{\varPhi, u}$).
	
	Given a property $\varPhi : \Sigma^\omega \to \DD$ and a value $v \in \DD$, we define $\varPhi_{\sim v} = \{ w \in \Sigma^\omega \st \varPhi(w) \sim v \}$ for ${\sim} \in \{\leq, \geq, \not\leq, \not\geq\}$.
	The \emph{top value} of a property $\varPhi$ is $\sup_{w \in \Sigma^\omega} \varPhi(w)$, which we denote  by $\top_{\varPhi}$.
	For all properties $\varPhi_1,\varPhi_2$ on a value domain $\DD$ and all words $w \in \Sigma^\omega$, we let $\min(\varPhi_1,\varPhi_2)(w) = \min(\varPhi_1(w),\varPhi_2(w))$ and $\max(\varPhi_1,\varPhi_2)(w) = \max(\varPhi_1(w),\varPhi_2(w))$.
	For a value domain $\DD$, the \emph{inverse} of $\DD$ is the domain $\overline{\DD}$ that contains the same elements as $\DD$ but with the ordering reversed.
	For a property $\varPhi$, we define its \emph{complement} $\overline{\varPhi} : \Sigma^\omega \to \overline{\DD}$ by $\overline{\varPhi}(w) = \varPhi(w)$ for all $w \in \Sigma^\omega$.
	
	Some properties can be defined as limits of value sequences. 
	A \emph{finitary property} $\pi \colon \Sigma^* \rightarrow \DD$ associates a value with each finite word.
	A \emph{value function} $\Val \colon \DD^\omega\rightarrow\DD$ condenses an infinite sequence of values to a single value.
	Given a finitary property $\pi$, a value function $\Val$, and a word $w \in \Sigma^\omega$, we write $\Val_{u \prefix w} \pi(u)$ instead of $\Val(\pi(u_0)\pi(u_1)\ldots)$, where each $u_i$ satisfies $u_i \prefix w$ and $|u_i|=i$.

	\section{Quantitative Safety}\label{sec:QuantitativeSafety}
	A boolean property $P \subseteq \Sigma^\omega$ is safe in the boolean sense iff for every $w \notin P$ there is a prefix $u \prefix w$ with $uw' \notin P$ for all $w' \in \Sigma^\omega$~\cite{DBLP:journals/ipl/AlpernS85}, in other words, every wrong membership hypothesis has a finite witness.
	Given a property $\varPhi : \Sigma^\omega \to \DD$, a trace $w \in \Sigma^\omega$, and a value $v \in \DD$, the quantitative membership problem~\cite{DBLP:journals/tocl/ChatterjeeDH10} asks whether $\varPhi(w) \geq v$.
	We define quantitative safety as follows: the property $\varPhi$ is safe iff every wrong hypothesis of the form $\varPhi(w) \geq v$ has a finite witness $u \prec w$.
	
	\begin{defi}[Safety]
		A property $\varPhi : \Sigma^\omega \to \DD$ is \emph{safe} when for every $w \in \Sigma^\omega$ and value $v \in \DD$ with $\varPhi(w) \not \geq v$, there is a prefix $u \prefix w$ such that $\sup_{w' \in \Sigma^\omega} \varPhi(uw') \not \geq v$.
	\end{defi}
	
	Let us illustrate this definition with the \emph{minimal response-time} property.
	
	\begin{exa} \label{ex:minresp}
		Let $\Sigma = \{\req, \gra, \tick, \other\}$ and $\DD = \NN \cup \{\infty\}$.
		We define the minimal response-time property $\varPhi_{\min}$ through an auxiliary finitary property $\pi_{\min}$ that computes the minimum response time so far.
		In a finite or infinite trace, an occurrence of $\req$ is \emph{granted} if it is followed, later, by a $\gra$, and otherwise it is \emph{pending}.
		Let $\pi_{\text{last}}(u) = \infty$ if the finite trace $u$ contains a pending $\req$, or no $\req$, and $\pi_{\text{last}}(u) = |u'|_\tick - |u''|_\tick$ otherwise, where $u' \prec u$ is the longest prefix of $u$ with a pending $\req$, and $u'' \prec u'$ is the longest prefix of $u'$ without pending $\req$.
		Intuitively, $\pi_{\text{last}}$ provides the response time for the last request when all requests are granted, and $\infty$ when there is a pending request or no request.
		Given $u \in \Sigma^*$, taking the minimum of the values of $\pi_{\text{last}}$ over the prefixes $u' \preceq u$ gives us the minimum response time so far.
		Let $\pi_{\min}(u) = \min_{u' \preceq u} \pi_{\text{last}}(u')$ for all $u \in \Sigma^*$, and $\varPhi_{\min}(w) = \lim_{u \prec w} \pi_{\min}(u)$ for all $w \in \Sigma^\omega$.
		The limit always exists because $\pi_{\min}$ is nonincreasing.
		
		The minimal response-time property is safe.
		Let $w \in \Sigma^\omega$ and $v \in \DD$ such that $\varPhi_{\min}(w) < v$.
		Then, some prefix $u \prec w$ contains a $\req$ that is granted after $v' < v$ ticks, in which case, no matter what happens in the future, the minimal response time is guaranteed to be at most $v'$; that is, $\sup_{w' \in \Sigma^\omega} \varPhi_{\min}(uw') \leq v' < v$.
		Recalling from the introduction the ghost monitor that maintains the $\sup$ of possible prediction values for the minimal response-time property, that value is always $\pi_{\min}$; that is, $\sup_{w' \in \Sigma^\omega} \varPhi_{\min}(uw') = \pi_{\min}(u)$ for all $u \in \Sigma^*$.
		Note that in the case of minimal response time, the $\sup$ of possible prediction values is always realizable; that is, for all $u \in \Sigma^*$, there exists $w \in \Sigma^\omega$ such that $\sup_{w' \in \Sigma^\omega} \varPhi_{\min}(uw') = \varPhi_{\min}(uw)$.
	\end{exa}

	We first show that our definition of safety generalizes the boolean one.
	
	\begin{prop} \label{prop:safetyboolean}
		Quantitative safety generalizes boolean safety.
		In particular, for every boolean property $P \subseteq \Sigma^\omega$, the following statements are equivalent:
		\begin{enumerate}
			\item $P$ is safe according to the classical definition~\cite{DBLP:journals/ipl/AlpernS85}.
			\item The characteristic property $\varPhi_P$ is safe.
			\item For every $w \in \Sigma^\omega$ and $v \in \BB$ with $\varPhi_P(w) < v$, there exists a prefix $u \prefix w$ such that for all $w' \in \Sigma^\omega$, we have $\varPhi_P(uw') < v$.
		\end{enumerate}
	\end{prop}
	\begin{proof}
		Recall that (1) means the following: for every $w \notin P$ there exists $u \prec w$ such that for all $w' \in \Sigma^\omega$ we have $uw' \notin P$.
		Expressing the same statement with the characteristic property $\varPhi_P$ of $P$ gives us for every $w \in \Sigma^\omega$ with $\varPhi_P(w) = 0$ there exists $u \prec w$ such that for all $w' \in \Sigma^\omega$ we have $\varPhi_P(uw') = 0$.
		In particular, since $\BB = \{0,1\}$ and $0 < 1$, we have for every $w \in \Sigma^\omega$ with $\varPhi_P(w) < 1$ there exists $u \prec w$ such that for all $w' \in \Sigma^\omega$ we have $\varPhi_P(uw') < 1$.
		Moreover, since there is no $w \in \Sigma^\omega$ with $\varPhi_P(w) < 0$, we get the equivalence between (1) and (3).
		Now, observe that for every $u \in \Sigma^*$, we have $\varPhi_P(uw') < 1$ for all $w' \in \Sigma^\omega$ iff $\sup_{w' \in \Sigma^\omega} \varPhi_P(uw') < 1$, simply because the domain $\BB$ is a finite total order.
		Therefore, (2) and (3) are equivalent as well.
	\end{proof}
	\noindent 
	Next, we show that safety properties are closed under pairwise $\min$ and $\max$.

	\begin{prop}\label{prop:safe:clorure}
		For every value domain $\DD$, the set of safety properties over $\DD$ is closed under $\min$ and $\max$.
	\end{prop}
	\begin{proof}
		First, consider the two safety properties $\varPhi_1$, $\varPhi_2$ and let $\varPhi$ be their pairwise minimum, i.e., $\varPhi(w) = \min(\varPhi_1(w),\varPhi_2(w))$ for all $w\in \Sigma^\omega$.
		Suppose towards contradiction that $\varPhi$ is not safe, i.e., for some $w \in \Sigma^\omega$ and $v \in \DD$ such that $\varPhi(w) \not \geq v$ and $\sup_{w' \in \Sigma^\omega} \varPhi(uw') \geq v$ for all $u \prec w$.
		Observe that $\varPhi(w) \not \geq v$  implies $\varPhi_1(w) \not \geq v$ or $\varPhi_2(w) \not \geq v$.
		We assume without loss of generality that $\varPhi_1(w) \not \geq v$ holds.
		Thanks to the safety of $\varPhi_1$, there exists $u' \prec w$ such that $\sup_{w' \in \Sigma^\omega} \varPhi_1(u'w') \not \geq v$.
		Since $\varPhi_1(u'w') \geq \varPhi(u'w')$ for all $w' \in \Sigma^\omega$, we have that $\sup_{w' \in \Sigma^\omega} \varPhi_1(u'w') \geq \sup_{u'w' \in \Sigma^\omega}\varPhi(u'w') \geq v$.
		This implies that $\sup_{w' \in \Sigma^\omega} \varPhi_1(u'w') \geq v$, which yields a contradiction.
		
		Now, consider the two safety properties $\varPhi_1$, $\varPhi_2$ and let $\varPhi$ be their pairwise maximum, i.e., $\varPhi(w) = \max(\varPhi_1(w),\varPhi_2(w))$ for all $w\in \Sigma^\omega$.
		Suppose towards contradiction that $\varPhi$ is not safe, i.e., for some $w \in \Sigma^\omega$ and $v \in \DD$, we have $\varPhi(w) \not \geq v$ and $\sup_{w' \in \Sigma^\omega} \varPhi(uw') \geq v$ for all $u \prefix w$.
		Due to safety of both $\varPhi_1$ and $\varPhi_2$, we get for each $i\in\{1,2\}$ the following: for all $w \in \Sigma^\omega$ and $v \in \DD$ if $\varPhi_i(w) \not \geq v$ there is $u_i \prec w$ such that $\sup_{w' \in \Sigma^\omega} \varPhi(u_i w') \not \geq v$.
		Combining the two statements, we get for all $w \in \Sigma^\omega$ and $v \in \DD$ if $\max(\varPhi_1(w), \varPhi_2(w))\not \geq v$, then there exists $u \prec w$ such that $\max(\sup_{w' \in \Sigma^\omega} \varPhi_1(uw'), \sup_{w' \in \Sigma^\omega} \varPhi_2(uw')) \not \geq v$.
		In particular, $\max(\sup_{w' \in \Sigma^\omega} \varPhi_1(uw'), \sup_{w' \in \Sigma^\omega} \varPhi_2(uw')) \not \geq v$ holds since $\max(\varPhi_1(w), \varPhi_2(w)) = \varPhi(w) \not \geq v$.
		Since $\sup (X \cup Y) = \max (\sup X, \sup Y)$ for all $X, Y \subseteq \DD$, we get $$\sup_{w' \in \Sigma^\omega} (\max(\varPhi_1(uw'),\varPhi_2(uw'))) = \max \left( \sup_{w' \in \Sigma^\omega} \varPhi_1(uw'), \sup_{w' \in \Sigma^\omega} \varPhi_2(uw') \right).$$
		Consequently, $$\sup_{w' \in \Sigma^\omega} \max(\varPhi_1(uw'),\varPhi_2(uw')) = \sup_{w' \in \Sigma^\omega} \varPhi(uw') \not \geq v,$$ thus, a contradiction.	
	\end{proof}
	\noindent 
	We now generalize the notion of safety closure and present an operation that makes a property safe by increasing the value of each trace as little as possible.
	
	\begin{defi}[Safety closure]
		The \emph{safety closure} of a property $\varPhi$ is the property $\safe{\varPhi}$ defined by $\safe{\varPhi}(w) = \inf_{u \prec w} \sup_{w' \in \Sigma^\omega} \varPhi(uw')$ for all $w \in \Sigma^\omega$. 
	\end{defi}
	
	We can say the following about the safety closure operation.
	
	\begin{thm}\label{proposition:safe:closure} \label{thm:safe:main}
		For every property $\varPhi : \Sigma^\omega \to \DD$, the following statements hold.
		\begin{enumerate}
			\item $\safe{\varPhi}$ is safe.
			\item $\safe{\varPhi}(w) \geq \varPhi(w)$ for all $w \in \Sigma^\omega$.
			\item $\safe{\varPhi}(w) = \safe{\safe{\varPhi}}(w)$ for all $w \in \Sigma^\omega$.
			\item $\varPhi$ is safe iff $\varPhi(w) = \safe{\varPhi}(w)$ for all $w \in \Sigma^\omega$.
			\item For every safety property $\varPsi : \Sigma^\omega \to \DD$, if $\varPhi(w) \leq \varPsi(w)$ for all $w \in \Sigma^\omega$, then $\safe{\varPhi}(w) \leq \varPsi(w)$ for all $w \in \Sigma^\omega$.
		\end{enumerate}
	\end{thm}
	\begin{proof}
		We first prove that $\sup_{w' \in \Sigma^\omega} \safe{\varPhi}(uw') \leq \sup_{w' \in \Sigma^\omega} \varPhi(uw')$ for all $u\in\Sigma^*$, in other words, $\sup_{w' \in \Sigma^\omega} \inf_{u' \prefix uw'} \sup_{w'' \in \Sigma^\omega} \varPhi(u'w'') \leq \sup_{w' \in \Sigma^\omega} \varPhi(uw')$ for all $u \in \Sigma^*$.
		This will be useful for the proofs of the first and the third items above.
		$$\begin{array}{l}
			\forall u : \sup_{w' \in \Sigma^\omega} \varPhi(uw') \in \{ \sup_{w'' \in \Sigma^\omega} \varPhi(u'w'') \st u' \prefixeq u \}
			\\
			\implies \forall u : \sup_{w' \in \Sigma^\omega} \varPhi(uw') \geq \inf_{u' \prefixeq u} \sup_{w'' \in \Sigma^\omega} \varPhi(u'w'')
			\\
			\implies \forall u : \sup_{w' \in \Sigma^\omega} \varPhi(uw') \geq  \sup_{w'\in\Sigma^\omega} \inf_{u' \prefixeq u} \sup_{w'' \in \Sigma^\omega} \varPhi(u'w'') \hfill (\dagger)
			\medskip\\
			\forall u,t :  \sup_{w' \in \Sigma^\omega} \varPhi(uw') \geq \sup_{w'' \in \Sigma^\omega} \varPhi(u t w'')
			\\
			\implies \forall u : \sup_{w' \in \Sigma^\omega} \varPhi(uw') \geq  \sup_{w'\in\Sigma^\omega} \inf_{t \prefix w'} \sup_{w'' \in \Sigma^\omega} \varPhi(u t w'') \hfill(\ddagger)
			\medskip\\
			(\dagger) \land (\ddagger) \implies  \forall u : \sup_{w' \in \Sigma^\omega} \varPhi(uw') \geq  \sup_{w'\in\Sigma^\omega} \inf_{u' \prefix uw'} \sup_{w'' \in \Sigma^\omega} \varPhi(u'w'')
		\end{array}$$
		
		\begin{enumerate}
		\item
		Now, we prove that $\safe{\varPhi}$ is safe.
		Suppose $\safe{\varPhi}$ is not safe, i.e., there exist $w$ and $v$ for which $\safe{\varPhi}(w) \ngeq v$ and $\sup_{w' \in \Sigma^\omega} \safe{\varPhi}(uw') \geq v$ for all $u \prefix w$.	
		As a direct consequence of the fact that $\sup_{w' \in \Sigma^\omega} \safe{\varPhi}(uw') \leq \sup_{w' \in \Sigma^\omega} \varPhi(uw')$ for all $u\in\Sigma^*$, we have that $\inf_{u \prefix w}\sup_{w' \in \Sigma^\omega} \varPhi(uw') \geq v$.
		It implies that $\safe{\varPhi}(w) \geq v$, which contradicts the hypothesis $\safe{\varPhi}(w) \ngeq v$.
		Hence $\safe{\varPhi}$ is safe.
		
		\item
		Next, we prove that $\safe{\varPhi}(w) \geq \varPhi(w)$ for all $w\in\Sigma^\omega$.
		Given $u \in \Sigma^*$, let $P_{\varPhi,u} = \{\varPhi(uw') \st w' \in \Sigma^\omega\}$.
		Observe that $\safe{\varPhi}(w) = \lim_{u \prec w} (\sup P_{\varPhi,u})$ for all $w\in\Sigma^\omega$.
		Moreover, $\varPhi(w) \in P_{\varPhi,u}$ for each $u \prec w$, and thus $\sup P_{\varPhi,u} \geq \varPhi(w)$ for each $u \prec w$, which implies $\lim_{u \prec w} (\sup P_{\varPhi,u}) \geq \varPhi(w)$, since the sequence of suprema is nonincreasing.
		
		\item
		Next, we prove that $\safe{\varPhi}(w) = \safe{\safe{\varPhi}}(w)$ for all $w\in\Sigma^\omega$.
		Recall from the first paragraph that $\sup_{w' \in \Sigma^\omega} \safe{\varPhi}(uw') \leq \sup_{w' \in \Sigma^\omega} \varPhi(uw')$ for all $u\in\Sigma^*$.
		So, for every $w\in\Sigma^\omega$, we have $\inf_{u \prefix w} \sup_{w' \in \Sigma^\omega} \safe{\varPhi}(uw') \leq \inf_{u \prefix w} \sup_{w' \in \Sigma^\omega} \varPhi(uw')$ and thus $\safe{\safe{\varPhi}}(w) \leq \safe{\varPhi}(w)$ for all $w\in\Sigma^\omega$.
		Since we also have $\safe{\safe{\varPhi}}(w) \geq \safe{\varPhi}(w)$, then the equality holds for all $w\in\Sigma^\omega$.
		
		\item
		Next, we prove that $\varPhi$ is safe iff $\varPhi(w) = \safe{\varPhi}(w)$ for all $w \in \Sigma^\omega$.
		The right-to-left implication follows from the fact that $\safe{\varPhi}$ is safe, as proved above in item (1).
		Now, assume $\varPhi$ is safe, i.e., for all $w \in \Sigma^\omega$ and $v \in \DD$ if $\varPhi(w) \not \geq v$ then there exists $u \prec w$ with $\sup_{w' \in \Sigma^\omega} \varPhi(uw') \not \geq v$.
		Suppose towards contradiction that for some $x \in \Sigma^\omega$ we have $\varPhi(x) < \safe{\varPhi}(x) = \inf_{u \prec x} \sup_{w' \in \Sigma^\omega} \varPhi(uw')$.
		Let $v = \inf_{u \prec x} \sup_{w' \in \Sigma^\omega} \varPhi(uw')$.	
		Since $\varPhi$ is safe and $\varPhi(x) \not \geq v$, there exists $u' \prec x$ such that $\sup_{w' \in \Sigma^\omega} \varPhi(u'w') \not \geq v$.
		Observe that for all $x \in \Sigma^\omega$ and $u_1 \prec u_2 \prec x$ we have $\sup_{w' \in \Sigma^\omega} \varPhi(u_2 w') \leq \sup_{w' \in \Sigma^\omega} \varPhi(u_1 g)$, i.e., the supremum is nonincreasing with longer prefixes.
		Therefore, we have $\inf_{u \prec x} \sup_{w' \in \Sigma^\omega} \varPhi(uw') \leq \sup_{w' \in \Sigma^\omega} \varPhi(u'w')$.
		But since $\sup_{w' \in \Sigma^\omega} \varPhi(u'w') \not \geq v$, we get a contradiction.
		
		\item
		Finally, we prove that $\safe{\varPhi}$ is the least safety property that bounds $\varPhi$ from above.
		Assume there exists a safety property $\varPsi$ such that $\varPhi(w)\leq \varPsi(w)$ holds for all $w\in\Sigma^\omega$.
		Then, for every infinite word $w \in \Sigma^\omega$ and all of its prefixes $u \prec w$ we have $\varPhi(uw') \leq \varPsi(uw')$ for all $w' \in \Sigma^\omega$.
		It implies for every $w \in \Sigma^\omega$ and every $u \prec w$, we have $\sup_{w' \in \Sigma^\omega} \varPhi(uw') \leq \sup_{w' \in \Sigma^\omega} \varPsi(uw')$.
		Then, for every $w \in \Sigma^\omega$, we have $\inf_{u \prefix w} \sup_{w' \in \Sigma^\omega} \varPhi(uw') \leq \inf_{u \prefix w} \sup_{w' \in \Sigma^\omega} \varPsi(uw')$.
		By definition, this is the same as $\safe{\varPhi}(w) \leq \safe{\varPsi}(w)$ for all $w \in \Sigma^\omega$.
		Moreover, since $\varPsi$ is safe, it is equivalent to its safety closure as we proved above, and thus $\safe{\varPhi}(w) \leq \varPsi(w)$ for all $w \in \Sigma^\omega$.
		\qedhere
		\end{enumerate}
	\end{proof}
\noindent 
	We note that a property's safety remains unaffected by the top value of its domain.
	
	\begin{rem} \label{rem:safetydomain}
		Consider a property $\varPhi : \Sigma^\omega \to \DD$.
		If $\varPhi$ is safe, it remains safe after removing (resp. adding) values greater than $\top_{\varPhi}$ from $\DD$ (resp. to $\DD$).
		In particular, consider the value domains $\DD_{\varPhi} = \{ v \in \DD \st v \leq \top_{\varPhi} \}$ and $\DD' = \DD \cup \{\top'\}$ with $v < \top'$ for all $v \in \DD$.
		It is easy to see that if $\varPhi$ is safe, then $\varPhi_1 : \Sigma^\omega \to \DD_{\varPhi}$ and $\varPhi_2 : \Sigma^\omega \to \DD'$ where $\varPhi(w) = \varPhi_1(w) = \varPhi_2(w)$ for all $w \in \Sigma^\omega$ are also safe.
	\end{rem}

	Recall that a safety property allows rejecting wrong lower-bound hypotheses with a finite witness, by assigning a tight upper bound to each trace.
	We define co-safety properties symmetrically: a property $\varPhi$ is co-safe iff every wrong hypothesis of the form $\varPhi(w) \leq v$ has a finite witness $u \prec w$.
	
	\begin{defi}[Co-safety]
		A property $\varPhi : \Sigma^\omega \rightarrow \DD$ is \emph{co-safe} when for every $w\in\Sigma^\omega$ and value $v\in\DD$ with $\varPhi(w) \not\leq v$, there exists a prefix $u \prefix w$ such that $\inf_{w' \in \Sigma^\omega} \varPhi(uw') \not\leq v$.
	\end{defi}
	
	\begin{defi}[Co-safety closure]
		The \emph{co-safety closure} of a property $\varPhi$ is the property $\cosafe{\varPhi}(w)$ defined by $\cosafe{\varPhi}(w) = \sup_{u \prec w} \inf_{w' \in \Sigma^\omega} \varPhi(uw')$ for all $w\in\Sigma^\omega$. 
	\end{defi}

	It is easy to see that safety and co-safety are duals in the following sense.
	
	\begin{thm} \label{thm:cosafe}
		A property $\varPhi : \Sigma^\omega \to \DD$ is safe iff $\overline{\varPhi}$ is co-safe.
	\end{thm}
	
\noindent 
	Thanks to~\cref{thm:cosafe}, the duals of the results above for safety properties and the safety closure operation hold for co-safety properties and the co-safety closure operation.
	To demonstrate, let us define and investigate the \emph{maximal response-time} property.
	
	\begin{exa} \label{ex:maxresp}
		Let $\Sigma = \{\req, \gra, \tick, \other\}$ and $\DD = \NN \cup \{\infty\}$.
		We define the maximal response-time property $\varPhi_{\max}$ through an auxiliary property that computes the current response time for each finite trace.
		In particular, for all $u \in \Sigma^*$, let $\pi_{\text{curr}}(u) = |u|_\tick - |u'|_\tick$, where $u' \preceq u$ is the longest prefix of $u$ without pending $\req$.
		Then, let $\pi_{\max}(u) = \max_{u' \preceq u} \pi_{\text{curr}}(u')$ for all $u \in \Sigma^*$, and $\varPhi_{\max}(w) = \lim_{u \prec w} \pi_{\text{curr}}(u)$ for all $w \in \Sigma^\omega$.
		The limit always exists because $\pi_{\max}$ is nondecreasing.
		Note the contrast between $\pi_{\text{curr}}$ and $\pi_{\text{last}}$ from \Cref{ex:minresp}.
		While $\pi_{\text{curr}}$ takes an optimistic view of the future and assumes the $\gra$ will follow immediately, $\pi_{\text{last}}$ takes a pessimistic view and assumes the $\gra$ will never follow.
		Now, let $w \in \Sigma^\omega$ and $v \in \DD$. 
		If the maximal response time of $w$ is strictly greater than $v$, then for some prefix $u \prec w$ the current response time is strictly greater than $v$ also, which means that, no matter what happens in the future, the maximal response time is strictly greater than $v$ after observing $u$.
		Therefore, $\varPhi_{\max}$ is co-safe.
		By a similar reasoning, the sequence of greatest lower bounds of possible prediction values over the prefixes converges to the property value.
		In other words, we have $\sup_{u \prec w} \inf_{w' \in \Sigma^\omega} \varPhi_{\max}(uw') = \varPhi_{\max}(w)$ for all $w \in \Sigma^\omega$, thus $\varPhi_{\max}$ equals its co-safety closure.
		Now, consider the property $\overline{\varPhi_{\max}}$, which maps every trace to the same value as $\varPhi_{\max}$ on a value domain where the order is reversed.
		It is easy to see that $\overline{\varPhi_{\max}}$ is safe.
		Finally, recall the ghost monitor from the introduction, which maintains the infimum of possible prediction values for the maximal response-time property.
		Since the maximal response-time property is $\inf$-closed, the output of the ghost monitor after every prefix is realizable by some future continuation, and that output is $\pi_{\max}(u) = \max_{u' \preceq u} \pi_{\text{curr}}(u')$ for all $u \in \Sigma^*$.
	\end{exa}
	
	Although minimal and maximal response-time properties are $\sup$- and $\inf$-closed, let us note that safety and co-safety are independent of $\sup$- and $\inf$-closedness.
	
	\begin{prop} \label{cl:SafeCosafeNotSupInfClosed}
		There is a property $\varPhi$ that is safe and co-safe but neither $\sup$- nor $\inf$-closed.
	\end{prop}
	\begin{proof}
		Let $\Sigma = \{a,b\}$ be an alphabet and $\DD = \{v_1, v_2, \bot, \top\}$ be a lattice where $v_1$ and $v_2$ are incomparable.
		Let $\varPhi(w) = v_1$ if $a \prefix w$ and $\varPhi(w)= v_2$ if $b \prefix w$.
		The property $\varPhi$ is safe and co-safe because after observing the first letter, we know the value of the infinite word.
		However, it is not $\sup$-closed since $\sup_{w \in \Sigma^\omega} \varPhi(w) = \top$ but no infinite word has the value $\top$.
		Similarly, it is not $\inf$-closed either.
	\end{proof}

	\subsection{Threshold Safety} \label{sec:thresholdsafety}
	
	In this section, we define threshold safety to connect the boolean and the quantitative settings.
	It turns out that quantitative safety and threshold safety coincide on totally-ordered value domains.
	
	\begin{defi}[Threshold safety]
		A property $\varPhi : \Sigma^\omega \to \DD$ is \emph{threshold safe} when for every $v \in \DD$ the boolean property $\varPhi_{\geq v}$ is safe (and thus $\varPhi_{\not \geq v}$ is co-safe).
		Equivalently, for every $w \in \Sigma^\omega$ and $v \in \DD$ if $\varPhi(w) \not \geq v$ then there exists $u \prec w$ such that for all $w' \in \Sigma^\omega$ we have $\varPhi(uw') \not \geq v$.
	\end{defi}
	
	\begin{defi}[Threshold co-safety]
		A property $\varPhi : \Sigma^\omega \to \DD$ is \emph{threshold co-safe} when for every $v \in \DD$ the boolean property $\varPhi_{\not \leq v}$ is co-safe (and thus $\varPhi_{\leq v}$ is safe).
		Equivalently, for every $w \in \Sigma^\omega$ and $v \in \DD$ if $\varPhi(w) \not \leq v$ then there exists $u \prec w$ such that for all $w' \in \Sigma^\omega$ we have $\varPhi(uw') \not \leq v$.
	\end{defi}

	In general, quantitative safety implies threshold safety, but the converse need not hold with respect to partially-ordered value domains.
	
	\begin{prop} \label{cl:SafeImpliesThresholdSafe} \label{cl:ThresholdSafeDoesNotImplySafe}
		Every safety (resp. co-safety) property is threshold safe (resp. threshold co-safe), but not vice versa.
	\end{prop}
	\begin{proof}
		Consider a property $\varPhi$ over the value domain $\DD$.
		Observe that for all $u \in \Sigma^*$ and all $v \in \DD$, we have that $\sup_{w' \in \Sigma^\omega} \varPhi(uw') \not \geq v$ implies $\varPhi(uw) \not \geq v$ for all $w \in \Sigma^\omega$.
		If $\varPhi$ is safe then, by definition, for every $w \in \Sigma^\omega$ and value $v \in \DD$ if $\varPhi(w) \not \geq v$, there is a prefix $u \prefix w$ such that $\sup_{w' \in \Sigma^\omega} \varPhi(uw') \not \geq v$.
		Thanks to the previous observation, for every $w \in \Sigma^\omega$ and value $v \in \DD$ if $\varPhi(w) \not \geq v$ then there exists $u \prec w$ such that $\varPhi(uw') \not \geq v$ for all $w' \in \Sigma^\omega$.
		Hence $\varPhi$ is threshold safe.
		Proving that co-safety implies threshold co-safety can be done similarly.

		Consider the value domain $\DD = [0,1] \cup \{x\}$ where $x$ is such that $0 < x$ and $x < 1$, but it is incomparable with all $v \in (0,1)$, while within $[0,1]$ there is the standard order.
		Let $\varPhi$ be a property defined over $\Sigma = \{a,b\}$ as follows:
		$\varPhi(w) = x$ if $w=a^\omega$,
		$\varPhi(w) = 2^{-|w|_a}$ if $w \in \Sigma^* b^\omega$, and
		$\varPhi(w) = 0$ otherwise.
		
		First, we show that $\varPhi$ is threshold safe.
		Let $w \in \Sigma^\omega$ and $v \in \DD$.
		If $v = x$, then $\varPhi_{\geq v} = \{a^\omega, b^\omega\}$, which is safe.
		If $v = 0$, then $\varPhi_{\geq v} = \Sigma^\omega$, which is safe as well.
		Otherwise, if $v \in (0,1]$, there exists $n \in \NN$ such that the boolean property $\varPhi_{\geq v}$ contains exactly the words $w'$ such that $|w'|_a \leq n$, which is again safe.
		Therefore $\varPhi$ is threshold safe.
		
		Now, we show that $\varPhi$ is not safe.
		To witness, let $w = a^\omega$ and $v \in (0,1)$.
		Observe that $\varPhi(w) \not \geq v$.
		Moreover, for every prefix $u \prefix w$, there exist continuations $w_1 = a^\omega$ and $w_2 = b^\omega$ such that $\varPhi(u w_1) = x$ and $\varPhi(u w_2) \in (0,1)$.
		Then, it is easy to see that for every prefix $u \prefix w$ we have $\sup_{w' \in \Sigma^\omega} \varPhi(u w') = 1 \geq v$.
		Therefore, $\varPhi$ is not safe.
		Moreover, its complement $\overline{\varPhi}$ is threshold co-safe but not co-safe.
	\end{proof}
	\noindent 
	While safety and threshold safety can differ when considering a single fixed threshold, the two definitions are equivalent on totally-ordered domains since both inherently quantify over all thresholds.
	
	\begin{thm} \label{cl:TotalOrderSafeThresholdSafe}
		Let $\DD$ be a totally-ordered value domain.
		A property $\varPhi : \Sigma^\omega \to \DD$ is safe (resp. co-safe) iff it is threshold safe (resp. threshold co-safe).
	\end{thm}
	\begin{proof}
		We prove only the safety case; the co-safety case follows by duality.
		Consider a property $\varPhi : \Sigma^\omega \to \DD$ where $\DD$ is totally ordered.
		By \cref{cl:SafeImpliesThresholdSafe}, if $\varPhi$ is safe then it is also threshold safe.
		
		For the other direction, having that $\varPhi$ is not safe, i.e., for some $w_1 \in \Sigma^\omega$ and $v_1 \in \DD$ for which $\varPhi(w_1) < v_1$, and every prefix $u_1 \prefix w_1$ satisfies that $\sup_{w \in \Sigma^\omega} \varPhi(u_1 w) \geq v_1$, we exhibit $w_2 \in \Sigma^\omega$ and $v_2 \in \DD$ for which $\varPhi(w_2) < v_2$, and every prefix $u_2 \prec w_2$ admits a continuation $w \in \Sigma^\omega$ such that $\varPhi(u_2 w) \geq v_2$.
		We proceed case by case depending on how $\sup_{w \in \Sigma^\omega} \varPhi(u_1 w) \geq v_1$ holds.
		\begin{itemize}
			\item
			Suppose $\sup_{w \in \Sigma^\omega} \varPhi(u_1 w) > v_1$ for all $u_1 \prefix w_1$.
			Then, let $w_2 = w_1$ and $v_2 = v_1$, and observe that the claim holds since the supremum is either realizable by an infinite continuation or it can be approximated arbitrarily closely.
			
			\item
			Suppose $\sup_{w \in \Sigma^\omega} \varPhi(u_1 w) = v_1$ for some $u_1 \prefix w_1$,
			and for every finite continuation $u_1 \prefixeq r \prefix w_1$ there exists an infinite continuation $w'\in\Sigma^\omega$ such that $\varPhi(r w') = v_1$.
			Then, let $w_2 = w_1$ and $v_2 = v_1$, and observe that the claim holds since the supremum is realizable by some infinite continuation.
			
			\item
			Suppose $\sup_{w \in \Sigma^\omega} \varPhi(u_1 w) = v_1$ for some $u_1 \prefix w_1$,
			and for some finite continuation $u_1 \prefixeq r \prefix w_1$, every infinite continuation $w'\in\Sigma^\omega$ satisfies $\varPhi(r w') < v_1$.
			Let $\hat{r}$ be the shortest finite continuation for which $\varPhi(r w') < v_1$ for all $w'\in\Sigma^\omega$.
			Since $\varPhi(w_1) < v_1$ and $\DD$ is totally ordered, there exists $v_2$ such that $\varPhi(w_1) < v_2 < v_1$.
			We recall that, from the nonsafety of $\varPhi$, all prefixes $u_1 \prefix w_1$ satisfy $\sup_{w \in \Sigma^\omega} \varPhi(u_1 w) \geq v_1 > v_2$.
			Then, let $w_2=w_1$ and $\varPhi(w_1) < v_2 < v_1$, and observe that the claim holds since the supremum can be approximated arbitrarily closely. \qedhere
		\end{itemize}
	\end{proof}
	\noindent 
	Finally, we also show that the two definitions coincide for $\sup$-closed properties.
	
	\begin{prop} \label{cl:SupClosedSafeThresholdSafe}
		Let $\varPhi : \Sigma^\omega \to \DD$ be a $\sup$-closed (resp. $\inf$-closed) property.
		Then, $\varPhi$ is safe (resp. co-safe) iff it is threshold safe (resp. threshold co-safe).
	\end{prop}
	\begin{proof}
		We prove only the safety case; the co-safety case follows by duality.
		Consider a $\sup$-closed property $\varPhi : \Sigma^\omega \to \DD$.
		By \cref{cl:SafeImpliesThresholdSafe}, if $\varPhi$ is safe then it is also threshold safe.
		For the other direction, suppose $\varPhi$ is threshold safe.
		Let $w \in \Sigma^\omega$ and $v \in \DD$ be such that $\varPhi(w) \not\geq v$.
		Then, there exists $u \prec w$ such that $\varPhi(uw') \not\geq v$ for all $w' \in \Sigma^\omega$.
		Since $\varPhi$ is $\sup$-closed, there exists $\hat{w} \in \Sigma^\omega$ with $\varPhi(u\hat{w}) = \sup_{w' \in \Sigma^\omega} \varPhi(uw')$.
		Therefore, we have $\sup_{w' \in \Sigma^\omega} \varPhi(uw') \not\geq v$, implying that $\varPhi$ is safe.
	\end{proof}
	\noindent 

	\subsection{Continuity and Discounting} \label{sec:continuity}
	
	We move next to the relation between safety and continuity.
	We recall some standard definitions; more about them can be found in textbooks, e.g.,~\cite{DBLP:books/sp/trends86/HoogeboomR86,gamelin1999introduction,gierz2003continuous}.
	
	A \emph{topology} of a set $X$ can be defined to be its collection $\tau$ of open subsets, and the pair $(X,\tau)$ stands for a \emph{topological space}.
	It is \emph{metrizable} when there exists a distance function (metric) $d$ on $X$ such that the topology induced by $d$ on $X$ is $\tau$. 
	
	Given a topological space $(X,\tau)$, a set $S \subseteq X$ is closed in $(X,\tau)$ iff its complement $\overline{S} = X \setminus S$ is open in $(X,\tau)$.
	Moreover, given a set $S \subseteq X$, the topological closure $\TopolClosure(S)$ of $S$ is the smallest closed set that contains $S$, and the topological interior $\TopolInterior(S)$ of $S$ is the greatest open set that is contained in $S$.
	
	The \emph{Cantor space} of infinite words is the set $\Sigma^\omega$ with the metric $\mu : \Sigma^\omega \times \Sigma^\omega \to [0,1]$ such that $\mu(w,w) = 0$ and $\mu(w,w') = 2^{-|u|}$ where $u \in \Sigma^*$ is the longest common prefix of $w,w' \in \Sigma^\omega$ with $w \neq w'$.
	Accordingly, a set $P \subset \Sigma^\omega$ is \emph{open} in the Cantor space of infinite words iff for every $w \in P$ there exists a prefix $u \prec w$ such that $u\Sigma^\omega \subseteq P$.
	Let $\DD$ be a value domain and $S \subseteq \DD$ a subset.
	Let ${\uparrow\!S} = \{ y \in \DD \st \exists x \in S : x \leq y \}$ and ${\downarrow\!S} = \{ y \in \DD \st \exists x \in S : y \leq x \}$.
	
	A set $S \subseteq \DD$ is \emph{upward directed} iff for every $x,y \in S$ there is $z \in S$ such that $x \leq z$ and $y \leq z$.
	A set $S \subseteq \DD$ is \emph{Scott open} iff (i) $S = {\uparrow\!S}$, and (ii) $\sup V \in S$ implies $V \cap S \neq \emptyset$ for all upward-directed sets $V \subseteq \DD$.
	A set $S \subseteq \DD$ is \emph{Scott closed} iff its complement $\overline{S}$ is Scott open.
	The \emph{Scott topology} on a complete lattice $\DD$ is the topology induced by the Scott open sets of $\DD$.
	Considering the Scott topology on $\DD$, we have $\TopolClosure(\{v\}) = {\downarrow\!\{v\}}$ for every $v \in \DD$.
	
	The \emph{dual Scott topology} on $\DD$ is the Scott topology on the inverse $\overline{\DD}$ of $\DD$.
	An equivalent definition can be obtained by using the duals of above notions as follows.
	A set $S \subseteq \DD$ is \emph{downward directed} iff for every $x,y \in S$ there is $z \in S$ such that $z \leq x$ and $z \leq y$.
	A set $S \subseteq \DD$ is \emph{dual Scott open} iff (i) $S = {\downarrow\!S}$, and (ii) $\inf V \in S$ implies $V \cap S \neq \emptyset$ for all downward-directed sets $V \subseteq \DD$.
	A set $S \subseteq \DD$ is \emph{dual Scott closed} iff its complement $\overline{S}$ is dual Scott open.
	Then, the \emph{dual Scott topology} on a complete lattice $\DD$ is the topology induced by the dual Scott open sets of $\DD$.
	Considering the dual Scott topology on $\DD$, we have $\TopolClosure(\{v\}) = {\uparrow\!\{v\}}$ for every $v \in \DD$.	
	
	Consider a totally-ordered value domain $\DD$.
	For each element $v \in \DD$, let $L_v = \{v' \in \DD \st v' < v\}$ and $R_v = \{v' \in \DD \st v < v'\}$.
	The \emph{order topology} on $\DD$ is generated by the set $\{L_v \st v \in \DD\} \cup \{R_v \st v \in \DD\}$.
	Moreover, the \emph{left order topology} (resp. \emph{right order topology}) is generated by the set $\{L_v \st v \in \DD\}$ (resp. $\{R_v \st v \in \DD\}$).
	
	For a given property $\varPhi : \Sigma^\omega \to \DD$ and a set $V \subseteq \DD$ of values, the \emph{preimage} of $V$ on $\varPhi$ is defined as $\varPhi^{-1}(V) = \{ w \in \Sigma^\omega \st \varPhi(w) \in V\}$.
	A property $\varPhi : \Sigma^\omega \to \DD$ on a topological space $\DD$ is \emph{continuous} when for every open subset $V \subseteq \DD$ the preimage $\varPhi^{-1}(V) \subseteq \Sigma^\omega$ is open.
	
	In~\cite{DBLP:conf/lics/HenzingerS21}, a property $\varPhi$ is defined as co-continuous when $\varPhi(w) = \lim_{u \prefix w} \sup_{w' \in \Sigma^\omega} \varPhi(u w')$ and as continuous when $\varPhi(w) = \lim_{u \prefix w} \inf_{w' \in \Sigma^\omega} \varPhi(u w')$ for all $w \in \Sigma^\omega$, extending the standard definitions of upper semicontinuity and lower semicontinuity for functions on extended reals to functions from infinite words to complete lattices.	
	Co-continuity and continuity respectively coincide with safety and co-safety properties.
	This characterization holds because each definition is equivalent to a property expressing the same function as its corresponding closure (see \cref{thm:safe:main}).
	We complete the picture by providing a purely topological characterization of safety and co-safety properties in terms of their continuity.

	\begin{thm} \label{cl:SafetyImpliesDSTCont}
		Consider a property $\varPhi : \Sigma^\omega \to \DD$.
		If $\varPhi$ is safe (resp. co-safe), then it is continuous with respect to the dual Scott topology (resp. Scott topology) on $\DD$.
	\end{thm}
	\begin{proof}
		Let $\varPhi : \Sigma^\omega \to \DD$ be a property.
		We prove the statement for safety properties.
		The case of co-safety is dual.
		
		Assume $\varPhi$ is safe.
		Let $S \subseteq \DD$ be an open set and suppose towards contradiction that $\varPhi^{-1}(S) \subseteq \Sigma^\omega$ is not open.
		There exists a word $w \in \varPhi^{-1}(S)$ such that for every prefix $u \prec w$ there exists a continuation $w'$ such that $uw' \notin \varPhi^{-1}(S)$.
		It implies that for each such prefix $u$, we have $\sup_{w' \in \Sigma^\omega} \varPhi(uw') \notin S$.
		For each $i \geq 1$, let $u_i \prec w$ be of length $i$, and consider the set $V = \{ \sup_{w' \in \Sigma^\omega} \varPhi(u_i w') \st u_i \prec w \}$.
		Observe that $V$ is a downward-directed set.
		If $\inf V \in S$, since $S$ is open, we have $V \cap S \neq \emptyset$, i.e., $\sup_{w' \in \Sigma^\omega} \varPhi(u_i w') \in S$ for some $u_i \prec w$. 
		Then, we have $\varPhi(u_i w') \in S$ for all $w' \in \Sigma^\omega$ since $S = {\downarrow\!S}$, which contradicts the supposition that $\varPhi^{-1}(S) \subseteq \Sigma^\omega$ is not open.
		If $\inf V \notin S$, then observe that $\inf V = \safe{\varPhi}(w)$.
		Moreover, $\safe{\varPhi}(w) = \varPhi(w)$ since $\varPhi$ is safe, which implies $\inf V \in S$ since $\varPhi(w) \in S$, which is a contradiction.
		Therefore, $\varPhi^{-1}(S) \subseteq \Sigma^\omega$ is open, and thus $\varPhi$ is continuous.
	\end{proof}

 \noindent 
	The converse does not hold in general essentially due to the fact that the safety closure values may be unrealizable.
	
	\begin{prop} \label{cl:DSTContDoesNotImplySafety}
		There exists a property $\varPhi : \Sigma^\omega \to \DD$ that is continuous with respect to the dual Scott topology (resp. Scott topology) on $\DD$ but not safe (resp. co-safe).
	\end{prop}
	\begin{proof}
		Let us recall the property $\varPhi$ from the proof of \cref{cl:ThresholdSafeDoesNotImplySafe}:
		Consider the value domain $\DD = [0,1] \cup \{x\}$ where $x$ is such that $0 < x$ and $x < 1$, but it is incomparable with all $v \in (0,1)$, while within $[0,1]$ there is the standard order.
		Let $\varPhi$ be a property defined over $\Sigma = \{a,b\}$ as follows:
		$\varPhi(w) = x$ if $w=a^\omega$,
		$\varPhi(w) = 2^{-|w|_a}$ if $w \in \Sigma^* b^\omega$, and
		$\varPhi(w) = 0$ otherwise.
		We showed in the proof of \cref{cl:ThresholdSafeDoesNotImplySafe} that $\varPhi$ is not safe.
		Below, we show that $\varPhi$ is continuous with respect to the dual Scott topology on $\DD$.
		One can symmetrically show that $\overline{\varPhi}$ is continuous with respect to the Scott topology on $\DD$ but not co-safe.
		
		Let us identify the open subsets of $\DD$.
		The sets $\emptyset$ and $\DD$ are open in $\DD$ as they are open in any topology.
		Moreover, notice that every open subset containing $1$ is exactly the entire value domain due to the downward closure requirement.
		Now, consider a subset $S \subseteq \DD$ with $1 \notin S$.
		We argue that if $S$ is open, it is either of the form $[0,r)$ or $[0,r) \cup \{x\}$ for some $r \in (0,1]$.
		
		First, consider the case when $x \notin S$.
		Notice that again due to the downward closure requirement the set $S$ must contain an interval $I \subseteq [0,1]$ with  $0 \in I$.
		Moreover, the interval $I$ cannot contain its upper bound.
		Suppose towards contradiction that $I = [0,r]$ for some $r \in [0,1]$.
		If $r = 1$, then $S = [0,1]$, which is not open because it violates the downward closure requirement since $x \notin S$.
		If $r < 1$, then $S = [0,r]$, which is not open because $V = (r,1]$ is a downward-directed set with $\inf V = r \in S$ but $V \cap S = \emptyset$.
		Therefore, if $x \notin S$, then $S$ is of the form $[0,r)$ for some $r \in (0,1]$.
		For the case of $x \in S$, notice that the inclusion of $x$ in $S$ does not affect the downward closure requirement.
		Moreover, the only downward-directed sets whose infimum is $x$ are $\{x\}$ and $\{x,1\}$, and their intersection with $S$ is not empty as $x \in S$.
		Therefore, if $x \in S$, then $S$ is of the form $[0,r) \cup \{x\}$ for some $r \in (0,1]$.
		
		Now, let us show that $\varPhi$ is continuous.
		If $S = \emptyset$ (resp. $\DD$), then we have $\varPhi^{-1}(S) = \emptyset$ (resp. $\Sigma^\omega$), which is evidently open in the Cantor topology of $\Sigma^\omega$.
		Suppose $S = [0,r)$ for some $r \in (0,1]$.
		Let $k_r = \min \{ k \in \NN \st 2^{-k} < r \}$.
		Then, observe that $\varPhi^{-1}(S)$ is exactly the set of infinite words $w$ where $w$ contains at least $k_r$ occurrences of $a$ and at least one $b$, which is an intersection of two open sets, and thus open.
		Finally, suppose $S = [0,r) \cup \{x\}$ for some $r \in (0,1]$.
		Let $k_r$ be as above, and notice that $\varPhi^{-1}(S)$ is exactly the set of infinite words $w$ where $w$ contains at least $k_r$ occurrences of $a$, which is open.
		Therefore, $\varPhi$ is continuous.
	\end{proof}
	\noindent 
	Next, we examine the relation between threshold safety and continuity with respect to the dual Scott topology. We show in particular that continuity implies threshold safety.
	\begin{thm} \label{cl:DSTContImpliesThresholdSafety}
		Consider a property $\varPhi : \Sigma^\omega \to \DD$.
		If $\varPhi$ is continuous with respect to the dual Scott topology (resp. Scott topology) on $\DD$, then it is threshold safe (resp. threshold co-safe).
	\end{thm}
	\begin{proof}
		Let $\varPhi : \Sigma^\omega \to \DD$ be a property.
		We prove the statement for safety properties.
		The case of co-safety is dual.
		
		Assume $\varPhi$ is continuous, i.e., for every open set $S \subseteq \DD$ the preimage $\varPhi^{-1}(S)$ is open.
		We want to show that $\varPhi$ is threshold safe, i.e., $\varPhi_{\not\geq v} = \{w \in \Sigma^\omega \st \varPhi(w) \not \geq v\}$ is co-safe in the boolean sense for every $v \in \DD$.
		Let $v \in \DD$ and notice that $\varPhi_{\not\geq v} = \varPhi^{-1}(\overline{\uparrow\!\{v\}})$.
		Since the set $\overline{\uparrow\!\{v\}}$ is open in $\DD$ and $\varPhi$ is continuous, its preimage $\varPhi_{\not\geq v}$ is open in $\Sigma^\omega$, i.e., co-safe in the boolean sense.
		Therefore, $\varPhi$ is threshold safe.
	\end{proof}
	\noindent 
	Moreover, we establish that the inclusion is strict: there is a threshold safety property that is not continuous with respect to the dual Scott topology.
	
	\begin{prop} \label{cl:ThresholdSafetyDoesNotImplyDSTCont}
		There exists a property $\varPhi : \Sigma^\omega \to \DD$ that is threshold safe (resp. threshold co-safe) but not continuous with respect to the dual Scott topology (resp. Scott topology) on~$\DD$.
	\end{prop}
	\begin{proof}
		Let $\Sigma = \{a,b\}$ be a finite alphabet.
		Consider the value domain $\DD = \Sigma^\omega \cup \{\bot, \top\}$ where for every $x \in \DD$ we have $\top \geq x$ and $x \geq \bot$, but the elements from $\Sigma^\omega$ are incomparable with each other.
		Let $\varPhi : \Sigma^\omega \to \DD$ be such that $\varPhi(w) = w$ for all $w \in \Sigma^\omega$.
		
		We show that $\varPhi$ is threshold safe, i.e., for every $w \in \Sigma^\omega$ and every $v \in \DD$ with $\varPhi(w) \not\geq v$ there exists $u \prec w$ such that for every $w' \in \Sigma^\omega$ we have $\varPhi(uw') \not\geq v$.
		Let $w \in \Sigma^\omega$ and $v \in \DD$.
		If $v = \top$, the finite witness for $\varPhi(w) \not\geq \top$ is the empty word since no infinite word has the value $\top$.
		If $v < \top$, we have $\varPhi(w) \not\geq v$ iff $v \in \Sigma^\omega$ and $w \neq v$ since $\varPhi$ is the identity function on $\Sigma^\omega$ and the elements from $\Sigma^\omega$ are incomparable.
		Observe that two infinite words are distinct iff there is a finite word that is a prefix of one and not the other.
		Then, such a prefix of $w$ is the finite witness for $\varPhi(w) \not\geq v$.
		Therefore, $\varPhi$ is threshold safe.
		
		We show that $\varPhi$ is not continuous with respect to the dual Scott topology on $\DD$.
		Let $P \subseteq \Sigma^\omega$ be a set of infinite words.
		First, we argue that $S = P \cup \{\bot\}$ is open in $\DD$.
		The set $S$ is downward closed because for every $w \in P$ the only element smaller than $w$ is $\bot$, which is in $S$.
		Let $V$ be a downward-directed subset of $\DD$.
		If $\bot \in V$, then $\inf V = \bot \in S$ and we have $V \cap S \neq \emptyset$.
		If $\bot \notin V$, then $V$ contains at most one element from $\Sigma^\omega$ (otherwise we would have $\bot \in V$ since $V$ is downward directed).
		If $V$ contains no elements from $\Sigma^\omega$, then it is either $\emptyset$ or $\{\top\}$, and thus $\inf V = \top \notin S$.
		If $V$ contains some element $w$ from $\Sigma^\omega$, we have $\inf V = w$.
		Moreover, if $\inf V \in S$, then clearly $w \in S$ and thus $V \cap S \neq \emptyset$.
		
		Now, let $P = \Sigma^* a^\omega$.
		As we proved above, the set $S = P \cup \{\bot\}$ is open in $\DD$.
		However, its preimage $\varPhi^{-1}(S)$ is exactly the set $P$, which is not open in $\Sigma^\omega$.
		Therefore, $\varPhi$ is not continuous.
	
		The property $\varPhi$ above and the same arguments also cover the case of co-safety.
	\end{proof}
	\noindent 
	An immediate result of \cref{cl:SafetyImpliesDSTCont,cl:DSTContImpliesThresholdSafety} is that whenever safety and threshold safety coincide, they also coincide with continuity with respect to the dual Scott topology.
	In particular, thanks to \cref{cl:SupClosedSafeThresholdSafe}, we obtain the following.
	
	\begin{cor} \label{cl:SupClosedDSTContImpliesSafety}
		Consider a $\sup$-closed (resp. $\inf$-closed) property $\varPhi : \Sigma^\omega \to \DD$.
		Then, $\varPhi$ is safe (resp. co-safe) iff it is continuous with respect to the dual Scott topology (resp. Scott topology) on $\DD$.
	\end{cor}
\noindent 
	Moreover, for totally-ordered value domains $\DD$, it is well known that a property is continuous with respect to the dual Scott topology (resp. Scott topology) on $\DD$ iff it is continuous with respect to the left order topology (resp. right order topology) on $\DD$, which coincides with upper semicontinuity (resp. lower semicontinuity) when $\DD = \RR \cup \{-\infty, +\infty\}$.
	Then, thanks to \cref{cl:TotalOrderSafeThresholdSafe}, we get the following.

	\begin{cor} \label{cl:TotalOrderDSTContImpliesSafety}
		Let $\DD$ be a totally-ordered value domain.
		A property $\varPhi : \Sigma^\omega \to \DD$ is safe (resp. co-safe) iff it is continuous with respect to the left order topology (resp. right order topology) on $\DD$.
	\end{cor}
	\noindent 

	Finally, since a property is continuous with respect to the order topology on $\DD$ iff it is continuous with respect to both left and right order topologies on $\DD$, we immediately obtain the following.
	
	\begin{cor} \label{cl:SafeAndCosafeIffContinuous}
		Let $\DD$ be a totally-ordered value domain.
		A property $\varPhi : \Sigma^\omega \to \DD$ is safe and co-safe iff it is continuous with respect to the order topology on $\DD$.
	\end{cor}
	\noindent 
	Now, we shift our focus to totally-ordered value domains whose order topology is metrizable.
	We provide a general definition of discounting properties on such domains.
	
	\begin{defi}[Discounting]\label{def:Discounting}
		Let $\DD$ be a totally-ordered value domain for which the order topology is metrizable with a metric $d$.
		A property $\varPhi : \Sigma^\omega \to \DD$ is \emph{discounting} when for every $\varepsilon > 0$ there exists $n \in \NN$ such that for every $u \in \Sigma^n$ and $w,w' \in \Sigma^\omega$ we have $d(\varPhi(uw),\varPhi(uw')) < \varepsilon$.
	\end{defi}
	
	Intuitively, a property is discounting when the range of potential values for every word converges to a singleton.
	As an example, consider the following discounted safety property:
	Given a boolean safety property $P$, let $\varPhi$ be a quantitative property such that $\varPhi(w) = 1$ if $w \in P$, and $\varPhi(w) = 2^{-|u|}$ if $w \notin P$, where $u \prefix w$ is the shortest bad prefix of $w$ for $P$. 
	We remark that our definition captures the previous definitions of discounting given in~\cite{DBLP:conf/icalp/AlfaroHM03,DBLP:conf/tacas/AlmagorBK14}.

	\begin{rem} \label{rem:heinecantor}
		Notice that the definition of discounting coincides with uniform continuity.
		Since $\Sigma^\omega$ equipped with Cantor distance is a compact space, every continuous property is also uniformly continuous by Heine-Cantor theorem, and thus discounting.
	\end{rem}
	
	As an immediate consequence, we obtain the following.
	
	\begin{cor} \label{cl:SafeAndCosafeIffDiscountingProp}
		Let $\DD$ be a totally-ordered value domain for which the order topology is metrizable.
		A property $\varPhi : \Sigma^\omega \to \DD$ is safe and co-safe iff it is discounting.
	\end{cor}
\noindent 
	Let $P \subseteq \Sigma^\omega$ be a boolean property.
	Recall that $\TopolClosure(P)$ is the smallest boolean safety property that contains $P$, and $\TopolInterior(P)$ of $P$ is the greatest boolean co-safety property that is contained in $P$.
	To conclude this subsection, we show the connection between the quantitative safety closure (resp. co-safety closure) and the topological closure (resp. topological interior) through $\sup$-closedness (resp. $\inf$-closedness).
	The $\sup$-closedness assumption makes the quantitative safety closure values realizable.
	This guarantees that for every value $v$, every word whose safety closure value is at least $v$ belongs to the topological closure of the set of words whose property values are at least $v$.
	Similarly, the $\inf$-closedness assumption helps in the case of co-safety and topological interior.
	
	\begin{thm} \label{cl:TopologicalClosureAndInterior}
		Consider a property $\varPhi : \Sigma^\omega \to \DD$ and a threshold $v \in \DD$.
		If $\varPhi$ is $\sup$-closed, then $(\safe{\varPhi})_{\geq v} = \TopolClosure(\varPhi_{\geq v})$.
		If $\varPhi$ is $\inf$-closed, then $(\cosafe{\varPhi})_{\leq v} = \TopolInterior(\varPhi_{\leq v})$.
	\end{thm}
	\begin{proof}
		First, we observe that for all $u \in \Sigma^*$, if $\sup_{w' \in \Sigma^\omega} \varPhi(uw') \not \geq v$ then for every $w \in \Sigma^\omega$, we have $\varPhi(uw) \not \geq v$.
		Next, we show that $\TopolClosure(\varPhi_{\geq v}) \subseteq (\safe{\varPhi})_{\geq v}$.
		Suppose towards contradiction that there exists $w \in  \TopolClosure(\varPhi_{\geq v}) \setminus (\safe{\varPhi})_{\geq v}$, that is, $\safe{\varPhi}(w) \not \geq v$ and $w \in \TopolClosure(\varPhi_{\geq v})$.
		This means that (i) $\inf_{u \prec w} \sup_{w' \in \Sigma^\omega} \varPhi(u w') \not \geq v$, and (ii) for every prefix $u \prefix w$ there exists $w' \in \Sigma^\omega$ such that $\varPhi(u w') \geq v$.
		By the above observation, (i) implies that there exists a prefix $u' \prefix w$ such that for all $w'' \in \Sigma^\omega$ we have $\varPhi(u' w'') \not \geq v$, which contradicts (ii).
		
		Now, we show that if $\varPhi$ is $\sup$-closed then $(\safe{\varPhi})_{\geq v} \subseteq \TopolClosure(\varPhi_{\geq v})$.
		Suppose towards contradiction that there exists $w \in (\safe{\varPhi})_{\geq v} \setminus \TopolClosure(\varPhi_{\geq v})$, that is, $\safe{\varPhi}(w) \geq v$ and $w \notin \TopolClosure(\varPhi_{\geq v})$.
		By the duality between closure and interior, we have $w \in \TopolInterior(\varPhi_{\not \geq v})$.
		Then, (i) $\inf_{u \prec w} \sup_{w' \in \Sigma^\omega} \varPhi(u w') \geq v$, and (ii) there exists $u' \prefix w$ such that for all $w'' \in \Sigma^\omega$ we have $\varPhi(u' w'') \not \geq v$.
		Since $\varPhi$ is $\sup$-closed, (i) implies that for every prefix $u \prefix w$ there exists $w' \in \Sigma^\omega$ such that $\varPhi(u w') \geq v$, which contradicts (ii).
		
		Proving that if $\varPhi$ is $\inf$-closed then $(\cosafe{\varPhi})_{\leq v} = \TopolInterior(\varPhi_{\leq v})$ can be done similarly, based on the observation that for all $u \in \Sigma^*$, if $\inf_{w' \in \Sigma^\omega} \varPhi(uw') \not \leq v$ then for every word $w \in \Sigma^\omega$, we have $\varPhi(uw) \not \leq v$.
	\end{proof}

	\subsection{Additional Notions Related to Quantitative Safety}
	
	In~\cite{DBLP:journals/isci/LiDL17}, the authors consider the model-checking problem for properties  on multi-valued truth domains.
	They introduce the notion of multi-safety through a closure operation that coincides with our safety closure.
	Formally, a property $\varPhi$ is \emph{multi-safe} iff $\varPhi(w) = \safe{\varPhi}(w)$ for every $w \in \Sigma^\omega$.
	By \cref{thm:safe:main}, we immediately obtain the following.
	
	\begin{prop} \label{cl:multisafe}
		A property is multi-safe iff it is safe.
	\end{prop}
	\noindent 
	Although the two definitions of safety are equivalent, our definition is consistent with the membership problem for quantitative properties and motivated by their monitoring.
	
	In~\cite{DBLP:conf/nfm/GorostiagaS22}, the authors extend a refinement of the safety-liveness classification for monitoring~\cite{DBLP:conf/birthday/PeledH18} to richer domains.
	They introduce the notion of verdict-safety through dismissibility of values not less than or equal to the property value.
	Formally, a property $\varPhi$ is \emph{verdict-safe} iff for every $w \in \Sigma^\omega$ and $v \not \leq \varPhi(w)$, there exists a prefix $u \prec w$ such that for all $w' \in \Sigma^\omega$, we have $\varPhi(uw') \neq v$. 
	
	We demonstrate that verdict-safety is weaker than safety.
	Moreover, we provide a condition under which the two definitions coincide.
	To achieve this, we reason about sets of possible prediction values:
	for a property $\varPhi$ and $u \in \Sigma^*$, let $P_{\varPhi,u} = \{\varPhi(uw) \st w \in \Sigma^\omega\}$.
	
	\begin{lem}\label{lem:GS:characterization}
		A property $\varPhi$ is verdict-safe iff $\varPhi(w) = \sup (\lim_{u \prec w} P_{\varPhi,u})$ for all $w \in \Sigma^\omega$. 
	\end{lem}
	\begin{proof}
		For all $w \in \Sigma^\omega$ let us define $P_w = \lim_{u \prec w} P_{\varPhi,u} = \bigcap_{u \prec w} P_{\varPhi,u}$.
		Assume $\varPhi$ is verdict-safe and suppose towards contradiction that $\varPhi(w) \neq \sup P_w$ for some $w \in \Sigma^\omega$.
		If $\varPhi(w) \not \leq \sup P_w$, then $\varPhi(w) \notin P_w$, which is a contradiction.
		Otherwise, if $\varPhi(w) < \sup P_w$, there exists $v \not \leq \varPhi(w)$ with $w \in P_w$.
		It means that there is no $u \prec w$ that dismisses the value $v \not \leq \varPhi(w)$, which contradicts the fact that $\varPhi$ is verdict-safe.
		Therefore, $\varPhi(w) = \sup P_w$ for all $w \in \Sigma^\omega$.
		
		We prove the other direction by contrapositive.
		Assume $\varPhi$ is not verdict-safe, i.e., for some $w \in \Sigma^\omega$ and $v \not \leq \varPhi(w)$, every $u \prec w$ has an extension $w' \in \Sigma^\omega$ with $\varPhi(uw') = v$.
		Equivalently, for some $w \in \Sigma^\omega$ and $v \not \leq \varPhi(w)$, every $u \prec w$ satisfies $v \in P_{\varPhi,u}$.
		Then, $v \in P_w$, but since $v \not \leq \varPhi(w)$, we have $\sup P_w > \varPhi(w)$.
	\end{proof}
	\noindent 
	Notice that $\varPhi$ is safe iff $\varPhi(w) = \lim_{u \prec w} (\sup P_{\varPhi,u})$ for all $w \in \Sigma^\omega$, thanks to \cref{thm:safe:main}.
	Below we describe a property that is verdict-safe but not safe.
	
	\begin{exa}
		Let $\Sigma = \{a,b\}$.
		Define $\varPhi$ by $\varPhi(w) = 0$ if $w = a^\omega$, and $\varPhi(w) = |u|$ otherwise, where $u \prec w$ is the shortest prefix in which $b$ occurs.
		The property $\varPhi$ is verdict-safe.
		First, observe that $\DD = \NN \cup \{\infty\}$.
		Let $w \in \Sigma^\omega$ and $v \in \DD$ with $v > \varPhi(w)$.
		If $\varPhi(w) > 0$, then $w$ contains $b$, and $\varPhi(w) = |u|$ for some $u \prec w$ in which $b$ occurs for the first time.
		After the prefix $u$, all $w' \in \Sigma^\omega$ yield $\varPhi(uw') = |u|$, thus all values above $|u|$ are rejected.
		If $\varPhi(w) = 0$, then $w = a^\omega$.
		Let $v \in \NN$ with $v > 0$, and consider the prefix $a^v \prec w$.
		Observe that the set of possible prediction values after reading $a^v$ is $\{0, v+1, v+2, \ldots\}$, therefore $a^v$ allows the ghost monitor to reject the value $v$.
		However, $\varPhi$ is not safe because, although $\varPhi(a^\omega) = 0$, for every $u \prec a^\omega$, we have $\sup_{w' \in \Sigma^\omega} \varPhi(uw') = \infty$.
		\end{exa}
	
	The separation is due to the fact that for some finite traces, the $\sup$ of possible prediction values cannot be realized by any future.
	This is not the case for the minimal response-time property $\varPhi_{\min}$ from \Cref{ex:minresp} because for every $u \in \Sigma^*$ the continuation $\gra^\omega$ realizes the value $\sup_{w' \in \Sigma^\omega} \varPhi_{\min}(uw')$, and thus $\varPhi_{\min}$ is $\sup$-closed.
	
	Recall from the introduction the ghost monitor that maintains the $\sup$ of possible prediction values.
	For monitoring $\sup$-closed properties this suffices; otherwise the ghost monitor also needs to maintain whether or not the supremum of the possible prediction values is realizable by some future continuation.
	In general, we have the following for every $\sup$-closed property.
	
	\begin{lem}\label{lemma:value:closed}
		Let $\varPhi$ be a $\sup$-closed property. Then, $\lim_{u \prec w} (\sup P_{\varPhi, u}) = \sup (\lim_{u \prec w} P_{\varPhi,u})$ for all $w \in \Sigma^\omega$.
	\end{lem}
	\begin{proof}
		Note that $\lim_{u \prec w} (\sup P_{\varPhi, u}) \geq \sup (\lim_{u \prec w} P_{\varPhi,u})$ holds in general, and we want to show that $\lim_{u \prec w} (\sup P_{\varPhi, u}) \leq \sup (\lim_{u \prec w} P_{\varPhi,u})$ holds for every $\sup$-closed $\varPhi$.
		Let $w \in \Sigma^\omega$.
		Since the sequence $(P_{\varPhi,u})_{u \prec w}$ of sets is nonincreasing and $\sup P_{\varPhi,u} \in P_{\varPhi,u}$ for every $u \in \Sigma^*$ (thanks to $\sup$-closedness of $\varPhi$), we have $\sup P_{\varPhi,u'} \in P_{\varPhi,u}$ for every $u,u' \in \Sigma^*$ with $u \preceq u'$. 
		Moreover, $\lim_{u \prec w} (\sup P_{\varPhi,u}) \in P_{\varPhi,u'}$ for every $u' \in \Sigma^*$ with $u' \prec w$.
		Then, by definition, we have $\lim_{u \prec w} (\sup P_{\varPhi,u}) \in \lim_{u \prec w} P_{\varPhi,u}$, and therefore $\lim_{u \prec w} (\sup P_{\varPhi,u}) \leq \sup (\lim_{u \prec w} P_{\varPhi,u})$.
	\end{proof}
	\noindent 
	As a consequence of the above, we get the following.
	
	\begin{thm}
		Every safety property is verdict-safe, but not vice versa.
		Moreover, a $\sup$-closed property is safe iff it is verdict-safe.
	\end{thm}
\noindent 
	Let us conclude with a remark on the form of hypotheses in our definition of safety.
	
	\begin{rem} \label{rem:strictsafety}
		Suppose we define safety with strict lower bound hypotheses instead of nonstrict: for every $w \in \Sigma^\omega$ and value $v \in \DD$ with $\varPhi(w) \not > v$, there is a prefix $u \prefix w$ such that $\sup_{w' \in \Sigma^\omega} \varPhi(uw') \not > v$.
		Let $w$ be an arbitrary word and consider $v = \varPhi(w)$.
		It is clear that this definition would require the $\sup$ of possible prediction values to converge to $\varPhi(w)$ after a finite prefix, which is too restrictive.
	\end{rem}
	
	\section{The Quantitative Safety-Progress Hierarchy}\label{sec:Hierarchy}

	The safety-progress classification of boolean properties~\cite{ChangMP93} is a Borel hierarchy built from the Cantor topology of traces.
	Safety and co-safety properties lie on the first level, respectively corresponding to closed sets and open sets.
	The second level is obtained through countable unions and intersections of properties from the first level:
	persistence properties are countable unions of closed sets, while response properties are countable intersections of open sets.
	We generalize this construction to the quantitative setting.
	
	In the boolean case, each property class is defined through an operation that takes a set $S \subseteq \Sigma^*$ of finite traces and produces a set $P \subseteq \Sigma^\omega$ of infinite traces.
	For example, to obtain a co-safety property from $S \subseteq \Sigma^*$, the corresponding operation yields $S\Sigma^\omega$.
	Similarly, we formalize each property class by a value function.
	
	\begin{defi}[Limit property]
		A property $\varPhi : \Sigma^\omega \to \DD$ is a \emph{limit property} when there exists a finitary property $\pi : \Sigma^* \rightarrow \DD$ and a value function $\Val : \DD^\omega \to \DD$ such that $\varPhi(w) = \Val_{u \prec w} \pi(u)$ for all $w \in \Sigma^\omega$.
		We denote this by $\varPhi = (\pi,\Val)$.
		In particular, if $\varPhi = (\pi,\Val)$ for $\Val \in \{ \Inf, \Sup, \LimInf, \LimSup\}$, then $\varPhi$ is a \emph{$\Val$-property}.
	\end{defi}

	\begin{rem}
		Every quantitative property $\varPhi : \Sigma^\omega \to \DD$ where $|\Sigma| \leq |\DD|$ is a limit property because $\pi$ can encode infinite words through their prefixes and $\Val$ can map each infinite sequence (corresponding to a unique infinite word) to the desired value.
		Below, we focus on particular value functions (namely $\Inf, \Sup, \LimInf, \LimSup$) for which this is not possible.
	\end{rem}
	
	To account for the value functions that construct the first two levels of the safety-progress hierarchy, we start our investigation with $\Inf$- and $\Sup$-properties and later focus on $\LimInf$- and $\LimSup$- properties.
	
	\subsection{Infimum and Supremum Properties}
	
	Let us start by showing that $\Inf$-properties are closed under countable infima.
	
	\begin{prop} \label{cl:ClosureOfInfProp}
		Every countable infimum of $\Inf$-properties is an $\Inf$-property.
	\end{prop}
	\begin{proof}
		Let $\varPhi_i = (\pi_i, \Inf)$ be for each $i \in \NN$.
		Let $\varPhi = (\pi, \Inf)$ where $\pi(u) = \inf_{i \in \NN} \pi_i(u)$ for all $u \in \Sigma^*$.
		Let $w \in \Sigma^\omega$ be arbitrary.
		We have $\varPhi(w) = \Inf_{u \prec w} \inf_{i \in \NN} \pi_i(u) = \inf_{i \in \NN} \Inf_{u \prec w} \pi_i(u) = \inf_{i \in \NN} \varPhi_i(w)$.
		
		We show below that $\Inf_{u \prec w} \inf_{i \in \NN} \pi_i(u) = \inf_{i \in \NN} \Inf_{u \prec w} \pi_i(u)$ holds.
		Note that we can assume without loss of generality that for each $i \in \NN$, the finitary property $\pi_i$ is nonincreasing.
		For each $i \in \NN$, let $x_i = \Inf_{u \prec w} \pi_i(u)$.
		For each $u \prec w$, let $y_{|u|} = \inf_{i \in \NN} \pi_i(u)$.
		Moreover, let $x = \inf_{i \in \NN} x_i$ and $y = \inf_{j \in \NN} y_j$.
		Let us denote by $u_j$ the prefix of $w$ of length $j$.
		For all $i,j \in \NN$, we have $x \leq x_i \leq \pi_i(u_j)$ and $y \leq y_j \leq \pi_i(u_j)$.
		Then, $x$ and $y$ are lower bounds on the set $P = \{ \pi_i(u_j) \st i,j \in \NN \}$.
		Now, let $z$ be another lower bound, i.e., $z \leq \pi_i(u_j)$ for all $i,j \in \NN$.
		For a fixed $i \in \NN$, we still have $z \leq \pi_i(u_j)$ for all $j \in \NN$.
		It means that $z$ is a lower bound on the sequence $(\pi_i(u))_{u \prec w}$ and since $x_i$ is the infimum of this sequence, we have $z \leq x_i$.
		Moreover, since this holds for any $i \in \NN$ and $x = \inf_{i \in \NN} x_i$, we have $z \leq x$.
		By similar arguments, we obtain $z \leq y$.
		It implies that both $x$ and $y$ are the greatest lower bound on $P$, which means $x = y$ due to the uniqueness of greatest lower bound.
	\end{proof}
	\noindent 
	Next, we demonstrate that the minimal response-time property is an $\Inf$-property.
	
	\begin{exa} \label{ex:infresp}
		Recall the safety property $\varPhi_{\min}$ of minimal response time from \Cref{ex:minresp}.
		We can equivalently define $\varPhi_{\min}$ as a limit property by taking the finitary property $\pi_{\text{last}}$ and the value function $\Inf$.
		As discussed in \Cref{ex:minresp}, the function $\pi_{\text{last}}$ outputs the response time for the last request when all requests are granted, and $\infty$ when there is a pending request or no request.
		Then $\Inf_{u \prec w} \pi_{\text{last}}(u) = \varPhi_{\min}(w)$ for all $w \in \Sigma^\omega$, and therefore $\varPhi_{\min} = (\pi_{\text{last}}, \Inf)$.
		\end{exa}
	
	In fact, the safety properties coincide with $\Inf$-properties.
	
	\begin{thm}\label{thm:safe:inf}
		A property $\varPhi$ is safe iff it is an $\Inf$-property.
	\end{thm}
	\begin{proof}
		Assume $\varPhi$ is safe.
		By \Cref{thm:safe:main}, we have $\varPhi(w) = \inf_{u \prec w} \sup_{w' \in \Sigma^\omega} \varPhi(uw')$ for all $w \in \Sigma^\omega$.
		Then, simply taking $\pi(u) = \sup_{w' \in \Sigma^\omega} \varPhi(uw')$ for all $u \in \Sigma^*$ yields that $\varPhi$ is an $\Inf$-property.
		
		Now, assume $\varPhi$ is an $\Inf$-property, and suppose towards contradiction that $\varPhi$ is not safe.
		In other words, let $\varPhi = (\pi, \Inf)$ for some finitary property $\pi : \Sigma^* \to \DD$ and suppose $\Inf_{u \prec x} \sup_{w' \in \Sigma^\omega} \varPhi(uw') > \varPhi(x) = \Inf_{u \prec x} \pi(u)$ for some $x \in \Sigma^\omega$.
		Let $u \in \Sigma^*$ and note that $\sup_{w' \in \Sigma^\omega} \varPhi(uw') = \sup_{w' \in \Sigma^\omega} (\Inf_{u' \prec uw'} \pi(u'))$ by definition.
		Moreover, for every $w' \in \Sigma^\omega$, notice that $\Inf_{u' \prec uw'} \pi(u') \leq \pi(u)$ since $u \prec uw'$.
		Then, we obtain $\sup_{w' \in \Sigma^\omega} \varPhi(uw') \leq \pi(u)$ for every $u \in \Sigma^*$.
		In particular, this is also true for all $u \prec x$.
		Therefore, we get $\Inf_{u \prec x} \sup_{w' \in \Sigma^\omega} \varPhi(uw') \leq \Inf_{u \prec x} \pi(u)$, which contradicts to our initial supposition.
	\end{proof}
\noindent 
	Notice that \cref{cl:ClosureOfInfProp,thm:safe:inf} imply a stronger closure result than \cref{prop:safe:clorure}: safety properties are closed under countable infima.
	
	Defining the minimal response-time property as a limit property, we observe the following relation between its behavior on finite traces and infinite traces.
	
	\begin{exa}
		Consider the property $\varPhi_{\min} = (\pi_{\text{last}}, \Inf)$ from \Cref{ex:infresp}.
		Let $w \in \Sigma^\omega$ and $v \in \DD$.
		Observe that if the minimal response time of $w$ is at least $v$, then the last response time for each prefix $u \prec w$ is also at least $v$.
		Conversely, if the minimal response time of $w$ is below $v$, then there is a prefix $u \prec w$ for which the last response time is also below $v$.
		\end{exa}
	
	In light of this observation, we provide another characterization of safety properties, explicitly relating the specified behavior of the limit property on finite and infinite traces.	

	\begin{thm}\label{thm:safe:inflooking}
		A property $\varPhi:\Sigma^\omega\rightarrow\DD$ is safe iff $\varPhi = (\pi, \Val)$ such that for every $w \in \Sigma^\omega$ and value $v \in \DD$, we have $\varPhi(w) \geq v$ iff $\pi(u) \geq v$ for all $u \prec w$.
	\end{thm}
	\begin{proof}
		Assume $\varPhi$ is safe.
		Then, we know by \Cref{thm:safe:inf} that $\varPhi$ is an $\Inf$-property, i.e., $\varPhi = (\pi, \Inf)$ for some finitary property $\pi : \Sigma^* \to \DD$, and thus a limit property.
		Suppose towards contradiction that for some $w \in \Sigma^\omega$ and $v \in \DD$ we have (i) $\varPhi(w) \geq v$ and $\pi(u) \not \geq v$ for some $u \prec w$, or (ii) $\varPhi(w) \not \geq v$ and $\pi(u) \geq v$ for every $u \prec w$.
		One can easily verify that (i) yields a contradiction, since if for some $u \prec w$ we have $\pi(u) \not \geq v$ then $\Inf_{u \prec w} \pi(u) = \varPhi(w) \not \geq v$.
		Similarly, (ii) also yields a contradiction, since if $\varPhi(w) = \Inf_{u \prec w} \pi(u) \not \geq v$ then there exists $u \prec w$ such that $\pi(u) \not \geq v$.
		
		Now, assume $\varPhi = (\pi,\Val)$ for some finitary property $\pi$ and value function $\Val$ such that for every $w \in \Sigma^\omega$ and value $v \in \DD$ we have $\varPhi(w) \geq v$ iff $\pi(u) \geq v$ for every $u \prec w$.
		We claim that $\varPhi(w) = \Inf_{u \prec w} \pi(u)$ for every $w \in \Sigma^\omega$.
		Suppose towards contradiction that the equality does not hold for some trace.
		If $\varPhi(w) \not \geq \Inf_{u \prec w} \pi(u)$ for some $w \in \Sigma^\omega$, let $v = \Inf_{u \prec w} \pi(u)$ and observe that (i) $\varPhi(w) \not \geq v$, and (ii) $\Inf_{u \prec w} \pi(u) \geq v$.
		However, while (i) implies $\pi(u) \not \geq v$ for some $u \prec w$ by hypothesis, (ii) implies $\pi(u) \geq v$ for all $u \prec w$, resulting in a contradiction.
		The case where $\varPhi(w) \not \leq \Inf_{u \prec w} \pi(u)$ for some $w \in \Sigma^\omega$ is similar.
		It means that $\varPhi$ is an $\Inf$-property.
		Therefore, $\varPhi$ is safe by \Cref{thm:safe:inf}.
	\end{proof}
	\noindent 
	Finally, observe that the maximal response-time property is a $\Sup$-property.
	As $\Sup$-properties and $\Inf$-properties are dual, $\Sup$-properties are closed under countable suprema (see \cref{cl:ClosureOfInfProp}).
	Thanks to the duality between safety and co-safety, we also obtain the following characterizations.
	
	\begin{thm} \label{thm:cosafeSup}
		For every property $\varPhi : \Sigma^\omega \to \DD$, the following are equivalent.
		\begin{enumerate}
			\item $\varPhi$ is co-safe.
			\item $\varPhi$ is a $\Sup$-property.
			\item $\varPhi = (\pi, \Val)$ such that for every $w \in \Sigma^\omega$ and value $v \in \DD$, we have $\varPhi(w) \leq v$ iff $\pi(u) \leq v$ for all $u \prec w$.
		\end{enumerate}
	\end{thm}

	\subsection{Limit Inferior and Limit Superior Properties}
	
	Let us start with an observation on the minimal response-time property.
	
	\begin{exa}
		Recall once again the minimal response-time property $\varPhi_{\min}$ from \Cref{ex:minresp}.
		In the previous subsection, we presented an alternative definition of $\varPhi_{\min}$ to establish that it is an $\Inf$-property.
		Observe that there is yet another equivalent definition of $\varPhi_{\min}$	which takes the nonincreasing finitary property $\pi_{\min}$ from \Cref{ex:minresp} and pairs it with either the value function $\LimInf$, or with $\LimSup$.
		Hence $\varPhi_{\min}$ is both a $\LimInf$- and a $\LimSup$-property.
		\end{exa}
	
	Before moving on to investigating $\LimInf$- and $\LimSup$-properties more closely, we show that the above observation can be generalized.
	
	\begin{thm}\label{thm:inf:liminf:limsup}
		For each $\Val\in\{\Inf, \Sup\}$, every $\Val$-property is both a $\LimInf$- and a $\LimSup$-property.
	\end{thm}
	\begin{proof}
		Let $\varPhi = (\pi, \Inf)$ and define an alternative finitary property as follows:
		$\pi'(u) = \min_{u' \preceq u} \pi(u)$.
		One can confirm that $\pi'$ is nonincreasing and thus $\lim_{u \prec w} \pi'(u) = \Inf_{u \prec w} \pi(u)$ for every $w \in \Sigma^\omega$.
		Then, letting $\varPhi_1 = (\pi',\LimInf)$ and $\varPhi_2 = (\pi',\LimSup)$, we obtain that $\varPhi(w) = \varPhi_1(w) = \varPhi_2(w)$ for all $w \in \Sigma^\omega$.
		For $\Val = \Sup$ we use $\max$ instead of $\min$.
	\end{proof}

	\noindent 
	An interesting response-time property beyond safety and co-safety arises when we remove extreme values: instead of minimal response time, consider the property that maps every trace to a value that bounds from below, not all response times, but all of them from a point onward (i.e., all but finitely many).
	We call this property  \emph{tail-minimal response time}.
	
	\begin{exa} \label{ex:liminfresp}
		Let $\Sigma = \{\req,\gra,\tick,\other\}$ and $\pi_{\text{last}}$ be the finitary property from \Cref{ex:minresp} that computes the last response time.
		We define the tail-minimal response-time property as $\varPhi_{\text{tmin}} = (\pi_{\text{last}},\LimInf)$.
		Intuitively, it maps each trace to the least response time over all but finitely many requests.
		This property is interesting as a performance measure, because it focuses on the long-term performance by ignoring finitely many outliers.	
		Consider $w \in \Sigma^\omega$ and $v \in \DD$.
		Observe that if the tail-minimal response time of $w$ is at least $v$, then there is a prefix $u \prec w$ such that for all longer prefixes $u \preceq u' \prec w$, the last response time in $u'$ is at least $v$, and vice versa.
		\end{exa}
	
	Similarly as for $\Inf$-properties, we characterize $\LimInf$-properties through a relation between property behaviors on finite and infinite traces.
	
	\begin{thm}\label{thm:liminf:liminflooking}
		A property $\varPhi:\Sigma^\omega\rightarrow\DD$ is a $\LimInf$-property iff $\varPhi = (\pi, \Val)$ such that for every $w \in \Sigma^\omega$ and value $v \in \DD$, we have $\varPhi(w) \geq v$ iff there exists $u \prec w$ such that for all $u \preceq u' \prec w$, we have $\pi(u') \geq v$.
	\end{thm}
	\begin{proof}
		Assume $\varPhi$ is a $\LimInf$-property, i.e., $\varPhi = (\pi,\LimInf)$ for some finitary property $\pi : \Sigma^* \to \DD$.
		Suppose towards contradiction that for some $w \in \Sigma^\omega$ and $v \in \DD$ we have
		(i) $\varPhi(w) \geq v$ and for all $u \prec w$ there exists $u \preceq u' \prec w$ such that $\pi(u') \not \geq v$, or
		(ii) $\varPhi(w) \not \geq v$ and there exists $u \prec w$ such that for all $u \preceq u' \prec w$ we have $\pi(u') \geq v$.
		One can easily verify that (i) yields a contradiction, since if for all $u \prec w$ there exists $u \preceq u' \prec w$ with $\pi(u') \not \geq v$, then $\LimInf_{u \prec w} \pi(u) = \varPhi(w) \not \geq v$.
		Similarly, (ii) also yields a contradiction, since if there exists $u \prec w$ such that for all $u \preceq u' \prec w$ we have $\pi(u') \geq v$ then $\LimInf_{u \prec w} \pi(u) = \varPhi(w) \geq v$.
		
		Now, assume $\varPhi = (\pi,\Val)$ for some finitary property $\pi$ and value function $\Val$ such that for every $w \in \Sigma^\omega$ and value $v \in \DD$ we have $\varPhi(w) \geq v$ iff there exists $u \prec w$ such that for all $u \preceq u' \prec w$ we have $\pi(u') \geq v$.
		We claim that $\varPhi(w) = \LimInf_{u \prec w} \pi(u)$ for every $w \in \Sigma^\omega$.
		Suppose towards contradiction that the equality does not hold for some trace.
		If $\varPhi(w) \not \geq \LimInf_{u \prec w} \pi(u)$ for some $w \in \Sigma^\omega$, let $v = \LimInf_{u \prec w} \pi(u)$ and observe that (i) $\varPhi(w) \not \geq v$, and (ii) $\LimInf_{u \prec w} \pi(u) \geq v$.
		However, by hypothesis, (i) implies that for all $u \prec w$ there exists $u \preceq u' \prec w$ with $\pi(u') \not \geq v$, which means that $\LimInf_{u \prec w} \pi(u) \not \geq v$, resulting in a contradiction to (ii).
		The case where $\varPhi(w) \not \leq \LimInf_{u \prec w} \pi(u)$ for some $w \in \Sigma^\omega$ is similar.
		Therefore, $\varPhi$ is a $\LimInf$-property.
	\end{proof}
	\noindent 
	Next, we show that $\LimInf$-properties are closed under pairwise minimum.
	
	\begin{prop}\label{prop:liminf:closure}
		For every value domain $\DD$, the set of $\LimInf$-properties over $\DD$ is closed under $\min$.
	\end{prop}
	\begin{proof}
		Consider two $\LimInf$-properties $\varPhi_1 = (\pi_1, \LimInf)$, $\varPhi_2 = (\pi_2, \LimInf)$ and let $\varPhi$ be as follows: $\varPhi = (\pi, \LimInf)$ where $\pi(u) = \min (\pi_1(u), \pi_2(u))$ for all $u\in \Sigma^*$.
		We now prove that $\varPhi(w) = \min(\varPhi_1(w), \varPhi_2(w))$ for all $w \in \Sigma^\omega$.
		
		Suppose towards contradiction that $\min(\varPhi_1(w), \varPhi_2(w)) \ngeq \varPhi(w)$ for some $w\in\Sigma^\omega$.
		Observe that for all $w'\in\Sigma^\omega$ and $v\in\DD$, if $\min(\varPhi_1(w'), \varPhi_2(w')) \ngeq v$ then $\varPhi_1(w') \ngeq v$ or $\varPhi_2(w') \ngeq v$.
		We assume without loss of generality that $\varPhi_1(w) \ngeq \varPhi(w)$.
		By \Cref{thm:liminf:liminflooking}, $\varPhi_1(w) \ngeq \varPhi(w)$ implies that for all $u' \prec w$ there exists $u' \prefixeq u'' \prefix w$ such that $\pi_1(u'') \ngeq \varPhi(w)$.
		Dually, since $\varPhi(w) \geq \varPhi(w)$, there exists $t \prec w$ such that $\pi(t') \geq \varPhi(w)$ for all $t \prefixeq t' \prefix w$.
		In particular, there exists $t \prefixeq t'' \prefix w$ such that $\pi_1(t'') \ngeq \varPhi(w)$ and $\pi(t'') \geq \varPhi(w)$.
		By the definition of $\min$, we have that $\pi_1(t'') \geq \pi(t'') \geq \varPhi(w)$ which contradicts that $\pi_1(t'') \ngeq \varPhi(w)$.
		Hence, we proved that $\min(\varPhi_1(w), \varPhi_2(w)) \geq \varPhi(w)$ for all $w\in\Sigma^\omega$.
		
		Suppose towards contradiction that $\varPhi(w) \ngeq \min(\varPhi_1(w), \varPhi_2(w))$ for some $w\in\Sigma^\omega$.
		In particular, $\LimInf_{u \prec w} \min(\pi_1(u), \pi_2(u)) \ngeq \min(\varPhi_1(w), \varPhi_2(w))$.
		Observe that for all $u\in\Sigma^*$ and $v\in\DD$, if $\min(\pi_1(u), \pi_2(u)) \ngeq v$ then $\pi_1(u) \ngeq v$ or $\pi_2(u) \ngeq v$.
		We assume without loss of generality that $|\{u \st \exists u \prefixeq u' \prefix w : \pi_1(u') \ngeq \min(\varPhi_1(w), \varPhi_2(w))\}| = \infty$, or equivalently for all $u \prec w$, there exists $u \prefixeq u' \prefix w$ such that $\pi_1(u') \ngeq \min(\varPhi_1(w), \varPhi_2(w))$.
		By \Cref{thm:liminf:liminflooking}, we get $\varPhi_1(w) \ngeq \min(\varPhi_1(w), \varPhi_2(w))$.
		By the definition of $\min$, we have that $\varPhi_1(w) \geq \min(\varPhi_1(w), \varPhi_2(w))$ which contradicts that $\varPhi_1(w) \ngeq \min(\varPhi_1(w), \varPhi_2(w))$.
		Hence, we proved that $\varPhi(w) \geq \min(\varPhi_1(w), \varPhi_2(w))$ for all $w\in\Sigma^\omega$.
	\end{proof}
	\noindent 
	Now, we show that the tail-minimal response-time property can be expressed as a countable supremum of $\Inf$-properties.
	
	\begin{exa}
		Let $i \in \NN$ and define $\pi_{i,\text{last}}$ as a finitary property that imitates $\pi_{\text{last}}$ from \Cref{ex:minresp}, but ignores the first $i$ observations of every finite trace.
		Formally, for all $u \in \Sigma^*$, we define $\pi_{i,\text{last}}(u) = \pi_{\text{last}}(u')$ if $u = u_i u'$ where $u_i \preceq u$ with $|u_i| = i$ and $u' \in \Sigma^*$, and $\pi_{i,\text{last}}(u) = \infty$ otherwise.
		Observe that an equivalent way to define $\varPhi_{\text{tmin}}$ from \Cref{ex:liminfresp} is $\sup_{i \in \NN} (\Inf_{u \prec w} (\pi_{i,\text{last}}(u)))$ for all $w \in \Sigma^\omega$.
		Intuitively, for each $i \in \NN$, we obtain an $\Inf$-property that computes the minimal response time of the suffixes of a given trace.
		Taking the supremum over these, we obtain the greatest lower bound on all but finitely many response times.
		\end{exa}
	
	We generalize this observation and show that every $\LimInf$-property is a countable supremum of $\Inf$-properties.
	
	\begin{thm}\label{thm:liminf:supinf}
		Every $\LimInf$-property is a countable supremum of $\Inf$-properties.
	\end{thm}
	\begin{proof}
		Let $\varPhi = (\pi,\LimInf)$.
		For each $i \in \NN$ let us define $\varPhi_i = (\pi_i,\Inf)$ where $\pi_i$ is as follows:
		$\pi_i(u) = \top$ if $|u| < i$, and $\pi_i(u) = \pi(u)$ otherwise.
		We claim that $\varPhi(w) = \sup_{i \in \NN} \varPhi_i(w)$ for all $w \in \Sigma^\omega$.
		Expanding the definitions, observe that the claim is $\LimInf_{u \prec w} \pi(u) = \sup_{i \in \NN} \Inf_{u \prec w} \pi_i(u)$.
		Due to the definition of $\LimInf$, the expression $\sup_{i \in \NN} \Inf_{u \prec w \land |u| \geq i} \pi(u)$ equals the left-hand side.
		Moreover, by the definition of $\pi_i$, it equals the right-hand side.
	\end{proof}
	\noindent 
	We would also like to have the converse of \Cref{thm:liminf:supinf}, i.e., that every countable supremum of $\Inf$-properties is a $\LimInf$-property.
	Currently, we are able to show only the following.
	
	\begin{prop}\label{cl:liminf:leq}
		Consider an infinite sequence $(\varPhi_i)_{i \in \NN}$ of properties with $\varPhi_i = (\pi_i, \Inf)$ for each $i \in \NN$.
		The property $\varPhi = (\pi, \LimInf)$ where $\pi(u) = \max_{i \leq |u|} \pi_i(u)$ for all $u \in \Sigma^*$ satisfies $\sup_{i \in \NN} \varPhi_i(w) \leq \varPhi(w)$ for all $w \in \Sigma^\omega$.
	\end{prop}
	\begin{proof}
		For each $i \in \NN$, assume without loss of generality that each $\pi_i$ is nonincreasing.
		Let $\varPhi = (\pi, \LimInf)$ be as in the statement.
		We want to show that $\sup_{i \in \NN} \varPhi_i(w) \leq \varPhi(w)$ for all $w \in \Sigma^\omega$.
		Expanding the definitions, observe that the claim is the following: $\sup_{i \in \NN}( \Inf_{u \prec w} \pi_i(u)) \leq \LimInf_{u \prec w} (\max_{i \leq |u|} \pi_i(u))$ for all $w \in \Sigma^\omega$.
		
		Let $w \in \Sigma^\omega$, and for each $k \in \NN$, let $n_k = \max_{i \leq k} \Inf_{u \prec w} \pi_i(u)$ and $m_k = \max_{i \leq k} \pi_i(u_k)$ where $u_k \prec w$ with $|u_k| = k$.
		Observe that we have $n_k \leq m_k$ for all $k \in \NN$.
		Then, we have $\liminf_{k \to \infty} n_k \leq \liminf_{k \to \infty} m_k$.
		Moreover, since the sequence $(n_k)_{k \in \NN}$ is nondecreasing, we can replace the $\liminf$ on the left-hand side with $\lim$ to obtain the following:
		$\lim_{k \to \infty} \max_{i \leq k} \Inf_{u \prec w} \pi_i(u) \leq \liminf_{k \to \infty} \max_{i \leq k} \pi_i(u_k)$. 
		Then, rewriting the expression concludes the proof by giving us  $\sup_{i \in \NN} (\Inf_{u \prec w} \pi_i(u)) \leq \LimInf_{u \prec w} (\max_{i \leq |u|} \pi_i(u))$. 
	\end{proof}
	
	\begin{rem}
		Consider an infinite sequence $(\varPhi_i)_{i \in \NN}$ of finitely-converging $\Inf$-properties, i.e., for every $i \in \NN$ and every infinite word $w$ there is a prefix $u \prec w$ such that $\varPhi_i(w) = \varPhi(u w')$ for all continuations $w'$.
		Evidently, each $\varPhi_i$ is also a $\Sup$ property.
		Moreover, since $\Sup$-properties are closed under countable suprema, $\sup_{i \in \NN} \varPhi_i$ is a $\Sup$-property, and thus a $\LimInf$-property by \cref{thm:inf:liminf:limsup}.
	\end{rem}
	
	We conjecture that some $\LimInf$-property that is an upper bound like in \Cref{cl:liminf:leq} is also a lower bound on the countable supremum that occurs in the theorem.
	(The property $\varPhi$ in \Cref{cl:liminf:leq} is not one.)
	This, together with \Cref{cl:liminf:leq}, would imply the converse of \Cref{thm:liminf:supinf}.
	Proving the converse of \Cref{thm:liminf:supinf} would give us, thanks to the following duality, that the $\LimInf$- and $\LimSup$-properties respectively characterize the countable suprema of $\Inf$-properties and countable infima of $\Sup$-properties, completing the picture for the generalization of the safety-progress hierarchy to the quantitative setting.

	\begin{prop}
		A property $\varPhi$ is a $\LimInf$-property iff its complement $\overline{\varPhi}$ is a $\LimSup$-property.
	\end{prop}
	
	\section{Quantitative Liveness}\label{sec:QuantitativeLiveness}
	
	A boolean property $P \subseteq \Sigma^\omega$ is live in the boolean sense iff for every $u \in \Sigma^*$ there exists $w \in \Sigma^\omega$ with $uw \in P$~\cite{DBLP:journals/ipl/AlpernS85}, in other words, a wrong membership hypothesis can never be dismissed by a finite prefix.
	Similarly as for safety, we take the perspective of the quantitative membership problem to define liveness:
	a property $\varPhi$ is live iff whenever a property value is less than $\top$, there exists a value $v$ for which the wrong hypothesis $\varPhi(w) \geq v$ can never be dismissed by any finite witness $u \prec w$.
	
	\begin{defi}[Liveness]
		A property $\varPhi : \Sigma^\omega \to \DD$ is \emph{live in $\DD$} when for all $w \in \Sigma^\omega$, if $\varPhi(w) < \top$, then there exists a value $v \in \DD$ such that $\varPhi(w) \not \geq v$ and for all prefixes $u \prec w$, we have $\sup_{w' \in \Sigma^\omega} \varPhi(uw') \geq v$.
	\end{defi}
	
	When we write that a property $\varPhi : \Sigma^\omega \to \DD$ is \emph{live} (instead of live in $\DD$), we mean that $\varPhi$ is live in the value domain $\DD_{\varPhi} = \{v \in \DD \st v \leq \top_\varPhi\}$, and we let $\top = \sup \DD_{\varPhi}$.
	This is motivated by the following remark showing that a property's liveness may be closely tied to its value domain.
	
	\begin{rem} \label{rem:livenessdomain}
		Liveness of a property may depend on the top value of its value domain.
		Consider a liveness property $\varPhi : \Sigma^\omega \to \DD$ and 
		the value domains $\DD_{\varPhi} = \{ v \in \DD \st v \leq \top_{\varPhi} \}$ and $\DD' = \DD \cup \{\top'\}$ with $v < \top'$ for all $v \in \DD$.
		
		The property $\varPhi_1 : \Sigma^\omega \to \DD_{\varPhi}$ where $\varPhi(w) = \varPhi_1(w)$ for all $w \in \Sigma^\omega$ is also live in $\DD_{\varPhi}$.
		This is easy to see by definition.
		For words $w \in \Sigma^\omega$ with $\varPhi(w) = \top_{\varPhi}$, the property is vacuously live in $\DD_{\varPhi}$, and those with $\varPhi(w) < \top_{\varPhi}$, it is live in $\DD_{\varPhi}$ thanks to its liveness in $\DD$.
		
		In contrast, $\varPhi_2 : \Sigma^\omega \to \DD'$ where $\varPhi(w) = \varPhi_2(w)$ for all $w \in \Sigma^\omega$ may be not live in $\DD'$.
		For example, consider $\Sigma = \{a,b\}$ and $\DD = \{0, x, y, 1\}$ where $0 < x < 1$ and $0 < y < 1$ but $x$ and $y$ are incomparable.
		Let $\varPhi(w) = x$ if $w \in \Sigma^* a^\omega$, let $\varPhi(w) = y$ if $w \in \Sigma^* b^\omega$, and let $\varPhi(w) = 1$ otherwise.
		The property $\varPhi$ is live in $\DD$ since $\top_{\varPhi} = \sup \DD = 1$ and $\sup_{w \in \Sigma^\omega} \varPhi(uw) = 1$ for every $u \in \Sigma^*$.
		However, considering the domain $\DD' = \DD \cup \{2\}$ with $1 < 2$, the same property is not live in $\DD'$ because $\varPhi((ab)^\omega) = 1$ and the only wrong lower bound hypothesis for $(ab)^\omega$ is $\sup \DD' = 2$, which can be dismissed as $\sup_{w \in \Sigma^\omega} \varPhi(u w) = 1$ for every prefix $u \prec (ab)^\omega$.
		In fact, for every property $\varPhi : \Sigma^\omega \to \DD$, if $\top_{\varPhi} < \top$ and $\top_{\varPhi}$ is attainable by some word, then $\varPhi$ is not live in $\DD$.
	\end{rem}

	Let us first show that our definition of liveness generalizes the boolean one.

	\begin{prop} \label{prop:livenessboolean}
		Quantitative liveness generalizes boolean liveness.
		In particular, for every boolean property $P \subseteq \Sigma^\omega$, the following statements are equivalent:
		\begin{enumerate}
			\item $P$ is live according to the classical definition~\cite{DBLP:journals/ipl/AlpernS85}.
			\item The characteristic property $\varPhi_P$ is live in $\BB$.
		\end{enumerate}
	\end{prop}
	\begin{proof}
		Recall that (1) means the following: for every $w \notin P$ and every $u \prec w$ there exists $w' \in \Sigma^\omega$ such that $uw' \in P$.
		Expressing the same statement with the characteristic property $\varPhi_P$ of $P$ gives us the following:
		for every $w \in \Sigma^\omega$ if $\varPhi_P(w) < 1$ then for every $u \prec w$ there exists $w' \in \Sigma^\omega$ such that $\varPhi_P(uw') = 1$.
		Since $\BB = \{0,1\}$ and $0 < 1$, it is easy to see that this statement is equivalent to the definition of liveness in $\BB$.
	\end{proof}
	\noindent 
	Next, we provide a characterization of liveness through the safety closure operation.
	
	\begin{thm} \label{thm:livenessclosure}
		A property $\varPhi$ is live iff $\safe{\varPhi}(w) > \varPhi(w)$ for every $w \in \Sigma^\omega$ with $\varPhi(w) < \top$.
	\end{thm}
	\begin{proof}
		First, suppose $\varPhi$ is live.
		Let $w \in \Sigma^\omega$ be such that $\varPhi(w) < \top$, and let $v$ be as in the definition of liveness.
		Since $\sup_{w' \in \Sigma^\omega} \varPhi(uw') \geq v$ for all prefixes $u \prec w$, we have that $\safe{\varPhi}(w) \geq v$. 
		Moreover, since $v \not \leq \varPhi(w)$, we are done.
		Now, suppose $\safe{\varPhi}(w) > \varPhi(w)$ for every $w \in \Sigma^\omega$ with $\varPhi(w) < \top$.
		Let $w \in \Sigma^\omega$ be such a trace, and let $v = \safe{\varPhi}(w)$.
		It is easy to see that $v$ satisfies the liveness condition since $\safe{\varPhi}(w) = \inf_{u \prec w} \sup_{w' \in \Sigma^\omega} \varPhi(uw') > \varPhi(w)$.
	\end{proof}
	\noindent 
	We show that liveness properties are closed under pairwise $\max$ considering totally-ordered value domains.
	
	\begin{prop}\label{prop:live:closuretotal}
		For every totally-ordered value domain $\DD$, the set of liveness properties over $\DD$ is closed under $\max$.
	\end{prop}
	\begin{proof}
		Consider two properties $\varPhi_1,\varPhi_2 : \Sigma^\omega \to \DD$ that are live in $\DD$.
		Let $\varPhi$ be their pairwise maximum, i.e., $\varPhi(w) = \max(\varPhi_1(w),\varPhi_2(w))$ for all $w\in \Sigma^\omega$.
		We show that $\varPhi$ fulfills the liveness definition for all $w \in \Sigma^\omega$.
		If $\varPhi_1(w) = \top$ or $\varPhi_2(w) = \top$ then $\varPhi(w) = \top$.
		Otherwise, for each $i \in \{1, 2\}$, there exists $v_i$ such that $\varPhi_i(w) < v_i$, and for all $u \prefix w$ we have $\sup_{w' \in \Sigma^\omega} \varPhi_i(uw') \geq v_i$.
		Hence, because $\DD$ is totally-ordered, defining $v = \max(v_1, v_2)$ implies $\varPhi(w) < v$ as well as $\sup_{w' \in \Sigma^\omega} \varPhi(uw') \geq v$ for all $u \prefix w$.
	\end{proof}
	\noindent 
	As in the boolean setting, the intersection of safety and liveness contains only the degenerate properties that are constant, i.e., always output $\top$.
	
	\begin{prop}\label{prop:top}
		A property $\varPhi$ is safe and live iff $\varPhi(w) = \top$ for all $w \in \Sigma^\omega$.
	\end{prop}
	\begin{proof}
		Observe that the constant function $\varPhi = \top$ is trivially safe and live.
		Now, let $\varPhi$ be a property that is both safe and live, and suppose towards contradiction that $\varPhi(w) < \top$ for some $w \in \Sigma^\omega$.
		Since $\varPhi$ is live, there exists a value $v$ with $\varPhi(w) \not \geq v$ such that for all $u \prec w$, we have $\sup_{w' \in \Sigma^\omega} \varPhi(uw') \geq v$.
		In particular, $\inf_{u \prefix w} \sup_{w' \in \Sigma^\omega} \varPhi(uw') \geq v$ and $\varPhi(w) \not \geq v$ hold, implying $\safe{\varPhi}(w) > \varPhi(w)$ by definition of safety closure and \Cref{thm:safe:main}.
		Then, again by \Cref{thm:safe:main}, this contradicts the assumption that $\varPhi$ is safe.
	\end{proof}
	\noindent 
	We define co-liveness symmetrically, and note that the duals of the statements above also hold for co-liveness.
	
	\begin{defi}[Co-liveness]
		A property $\varPhi : \Sigma^\omega \to \DD$ is \emph{co-live in $\DD$} when for all $w \in \Sigma^\omega$, if $\varPhi(w) > \bot$, then there exists a value $v \in \DD$ such that $\varPhi(w) \not \leq v$ and for all prefixes $u \prec w$, we have $\inf_{w' \in \Sigma^\omega} \varPhi(uw') \leq v$.
	\end{defi}
	
	Next, we present some examples of liveness and co-liveness properties.
	We start by showing that $\LimInf$- and $\LimSup$-properties can be live and co-live.
	
	\begin{exa}
		Let $\Sigma = \{a,b\}$ be an alphabet, and let $P = (\Sigma^* a)^\omega$ (infinitely many $a$'s) and $Q = \Sigma^\omega \setminus P$ (finitely many $a$'s) be boolean properties.
		Consider their characteristic properties $\varPhi_P$ and $\varPhi_Q$.
		As we pointed out earlier, our definitions generalize their boolean counterparts, therefore $\varPhi_P$ and $\varPhi_Q$ are both live and co-live.
		Moreover, $\varPhi_P$ is a $\LimSup$-property: define $\pi_P(u) = 1$ if $u \in \Sigma^* a$, and $\pi_P(u) = 0$ otherwise, and observe that $\varPhi_P(w) = \LimSup_{u \prec w} \pi_P(u)$ for all $w \in \Sigma^\omega$.
		Similarly, $\varPhi_Q$ is a $\LimInf$-property.
		\end{exa}
	
	Now, we show that the maximal response-time property is live, and the minimal response time is co-live.
	
	\begin{exa}
		Recall the co-safety property $\varPhi_{\max}$ of maximal response time from \Cref{ex:maxresp}.
		Let $w \in \Sigma^\omega$ such that $\varPhi_{\max}(w) < \infty$.
		We can extend every prefix $u \prec w$ with $w' = \req\, \tick^\omega$, which gives us $\varPhi_{\max}(uw') = \infty > \varPhi(w)$.
		Equivalently, for every $w \in \Sigma^\omega$, we have $\safe{\varPhi_{\max}}(w) = \infty > \varPhi_{\max}(w)$.
		Hence $\varPhi_{\max}$ is live and, analogously, the safety property $\varPhi_{\min}$ from \Cref{ex:minresp} is co-live.
		\end{exa}
	
	We next present the \emph{average response-time} property and show that it is live and co-live.
	
	\begin{exa} \label{ex:avgresp}
		Let $\Sigma = \{\req, \gra, \tick, \other\}$.
		For all $u \in \Sigma^*$, let $p(u) = 1$ if there is no pending $\req$ in $u$, and $p(u) = 0$ otherwise.
		Define $\pi_{\text{valid}}(u) = |\{u' \preceq u \st \exists u'' \in \Sigma^* : u' = u'' \req \,\land\, p(u'') = 1\}|$ as the number of valid requests in $u$, and define $\pi_{\text{time}}(u)$ as the total number of $\tick$ observations that occur after a valid $\req$ and before the matching $\gra$.
		Then, $\varPhi_{\avg} = (\pi_{\avg}, \LimInf)$, where $\pi_{\avg}(u) = \frac{\pi_{\text{time}}(u)}{\pi_{\text{valid}}(u)}$ for all $u \in \Sigma^*$ with $\pi_{\text{valid}}(u) > 0$, and $\pi_{\avg}(u) = \infty$ otherwise.
		For example, $\pi_{\avg}(u) = \frac{3}{2}$ for $u = \req\, \tick\, \gra\, \tick\, \req\, \tick\, \req\, \tick$.
		Note that $\varPhi_{\avg}$ is a $\LimInf$-property.
		
		The property $\varPhi_{\avg}$ is defined on the value domain $[0,\infty]$ and is both live and co-live.	
		To see this, let $w \in \Sigma^\omega$ such that $0 < \varPhi_{\avg}(w) < \infty$ and, for every prefix $u \prec w$, consider $w' = \req\, \tick^\omega$ and $w'' = \gra\, (\req\,\gra)^\omega$.
		Since $uw'$ has a pending request followed by infinitely many clock ticks, we have $\varPhi_{\avg}(uw') = \infty$.
		Similarly, since $uw''$ eventually has all new requests immediately granted, we get $\varPhi_{\avg}(uw'') = 0$.
		\end{exa}
	
	Notice that for the average response-time property $\varPhi_{\avg}$ in the example above, we have $\varPhi_{\avg}(w) = \varPhi_{\avg}(uw)$ for every $u \in \Sigma^*$ and $w \in \Sigma^\omega$.
	Such properties are called \emph{prefix independent}.
	Finally, we show that every prefix-independent property is both live and co-live.
	
	\begin{prop}
		Every prefix-independent property $\varPhi$ is live and co-live.
	\end{prop}
	\begin{proof}
		Consider a prefix-independent property $\varPhi$.
		We only show that $\varPhi$ is live as its co-liveness can be proved similarly.
		Suppose towards contradiction that $\varPhi$ is not live, and thus by \cref{thm:livenessclosure} that $\varPhi(w) = \safe{\varPhi}(w)$ for some $w \in \Sigma^\omega$ with $\varPhi(w) < \top$.
		Let $w$ be such a word, and consider two prefixes $u_1 \preceq u_2 \prec w$ such that $\sup_{w' \in \Sigma^\omega} \varPhi(u_2 w') < \sup_{w' \in \Sigma^\omega} \varPhi(u_1 w')$.
		Such prefixes exist because otherwise we have a contradiction to $\varPhi(w) < \top$.
		Then, there exists $w'' \in \Sigma^\omega$ such that $\varPhi(u_2 w'') < \varPhi(u_1 w'')$.
		Since $u_1 \preceq u_2$, there is a finite word $u_3$ with $u_2 = u_1 \cdot u_3$.
		Notice that, since $\varPhi$ is prefix independent, we have $\varPhi(w'') = \varPhi(u_1 w'') = \varPhi(u_1 u_3 w'')$, which contradicts $\varPhi(u_2 w'') < \varPhi(u_1 w'')$.		
	\end{proof}
	
	\subsection{The Quantitative Safety-Liveness Decomposition}
	A celebrated theorem states that every boolean property can be expressed as an intersection of a safety property and a liveness property~\cite{DBLP:journals/ipl/AlpernS85}. 
	In this section, we prove an analogous result in the quantitative setting. 
	
	\begin{exa}
		Let $\Sigma = \{\req, \gra, \tick, \other\}$.
		Recall the maximal response-time property $\varPhi_{\max}$ from \Cref{ex:maxresp}, and the average response-time property $\varPhi_{\avg}$ from \Cref{ex:avgresp}.
		Let $n > 0$ be an integer and define a new property $\varPhi : \Sigma^\omega \to [0,n]$ by $\varPhi(w) = \varPhi_{\avg}(w)$ if $\varPhi_{\max}(w) \leq n$, and $\varPhi(w) = 0$ otherwise.
		For the safety closure of $\varPhi$, we have $\safe{\varPhi}(w) = n$ if $\varPhi_{\max}(w) \leq n$, and $\safe{\varPhi}(w) = 0$ otherwise.
		Now, we further define $\varPsi(w) = \varPhi_{\avg}(w)$ if $\varPhi_{\max}(w) \leq n$, and $\varPsi(w) = n$ otherwise.
		Observe that $\varPsi$ is live, because every prefix of a trace whose value is less than $n$ can be extended to a greater value.
		Finally, note that for all $w \in \Sigma^\omega$, we can express $\varPhi(w)$ as the pointwise minimum of $\safe{\varPhi}(w)$ and $\varPsi(w)$.
		Intuitively, the safety part $\safe{\varPhi}$ of this decomposition checks whether the maximal response time stays below the permitted bound, and the liveness part $\varPsi$ keeps track of the average response time as long as the bound is satisfied.
		\end{exa}
	
	Following a similar construction, we show that a safety-liveness decomposition exists for every property.
	
	\begin{thm}\label{thm:decomp}
		For every property $\varPhi$, there exists a liveness property $\varPsi$ such that $\varPhi(w) = \min(\safe{\varPhi}(w), \varPsi(w))$ for all $w \in \Sigma^\omega$.
	\end{thm}
	\begin{proof}
		Let $\varPhi$ be a property and consider its safety closure $\safe{\varPhi}$.
		We define $\varPsi$ as follows:
		$\varPsi(w) = \varPhi(w)$ if $\safe{\varPhi}(w) \neq \varPhi(w)$, and $\varPsi(w) = \top$ otherwise.
		Note that $\safe{\varPhi}(w) \geq \varPhi(w)$ for all $w \in\Sigma^\omega$ by \Cref{proposition:safe:closure}.
		When $\safe{\varPhi}(w) > \varPhi(w)$, we have $\min(\safe{\varPhi}(w), \varPsi(w)) = \min(\safe{\varPhi}(w), \varPhi(w)) = \varPhi(w)$.
		When $\safe{\varPhi}(w) = \varPhi(w)$, we have $\min(\safe{\varPhi}(w), \varPsi(w)) = \min(\varPhi(w), \top) = \varPhi(w)$.
		
		Now, suppose towards contradiction that $\varPsi$ is not live, i.e., there exists $w \in \Sigma^\omega$ such that $\varPsi(w) < \top$ and for all $v \not \leq \varPhi(w)$, there exists $u \prec w$ satisfying $\sup_{w' \in \Sigma^\omega} \varPhi(uw') \not\geq v$.
		Let $w \in \Sigma^\omega$ be such that $\varPsi(w) < \top$.
		Then, by definition of $\varPsi$, we know that $\varPsi(w) = \varPhi(w) < \safe{\varPhi}(w)$.
		Moreover, since $\safe{\varPhi}(w) \not \leq \varPsi(w)$, there exists $u \prec w$ satisfying $\sup_{w' \in \Sigma^\omega} \varPhi(uw') \not\geq \safe{\varPhi}(w)$.
		In particular, we have $\sup_{w' \in \Sigma^\omega} \varPhi(uw') < \safe{\varPhi}(w)$.
		Since we have $\safe{\varPhi}(w) = \inf_{u' \prec w} \sup_{w' \in \Sigma^\omega} \varPhi(u'w')$ by definition and $u \prec w$, it yields a contradiction.
		Therefore, $\varPsi$ is live.
	\end{proof}
	\noindent 
	In particular, if the given property is safe or live, the decomposition is trivial.
	
	\begin{rem}\label{rem:trivialdecomp}
		Let $\varPhi$ be a property.
		If $\varPhi$ is safe, then the safety part of the decomposition is $\varPhi$ itself, and the liveness part is the constant property that maps every trace to $\top$.
		If $\varPhi$ is live, then the liveness part of the decomposition is $\varPhi$ itself, and the safety part is $\safe{\varPhi}$.
		Note that, in this case, $\safe{\varPhi}$ may differ from the constant function $\top$, but taking the safety part as constant function $\top$ is a valid decomposition.
	\end{rem}
	
	Another decomposition theorem is the one of boolean properties over nonunary alphabets into two liveness properties~\cite{DBLP:journals/ipl/AlpernS85}.
	We extend this result to the quantitative setting.
	
	\begin{thm}\label{thm:livedecomp}
		For every property $\varPhi$ over a nonunary alphabet $\Sigma$, there exist two liveness properties $\varPsi_1$ and $\varPsi_2$ such that $\varPhi(w) = \min(\varPsi_1(w), \varPsi_2(w))$ for all $w \in \Sigma^\omega$.
	\end{thm}
	\begin{proof}
		Let $\Sigma$ be a finite alphabet with $|\Sigma| \geq 2$ and $a_1, a_2 \in \Sigma$ be two distinct letters.
		Consider an arbitrary property $\varPhi$ over $\Sigma$.
		For $i \in \{1, 2\}$, we define $\varPsi_i$ as follows:
		$\varPsi_i(w) = \top$ if $w = u(a_i)^\omega$ for some $u\in\Sigma^*$, and $\varPsi_i(w) = \varPhi(w)$ otherwise.
		Note that, since $a_1$ and $a_2$ are distinct, whenever $w \in \Sigma^* (a_1)^\omega$ then $w \notin \Sigma^* (a_2)^\omega$, and vice versa.
		Then, we have that both $\varPsi_1$ and $\varPsi_2$ are $\top$ only when $\varPhi$ is $\top$.
		In the remaining cases, when at most one of $\varPsi_1$ and $\varPsi_2$ is $\top$, then either both equals $\varPhi$ or one of them is $\top$ and the other is $\varPhi$.
		As a direct consequence, $\varPhi(w) = \min(\varPsi_1(w), \varPsi_2(w))$ for all $w\in\Sigma^\omega$.
		
		Now, we show that $\varPsi_1$ and $\varPsi_2$ are both live.
		By construction, $\varPsi_i(u(a_i)^\omega)=\top$ for all $u\in\Sigma^*$.
		In particular, $\safe{\varPsi_i}(w) = \inf_{u \prefix w}\sup_{w' \in \Sigma^\omega}\varPsi_i(uw') = \top$ for all $w\in\Sigma^\omega$.
		We conclude that $\varPsi_i$ is live thanks to \Cref{thm:livenessclosure}.
	\end{proof}
	\noindent 
	For co-safety and co-liveness, the duals of \Cref{rem:trivialdecomp} and \Cref{thm:decomp,thm:livedecomp} hold.
	In particular, every property is the pointwise maximum of its co-safety closure and a co-liveness property.

	\subsection{Threshold Liveness and Top Liveness} \label{sec:notionsliveness}
	
	Threshold liveness connects a quantitative property and the boolean liveness of the sets of words whose values exceed a threshold value.
	
	\begin{defi}[Threshold liveness and co-liveness]
		A property $\varPhi : \Sigma^\omega \to \DD$ is \emph{threshold live} when for every $v \in \DD$ the boolean property $\varPhi_{\geq v}$ is live (and thus $\varPhi_{\not \geq v}$ is co-live).
		Equivalently, $\varPhi$ is threshold live when for every $u \in \Sigma^*$ and $v \in \DD$ there exists $w \in \Sigma^\omega$ such that $\varPhi(uw) \geq v$.
		Similarly, a property $\varPhi : \Sigma^\omega \to \DD$ is \emph{threshold co-live} when for every $v \in \DD$ the boolean property $\varPhi_{\not \leq v}$ is co-live (and thus $\varPhi_{\leq v}$ is live).
		Equivalently, $\varPhi$ is threshold co-live when for every $u \in \Sigma^*$ and $v \in \DD$ there exists $w \in \Sigma^\omega$ such that $\varPhi(uw) \leq v$.
	\end{defi}
	
	We relate threshold liveness with the boolean liveness of a single set of words.
	
	\begin{prop} \label{cl:ThresholdLiveIffDense}
		A property $\varPhi$ is threshold live iff the set $\varPhi_{\geq \top}$ is live in the boolean sense.
	\end{prop}
	\begin{proof}
		Consider a property $\varPhi : \Sigma^\omega \to \DD$.
		
		$(\Rightarrow)$:
		Assume $\varPhi$ to be threshold live, i.e., for every $v \in \DD$ the boolean property $\varPhi_{\geq v}$ is live.
		In particular, $\varPhi_{\geq \top}$ is also live.
		
		$(\Leftarrow)$:
		Assume $\varPhi_{\geq \top}$ to be live in the boolean sense.
		Observe that for every $v \leq \top$ we have $\varPhi_{\geq \top} \subseteq \varPhi_{\geq v}$.
		Since the union of a boolean liveness property with any boolean property is live~\cite{DBLP:journals/ipl/AlpernS85}, the boolean property $\varPhi_{\geq v}$ is also live for all $v \leq \top$, i.e., $\varPhi$ is threshold live.
	\end{proof}
	\noindent 
	Liveness is characterized by the safety closure being strictly greater than the property whenever possible (\cref{thm:livenessclosure}).
	Top liveness puts an additional requirement on liveness: the safety closure of the property should not only be greater than the original property but also equal to the top value.
	
	\begin{defi}[Top liveness and bottom co-liveness]
		A property $\varPhi$ is \emph{top live} when $\safe{\varPhi}(w) = \top$ for every $w \in \Sigma^\omega$. 
		Similarly, a property $\varPhi$ is \emph{bottom co-live} when $\cosafe{\varPhi}(w) = \bot$ for every $w \in \Sigma^\omega$.
	\end{defi}
	
	We provide a strict hierarchy of threshold liveness, top liveness, and liveness.
	
	\begin{prop} \label{cl:TopLivenessAndThresholdLiveness}
		Every threshold-live property is top live, but not vice versa; and every top-live property is live, but not vice versa. 
	\end{prop}
	\begin{proof}
		First, we show the strict inclusion of threshold liveness in top liveness.
		Let $\varPhi$ be a threshold-live property.
		In particular, taking the threshold $v = \top$ gives us that for every $u \in \Sigma^*$ there exists $w \in \Sigma^\omega$ such that $\varPhi(uw) = \top$.
		Then, $\sup_{w \in \Sigma^\omega} \varPhi(uw) = \top$ for all $u \in \Sigma^*$, which implies that $\varPhi$ is top live.
		Next, consider the property $\varPhi$ over the alphabet $\{a,b\}$, defined for all $w\in\Sigma^\omega$ as follows:
		$\varPhi(w) = |w|_a$ if $w$ has finitely many $a$'s, otherwise $\varPhi(w) = 0$.
		Observe that $\sup_{w \in \Sigma^\omega} \varPhi(uw) = \infty$ for all $u \in \Sigma^*$, therefore it is top live.
		However, for the threshold $v = \infty$, the set $\varPhi_{\geq v}$ is empty, implying that it is not threshold live.
		
		Now, we show that the strict inclusion of top liveness in liveness.
		Recall that, by \cref{thm:livenessclosure}, a property $\varPhi$ is live iff for every $w \in \Sigma^\omega$ if $\varPhi(w) < \top$ then $\varPhi(w) < \safe{\varPhi}(w)$.
		Then, notice that if a property $\varPhi$ is top live, it is obviously live.
		Next, let $\Sigma = \{a,b\}$ and consider the property $\varPhi : \Sigma^\omega \to [0,2]$ defined for all $w\in\Sigma^\omega$ as follows:
		$\varPhi(w) = 0$ if $w$ is of the form $\Sigma^* b^\omega$, otherwise $\varPhi(w) = \sum_{i \geq 0} 2^{-i} f(\sigma_i)$ where $w = \sigma_0 \sigma_1 \ldots$,  $f(a) = 0$, and $f(b) = 1$.
		Observe that $\varPhi$ is live since $\varPhi(w) < \safe{\varPhi}(w)$ for every word $w \in \Sigma^\omega$.
		However, it is not top live since $\safe{\varPhi}(a w) \leq 1 < 2 = \top$ for all $w \in \Sigma^\omega$.
	\end{proof}
	\noindent 
	Top liveness does not imply threshold liveness, but it does imply a weaker form of it.
	
	\begin{prop} \label{cl:WhenTopLivenessImpliesThresholdLiveness}
		For every top-live property $\varPhi$ and value $v < \top$, the set $\varPhi_{\geq v}$ is live in the boolean sense.
	\end{prop}
	\begin{proof}
		Let $\varPhi$ be top live property, i.e., $\inf_{u \prec w} \sup_{w' \in \Sigma^\omega} \varPhi(uw') = \top$ for all $w \in \Sigma^\omega$.
		Let $v < \top$ be a value.
		Suppose towards contradiction that $\varPhi_{\geq v}$ is not live in the boolean sense, i.e., there exists $\hat{u} \in \Sigma^*$ such that $\varPhi(\hat{u} w') \not \geq v$ for all $w' \in \Sigma^\omega$.
		Let $\hat{w} \in \Sigma^\omega$ be such that $\hat{u}\prefix \hat{w}$.
		Clearly $\inf_{u \prec \hat{w}} \sup_{w' \in \Sigma^\omega} \varPhi(uw') \not\geq v$.
		Either $\inf_{u \prec \hat{w}} \sup_{w' \in \Sigma^\omega} \varPhi(uw')$ is incomparable with $v$, or it is less than $v$.
		Since $\top$ compares with all values, we have that $\inf_{u \prec \hat{w}} \sup_{w' \in \Sigma^\omega} \varPhi(uw') < \top$, which contradicts the top liveness of $\varPhi$.
	\end{proof}
	\noindent 
	While the three liveness notions differ in general, they do coincide for $\sup$-closed properties.
	
	\begin{thm} \label{cl:LivenessCollapsingSupClosed}
		A $\sup$-closed property is live iff it is top live iff it is threshold live.
	\end{thm}
	\begin{proof}
		Notice that for every $\sup$-closed property $\varPhi$, top liveness means that for every $u \in \Sigma^*$ there is $w \in \Sigma^\omega$ such that $\varPhi(uw) = \top$.
		Let $\varPhi$ be a $\sup$-closed liveness property.
		Suppose towards contradiction that it is not top live, i.e., there is $u \in \Sigma^*$ such that for all $w \in \Sigma^\omega$ we have $\varPhi(uw) < \top$.
		Let $\sup_{w \in \Sigma^\omega} \varPhi(uw) = k < \top$, and note that since $\varPhi$ is $\sup$-closed, there exists an infinite continuation $w' \in \Sigma^\omega$ for which $\varPhi(u w') = k < \top$.
		As $\varPhi$ is live, there exists a value $v$ such that $k \not \geq v$ and for every prefix $u' \prefix uw'$ there exists $w'' \in \Sigma^\omega$ with $\varPhi(u'w'') \geq v$.
		However, letting $u' = u$ yields a contradiction to our initial supposition.
		
		Now, let $\varPhi$ be a $\sup$-closed top liveness property.
		Thanks to \cref{cl:ThresholdLiveIffDense}, it is sufficient to show that the boolean property $\varPhi_{\geq \top}$ is live in the boolean sense.
		Suppose towards contradiction that $\varPhi_{\geq \top}$ is not live, i.e., there exists $u \in \Sigma^*$ such that for all $w \in \Sigma^\omega$ we have $\varPhi(uw) < \top$.
		Due to $\sup$-closedness, we have $\sup_{w \in \Sigma^\omega} \varPhi(uw) < \top$ as well.
		Moreover, for every $w \in \Sigma^\omega$ such that $u \prefix w$, this means that $\inf_{u \prec w} \sup_{w' \in \Sigma^\omega} \varPhi(uw') < \top$, which is a contradiction.
	\end{proof}

	\subsection{Additional Notions
Related to Quantitative Liveness}

	In~\cite{DBLP:journals/isci/LiDL17}, the authors define a property $\varPhi$ as \emph{multi-live} iff $\safe{\varPhi}(w) > \bot$ for all $w\in \Sigma^\omega$.
	We show that our definition is more restrictive, resulting in fewer liveness properties while still allowing a safety-liveness decomposition.
	
	\begin{prop}\label{prop:multi:live}
		Every live property is multi-live, but not vice versa.
	\end{prop}
	\begin{proof}
		We prove that liveness implies multi-liveness.
		Suppose toward contradiction that some property $\varPhi$ is live, but not multi-live.
		Then, there exists $w\in\Sigma^\omega$ for which $\safe{\varPhi}(w) = \bot$, and therefore $\varPhi(w) = \bot$ too.
		Note that we assume $\DD$ is a nontrivial complete lattice, i.e., $\top \neq \bot$.
		Then, since $\varPhi$ is live, we have $\safe{\varPhi}(w) > \varPhi(w)$ by \Cref{thm:livenessclosure}, which yields a contradiction.
		
		Now, we provide a separating example on a totally ordered domain.
		Let $\Sigma = \{a,b, c\}$, and consider the following property:
		$\varPhi(w) = 0$ if $w = a^\omega$, and $\varPhi(w) = 1$ if $w \in \Sigma^* c \Sigma^\omega$, and $\varPhi(w) = 2$ otherwise (i.e., if $w$ has some $b$ and no $c$).
		For all $w\in\Sigma^\omega$ and prefixes $u\prefix w$, we have $\varPhi(u c^\omega) = 1$.
		Thus $\safe{\varPhi}(w) \neq \bot$, which implies that $\varPhi$ is multi-live.
		However, $\varPhi$ is not live.
		Indeed, for every $w\in\Sigma^\omega$ such that $w \in \Sigma^* c \Sigma^\omega$, we have $\varPhi(w) = 1 < \top$.
		Moreover, $w$ admits some prefix $u$ that contains an occurrence of $c$, thus satisfying $\sup_{w' \in \Sigma^\omega} \varPhi(uw') = 1$.
	\end{proof}
\noindent 
	Recall that a property is both safe and live iff it is the constant function $\top$ (\cref{prop:top}).
	For multi-safety and multi-liveness, this is not the case.
	
	\begin{exa}
		Let $\Sigma = \{a,b\}$ be an alphabet and $\DD = \{v_1, v_2, \bot, \top\}$ be a lattice where $v_1$ and $v_2$ are incomparable.
		Consider the property $\varPhi : \Sigma^\omega \to \DD$ that is defined as $\varPhi(w) = v_1$ if $a \prefix w$ and $\varPhi(w)= v_2$ if $b \prefix w$.
		Recall from \cref{cl:SafeCosafeNotSupInfClosed} that $\varPhi$ is safe, thus multi-safe by~\cref{cl:multisafe}.
		Clearly, $\safe{\varPhi}(w) > \bot$ for all $w \in \Sigma^\omega$, thus $\varPhi$ is multi-live.
		However, $\varPhi$ is not live as for all words $w$, we have $\varPhi(w) = \safe{\varPhi}(w) < \top$.
	\end{exa}
	
	In~\cite{DBLP:conf/nfm/GorostiagaS22}, the authors define a property $\varPhi$ as \emph{verdict-live} iff for every $w \in \Sigma^\omega$ and value $v \not \leq \varPhi(w)$, every prefix $u \prec w$ satisfies $\varPhi(uw') = v$ for some $w' \in \Sigma^\omega$.
	We show that our definition is more liberal.
	
	\begin{prop}\label{prop:verdict:live}
		Every verdict-live property is live, but not vice versa.
	\end{prop}
	\begin{proof}
		The implication holds trivially.
		We provide a separating example below, concluding that our definition is strictly more general even for totally ordered domains.
		Let $\Sigma = \{a,b\}$, and consider the following property:
		$\varPhi(w) = 0$ if $w = a^\omega$, and $\varPhi(w) = 1$ if $w \in \Sigma^* b \Sigma^* b \Sigma^\omega$, and $\varPhi(w) = 2^{-|u|}$ otherwise (if $w$ has exactly one $b$), where $u \prec w$ is the shortest prefix in which $b$ occurs.
		Consider an arbitrary $w \in \Sigma^\omega$.
		If $\varPhi(w) = 1$, then the liveness condition is vacuously satisfied.
		If $\varPhi(w) = 0$, then $w = a^\omega$, and every prefix $u \prec w$ can be extended with $w' = ba^\omega$ or $w'' = b^\omega$ to obtain $\varPhi(uw') = 2^{-(|u|+1)}$ and $\varPhi(uw'') = 1$.
		If $0 < \varPhi(w) < 1$, then $w$ has exactly one $b$, and every prefix $u \prec w$ can be extended with $b^\omega$ to obtain $\varPhi(u b^\omega) = 1$.
		Hence $\varPhi$ is live.
		However, $\varPhi$ is not verdict-live.
		To see this, consider the trace $w = a^k b a^\omega$ for some integer $k \geq 1$ and note that $\varPhi(w) = 2^{-(k+1)}$.
		Although all prefixes of $w$ can be extended to achieve the value 1, the value domain contains elements between $\varPhi(w)$ and 1, namely the values $2^{-m}$ for $1 \leq m \leq k$.
		Each of these values can be rejected after reading a finite prefix of $w$, because for $n \geq m$ it is not possible to extend $a^n$ to achieve $2^{-m}$. 
	\end{proof}
\noindent 
	Let us conclude with a remark on the form of hypotheses in our definition of liveness.
	
	\begin{rem} \label{rem:strictliveness}
		In the same vein as \cref{rem:strictsafety}, suppose we define liveness with strict lower bound hypotheses instead of nonstrict: for all $w \in \Sigma^\omega$, if $\varPhi(w) < \top$, then there exists a value $v \in \DD$ such that $\varPhi(w) \not > v$ and for all prefixes $u \prec w$, we have $\sup_{w' \in \Sigma^\omega} \varPhi(uw') > v$.
		Let $w$ be a word with $\varPhi(w) < \top$ and consider $v = \varPhi(w)$.
		Evidently, according to this definition, it would be permissible for the $\sup$ of possible prediction values to converge to $\varPhi(w)$, in other words, for the safety closure to have the same value as the property on a word whose value is less than $\top$, which is too lenient.
	\end{rem}

	\section{Quantitative Automata} \label{sec:preliminaries2}
	
	A \emph{nondeterministic quantitative\footnote{We speak of ``quantitative'' rather than ``weighted'' automata, following the distinction made in~\cite{Bok21} between the two.} automaton} (or just automaton from here on) on words is a tuple $\A=(\Sigma,Q,\iota,\delta)$, where $\Sigma$ is an alphabet; $Q$ is a finite nonempty set of states; $\iota\in Q$ is an initial state; and $\delta\colon Q\times \Sigma \to 2^{(\QQ \times Q)}$ is a finite transition function over weight-state pairs.
	A \emph{transition} is a tuple $(q,\sigma,x,q')\in Q \times \Sigma \times \QQ \times Q$, such that $(x,q')\in\delta(q,\sigma)$,  also written $\trans{q}{\sigma:x}{q'}$. (There might be finitely many transitions with different weights over the same letter between the same states.\footnote{The flexibility of allowing ``parallel'' transitions with different weights is often omitted, as it is redundant for some value functions, including the ones we focus on in the sequel, while important for others.})
	We write $\gamma(t)=x$ for the weight of a transition $t=(q,\sigma,x,q')$.
	$\A$ is  deterministic if for all $q\in Q$ and $\sigma \in \Sigma$, the set $\delta(q,\sigma)$ is a singleton.
	We require the automaton $\A$ to be \emph{total}, namely that for every state $q\in Q$ and letter $\sigma\in\Sigma$, there is at least one state $q'$ and a transition $\trans{q}{\sigma:x}{q'}$.
	For a state $q\in Q$, we denote by $\A^q$ the automaton that is derived from $\A$ by setting its initial state $\iota$ to $q$.

	A run of $\A$ on a word $w$ is a sequence $\rho = \trans{q_0}{w[0]:x_0}{q_1}\trans{}{w[1]:x_1}{q_2}\ldots$ of transitions where $q_0=\iota$ and $(x_i,q_{i+1})\in \delta(q_i,w[i])$.
	For $0 \leq i < |w|$, we denote the $i$th transition in $\rho$ by $\rho[i]$, and the finite prefix of $\rho$ up to and including the $i$th transition by $\rho[..i]$.
	As each transition $t_i$ carries a weight $\gamma(t_i)\in\QQ$, the sequence $\rho$ provides a weight sequence $\gamma(\rho) = \gamma(t_0) \gamma(t_1) \ldots$
	A $\Val$-automaton is one equipped with a value function $\Val:\QQ^\omega \to \RR$, which assigns real values to runs of $\A$. 
	We assume that $\Val$ is bounded for every finite set of rationals, i.e., for every finite $V \subset \QQ$ there exist $m,M \in \RR$ such that $m \leq \Val(x) \leq M$ for every $x \in V^\omega$.
	The finite set $V$ corresponds to transition weights of a quantitative automaton, and the concrete value functions we consider satisfy this assumption.

	Notice that while quantitative properties can be defined over arbitrary value domains, we restrict quantitative automata to totally-ordered numerical value domains (i.e., bounded subsets of $\mathbb{R}$) as this is the standard setting in the literature.
	
	The value of a run $\rho$ is $\Val(\gamma(\rho))$. 
	The value of a $\Val$-automaton $\A$ on a word $w$, denoted $\A(w)$, is the supremum of $\Val(\rho)$ over all runs $\rho$ of $\A$ on $w$, generalizing the standard approach in boolean automata where acceptance is defined through the existence of an accepting run.
	The \emph{top value} of a $\Val$-automaton $\A$ is $\top_{\A} = \sup_{w \in \Sigma^\omega} \A(w)$, which we denote by $\top$ when $\A$ is clear from the context.
	Note that when we speak of the top value of an automaton or a property expressed by an automaton, we always match its value domain to have the same top value.
	
	An automaton $\A$ is \emph{safe} (resp. \emph{live}) iff it expresses a safety (resp. liveness) property.
	Two automata $\A$ and $\A'$ are \emph{equivalent}, if they express the same function from words to reals.
	The size of an automaton consists of the maximum among the size of its alphabet, state-space, and transition-space, where weights are represented in binary.
	
	We list below the value functions for quantitative automata that we will use, defined over infinite sequences $v_0 v_1 \ldots$ of rational weights.
	
	\begin{multicols}{2}
		\begin{itemize}
			\item $\displaystyle \Inf(v) = \inf\{v_n \st n \geq 0\}$
			\item $\displaystyle \Sup(v) = \sup\{v_n \st n \geq 0\}$
		\end{itemize}
	\end{multicols}
	
	\begin{multicols}{2}
		\begin{itemize}
			\item $\displaystyle \LimInf(v) = \lim_{n\to\infty}\limits\inf\{v_i \st i \geq n\}$
			\item $\displaystyle \LimInfAvg(v) = \LimInf \left(\frac{1}{n} \sum_{i=0}^{n-1} v_i\right)$
			\item $\displaystyle \LimSup(v) = \lim_{n\to\infty}\limits\sup\{v_i \st i \geq n\}$
			\item $\displaystyle \LimSupAvg(v) = \LimSup \left(\frac{1}{n} \sum_{i=0}^{n-1} v_i\right)$
		\end{itemize}
	\end{multicols}
	
	\begin{itemize}
		\item For a discount factor $\lambda\in\QQ\cap(0,1)$, $\displaystyle \DSum_{\lambda}(v) = \sum_{i\geq 0} \lambda^i  v_i$
	\end{itemize}
	\noindent 
	Note that (i) when the discount factor $\lambda\in\QQ\cap(0,1)$ is unspecified, we write $\DSum$, and (ii) $\LimInfAvg$ and $\LimSupAvg$ are also called $\MPL$ and $\MPH$ in the literature.
	
	A value function $\Val$ is \emph{prefix independent} iff for all $x \in \QQ^*$ and all $y \in \QQ^\omega$ we have $\Val(y) = \Val(xy)$.
	The value functions $\LimInf$, $\LimSup$, $\LimInfAvg$, and $\LimSupAvg$ are prefix independent, while $\Inf$, $\Sup$, and $\DSum$ are not.
	
	The following statement allows us to consider $\Inf$- and $\Sup$-automata as only having runs with respectively nonincreasing and nondecreasing sequences of weights, and to also consider them as $\LimInf$- and $\LimSup$-automata.
	
	\begin{prop}\label{cl:InfAndSuptoLim}
		Let $\Val \in \{\Inf, \Sup\}$.
		Given a $\Val$-automaton, we can construct in \PTime an equivalent $\Val$-, $\LimInf$- or $\LimSup$-automaton whose runs yield monotonic weight sequences.
	\end{prop}
	\begin{proof}
		Consider a $\Sup$-automaton $\A = (\Sigma, Q, \iota, \delta)$.
		The idea is to construct an equivalent $\Sup$-automaton $\A'$ that memorizes the maximal visited weight, and optionally take it as a $\LimInf$- or $\LimSup$-automaton.
		A similar construction appears in~\cite[Lem.~1]{DBLP:journals/tocl/ChatterjeeDH10} where for every run of $\A$ there is a run of $\A'$ yielding a weight sequence that is eventually constant, but it is not necessarily the case that every run of $\A'$ has a monotonic weight sequence.
		Let $V$ be the set of weights on $\A$'s transitions.
		Since $|V| < \infty$, we can fix the minimal weight $v_0 = \min(V)$.
		We construct $\A' = (\Sigma, Q \times V, (\iota, v_0), \delta')$ where $\delta' \colon (Q \times V) \times \Sigma \to 2^{Q \times V}$ is defined as follows.
		Given $p \in Q$, $v,v' \in V$, and $\sigma \in \Sigma$, we have that $(v', (q, \max\{v,v'\})) \in \delta'((p,v), \sigma)$ if and only if $(v', q) \in \delta(p, \sigma)$.
		Notice that if $\A$ is deterministic, so is $\A'$.
		Clearly, the $\Sup$-automata $\A'$ and $\A$ are equivalent, and the construction of $\A'$ is in \PTime in the size of $\A$.  
		Observe that, by construction, every run $\rho$ of $\A'$ yields a nondecreasing weight sequence for which there exists $i \in \NN$ such that for all $j \geq i$ we have $\gamma(\rho[i]) = \gamma(\rho[j]) = \Sup(\gamma(\rho))$.
		Hence, $\A'$ can be equivalently interpreted as a $\Sup$-, $\LimInf$ or $\LimSup$-automaton.
		The construction for a given $\Inf$-automaton is dual as it consists in memorizing the minimal visited weight, therefore the weight sequences are nonincreasing.
	\end{proof}
	\noindent 
	We show that the common classes of quantitative automata always express $\sup$-closed properties, which will simplify the study of their safety and liveness.
	
	\begin{prop}\label{cl:ComputingTopValue}
		Let $\Val \in \{\Inf, \Sup, \LimInf, \LimSup, \LimInfAvg, \LimSupAvg,\DSum\}$.
		Every $\Val$-automaton expresses a property that is $\sup$-closed.
		Furthermore its top value is rational, attainable by a run, and can be computed in \PTime.
	\end{prop}
	\begin{proof}
		Observe that, by \cref{cl:InfAndSuptoLim} the cases of $\Val\in\{\Inf,\Sup\}$ reduce to $\Val\in\{\LimInf,\LimSup\}$.
		So, we can assume that $\Val \in \{\LimInf$, $\LimSup$, $\LimInfAvg$, $\LimSupAvg$, $\DSum\}$.
		
		It is shown in the proof of~\cite[Thm.\ 3]{DBLP:journals/tocl/ChatterjeeDH10} that the top value of every $\Val$-automaton $\A$ is attainable by a lasso run, and is therefore rational, and can be computed in \PTime.
		It is left to show that $\A$ is $\sup$-closed, meaning that for every finite word $u\in \Sigma^*$, there exists $\hat{w}\in\Sigma^\omega$, such that $\A(u\hat{w}) = \sup_{w'} \A(uw')$.
		
		Let $U$ be the set of states that $\A$ can reach running on $u$. 
		Observe that for every state $q\in U$, we have that $\A^q$ is also a $\Val$-automaton. Thus, by the above result, its top value $\top_q$ is attainable by a run on some word $w_q$. Hence, for $\Val \in \{\LimInf, \LimSup, \LimInfAvg, \LimSupAvg\}$, we have $\hat{w} = w_q$, such that $\top_q = \max (\top_{q'} \st q' \in U)$.
		For $\Val \in \{\DSum\}$ with a discount factor $\lambda$, let $P_q$ be the maximal accumulated value of a run of $\A$ on $u$ that ends in the state $q$. Then, we have $\hat{w} = w_q$, such that $P_q + \lambda^{|u|}\cdot \top_q = \max (P_{q'} + \lambda^{|u|}\cdot \top_{q'} \st q' \in U)$.
	\end{proof}

	\section{Subroutine: Constant-Function Check} \label{sec:ConstFuncProb}
	
	We will show that the problems of whether a given automaton is safe or live are closely related to the problem of whether an automaton expresses a constant function, motivating its study in this section.
	We first prove the problem hardness by reduction from the universality of nondeterministic finite-state automata (NFAs) and reachability automata.
	
	\begin{lem}\label{cl:ConstantCheckLowerBound}
		Let $\Val \in \{\Sup, \Inf, \LimInf, \LimSup, \LimInfAvg, \LimSupAvg, \DSum\}$.
		Deciding whether a $\Val$-automaton $\A$ expresses a constant function is \PSpaceH.
	\end{lem}
	\begin{proof}
		First, we prove the case where $\Val \in \{\Inf$, $\LimInf$, $\LimSup$, $\LimInfAvg$, $\LimSupAvg$, $\DSum\}$.
		The proof goes by reduction from the universality problem of nondeterministic finite-state automata (NFAs), which is known to be \PSpaceC.
		Consider an NFA $\A = (\Sigma, Q, \iota, F, \delta)$ over the alphabet $\Sigma = \{a, b\}$. We construct in \PTime a $\Val$-automaton $\A' = (\Sigma_{\#}, Q', \iota, \delta')$ over the alphabet $\Sigma_{\#} = \{a, b, \#\}$, such that $\A$ is universal if and only if $\A'$ is constant.
		$\A'$ has two additional states, $Q' = Q \uplus \{q_0, q_1\}$, and its transition function $\delta'$ is defined as follows:
		\begin{itemize}
			\item For every $(q, \sigma, p) \in \delta$, we have $\trans{q}{\sigma:1}{p}$.
			\item For every $q \in Q \setminus F$, we have $\trans{q}{\#:0}{q_0}$.
			\item For every $q \in F$, we have $\trans{q}{\#:1}{q_1}$.
			\item For every $\sigma \in \Sigma\cup \{\#\}$, we have $\trans{q_0}{\sigma:0}{q_0}$, and $\trans{q_1}{\sigma:1}{q_1}$.
		\end{itemize}
		Let $\top$ be the top value of $\A'$. (We have $\top=1$ in all cases, except for $\Val=\DSum$.)
		First, note that for every word $w$ with no occurrence of $\#$, we have that $\A'(w)=\top$, as all runs of $\A'$ visit only transitions with weight $1$.
		If $\A$ is not universal, then there exists a word $u\in\{a, b\}^*$ such that $\A$ has no run over $u$ from $\iota$ to some accepting state, and thus all runs of $\A'$ over $u\#$ from $\iota$ reach $q_0$.
		Hence, $\A'(u\#a^\omega) \neq \top$, while $\A'(a^\omega) = \top$, therefore $\A'$ is not constant.
		Otherwise, namely when $\A$ is universal, all infinite words with at least one occurrence of $\#$ can reach $q_1$ while only visiting $1$-weighted transitions, and thus $\A'(w) = \top$ for all $w\in\{a, b, \#\}^\omega$.
		
		Next, we prove the case where $\Val = \Sup$.
		The proof goes by reduction from the universality problem of a complete reachability automaton $\A'$ (i.e., a complete B\"uchi automaton all of whose states are rejecting, except for a single accepting sink).
		The problem is known to be \PSpaceH by a small adaptation to the standard reduction from the problem of whether a given Turing machine $T$ that uses a polynomial working space accepts a given word $u$ to NFA universality\footnote{Due to private communication with Christof L\"oding. See also~\cite[Thm. A.1]{Klein_2017}.}.
		By this reduction, if $T$ accepts $u$ then $\A'$ accepts all infinite words, and if $T$ does not accept $u$ then $\A'$ accepts some words, while rejecting others by arriving in all runs to a rejecting sink after a bounded number of transitions.
		As a complete reachability automaton $\A'$ can be viewed as special case of a $\Sup$-automaton $\A$, where transitions to nonaccepting states have weight $0$ and to accepting states have weight $1$, the hardness result directly follows to whether a $\Sup$-automaton is constant.	
	\end{proof}	
	\noindent 
	A simple solution to the problem is to check whether the given automaton $\A$ is equivalent to an automaton $\B$ expressing the constant top value of $\A$, which is computable in \PTime by \cref{cl:ComputingTopValue}.
	For some automata classes, it suffices for a matching upper bound.
	
	\begin{prop} \label{cl:ConstantCheckBasic}
		Deciding whether an $\Inf$-, $\Sup$-, $\LimInf$-, or $\LimSup$-automaton expresses a constant function is \PSpaceC. 
	\end{prop}
	\begin{proof}
		{\PSpaceH}ness is shown in \cref{cl:ConstantCheckLowerBound}.
		For the upper bound, we compute in \PTime, due to \cref{cl:ComputingTopValue}, the top value $\top$ of the given automaton $\A$, construct in constant time an automaton $\B$ of the same type as $\A$ that expresses the constant function $\top$, and check whether $\A$ and $\B$ are equivalent. This equivalence check is in \PSpace for arbitrary automata of the considered types~\cite[Thm.~4]{DBLP:journals/tocl/ChatterjeeDH10}.
	\end{proof}
	\noindent 
	Yet, this simple approach does not work for $\DSum$-automata, whose equivalence is an open problem, and for limit-average automata, whose equivalence is undecidable~\cite{DBLP:conf/csl/DegorreDGRT10,DBLP:conf/concur/ChatterjeeDEHR10,DBLP:journals/tcs/HunterPPR18}.
	
	For $\DSum$-automata, our alternative solution removes ``nonoptimal'' transitions from the automaton and then  reduces the problem to the universality problem of NFAs.
	
	\begin{thm}\label{cl:ConstantCheckForDSum}
		Deciding whether a $\DSum$-automaton expresses a constant function is \PSpaceC.
	\end{thm}
	\begin{proof}
		{\PSpaceH}ness is shown in \cref{cl:ConstantCheckLowerBound}.
		Consider a $\DSum$-automaton $\A$.
		By \cref{cl:ComputingTopValue}, for every state $q$ of $\A$ we can compute in \PTime the top value $\top_{q}$ of $\A^q$.
		We then construct in \PTime a $\DSum$-automaton $\A'$, by removing from $\A$ every transition $\trans{q}{\sigma:x}{q'}$, for which $x + \lambda\cdot \top_{q'}  < \top_{q}$.
		Finally, we consider $\A'$ as an (incomplete) NFA $\A''$ all of whose states are accepting.
		
		We claim that $\A''$ is universal, which is checkable in \PSpace, if and only if $\A$ expresses a constant function.
		Indeed, if $\A''$ is universal then for every word $w$, there is a run of $\A''$ on every prefix of $w$. Thus, by K\"onig's lemma there is also an infinite run on $w$ along the transitions of $\A''$.  Therefore, there is a run of $\A$ on $w$ that forever follows optimal transitions, namely ones that guarantee a continuation with the top value. Hence, by the discounting of the value function, the value of this run converges to the top value.
		If $\A''$ is not universal, then there is a finite word $u$ for which all runs of $\A$ on it reach a dead-end state. Thus, all runs of $\A$ on $u$ must have a transition $\trans{q}{\sigma:x}{q'}$, for which $x + \lambda \cdot \top_{q'} < \top_{q}$, implying that no run of $\A$ on a word $w$ for which $u$ is a prefix can have the top value.
	\end{proof}
	\noindent 
	The solution for limit-average automata is more involved.
	It is based on a reduction to the limitedness problem of distance automata, which is known to be in \PSpace~\cite{Has82,Sim94,Has00,LP04}.
	We start by presenting Johnson's algorithm, which we will use for manipulating the transition weights of the given automaton, and proving some properties of distance automata, which we will need for the reduction.
	
	A \emph{weighted graph} is a directed graph $G=\langle V,E \rangle$ equipped with a weight function $\gamma: E\to \ZZ$. The cost of a path $p=v_0,v_1,\ldots,v_k$ is $\gamma(p)=\sum_{i=0}^{k-1}\gamma(v_i,v_{i+1})$.
	
	\begin{prop}[{Johnson's Algorithm~\cite[Lem. 2 and Thms. 4 and 5]{Joh77}}]
		\label{cl:Johnson}
		Consider a weighted graph $G=\langle V,E \rangle$ with weight function $\gamma:E\to \ZZ$, such that $G$ has no negative cycles according to $\gamma$. We can compute in \PTime functions $h:V\to \ZZ$ and $\gamma':E\to \NN$ such that for every path $p=v_0,v_1,\ldots,v_k$ in $G$ it holds that $\gamma'(p)=\gamma(p)+h(v_0)-h(v_k)$. 
	\end{prop}
	
	\begin{rem}	
		\cref{cl:Johnson} is stated for graphs, while we will apply it for graphs underlying automata, which are multi-graphs, namely having several transitions between the same pairs of states. Nevertheless, to see that Johnson's algorithm holds also in our case, one can change every automaton to an equivalent one whose underlying graph is a standard graph, by splitting every state into several states, each having a single incoming transition.
	\end{rem}

	A \emph{distance automaton} is a weighted automaton over the tropical semiring (a.k.a., min-plus semiring) with weights in $\{0,1\}$. It can be viewed as a quantitative automaton over finite words with transition weights in $\{0,1\}$ and the value function of summation, extended with accepting states.
	A distance automaton is of \emph{limited distance} if there exists a bound on the automaton's values on all accepted words.
	
	Lifting limitedness to infinite words, we have by K\"onig's lemma that a total distance automaton of limited distance $b$, in which all states are accepting, is also guaranteed to have a run whose weight summation is bounded by $b$ on every infinite word.
	\begin{prop}\label{cl:BoundedDistanceAutomata}
		Consider a total distance automaton $\D$ of limited distance $b$, in which all states are accepting. Then, for every infinite word $w$, there exists an infinite run of $\D$ on $w$ whose summation of weights (considering only the transition weights and ignoring the final weights of states) is bounded by $b$.
	\end{prop}
	\begin{proof}
		Consider an infinite word $w$, and let $T$ be the tree of $\D$'s runs on prefixes of $w$ whose values are bounded by $b$. Notice that $T$ is an infinite tree since, by the totalness of $\D$ and the fact that all states are accepting, for every prefix of $w$ there is at least one such run. As the branching degree of $T$ is bounded by the number of states in $\D$, there exists by K\"onig's lemma an infinite branch $\rho$ in $T$. Observe that the summation of weights along $\rho$ is bounded by $b$---were it not the case, there would have been a position in $\rho$ up to which the summation has exceeded $b$, contradicting the definition of $T$.
	\end{proof}
	\noindent 
	Lifting nonlimitedness to infinite words, it may not suffice for our purposes to have an infinite word on which all runs of the distance automaton are unbounded, as their limit-average value might still be $0$.
	Yet, thanks to the following lemma, we are able to construct an infinite word on which the limit-average value is strictly positive.
	
	\begin{lem}\label{cl:UnBoundedDistanceAutomata}
		Consider a total distance automaton $\D$ of unlimited distance, in which all states are accepting. Then, there exists a finite nonempty word $u$ such that $\D(u)=1$ and the possible runs of $\D$ on $u$ lead to a set of states $U$ such that the distance automaton that is the same as $\D$ but with $U$ as the set of its initial states is also of unlimited distance. 
	\end{lem}
	\begin{proof}
		Let $Q$ be the set of states of $\D$. For a set $S\subseteq Q$, we denote by $\D^S$ the  distance automaton that is the same as $\D$ but with $S$ as the set of its initial states.
		Let $B$ be the set of sets of states from which $\D$ is of limited distance.
		That is, $B= \{S\subseteq Q \st \D^S \text{ is of limited distance}\}$.
		If $B=\emptyset$, the statement directly follows.
		
		Otherwise, $B\neq\emptyset$.
		Since for all $S\in B$, the distance automaton $\D^{S}$ is bounded, we can define $\hat{b}$ as the minimal number, such that for every $S\in B$ and finite word $u$, we have $\D^S(u)\leq \hat{b}$.
		Formally, $\hat{b}=\max_{S\in B} (\min \{b\in\NN \st \forall u \in \Sigma^*, \D^{S}(u)\leq b\}) $.
		Because $\D$ is of unlimited distance, we can exhibit a finite word mapped by $\D$ to an arbitrarily large value.
		In particular, there exists a word $z$ such that $\D(z)\geq \hat{b}+2$, i.e., the summation of the weights along every run of $\D$ on $z$ is at least $\hat{b}+2$.  
		Additionally, because transitions are weighted over $\{0, 1\}$, there exists at least one prefix $x\prefixeq z$ for which $\D(x)=1$.
		Let the finite word $y$ be such that $z=xy$.
		Next, we prove that $x$ fulfills the statement, namely that the distance automaton $\D^{X}$, where $X$ is the set of states that $\D$ can reach with runs on $x$, is also of unlimited distance.
		Assume towards contradiction that $X\in B$.
		By construction of $B$, we have that $\D^{X}$ is of limited distance.
		In fact, $\D^{X}(u)\leq \hat{b}$ for all finite words $u$, by the definition of $\hat{b}$.
		Hence $\D^{X}(y)\leq \hat{b}$, implying that $\D(z)=\D(xy)\leq \hat{b}+1$, leading to a contradiction, as $\D(z)\geq \hat{b}+2$.
	\end{proof}
	\noindent 
	Using \cref{cl:Johnson,cl:BoundedDistanceAutomata,cl:UnBoundedDistanceAutomata} we are in position to solve our problem by reduction to the limitedness problem of distance automata.
	
	\begin{thm}\label{cl:ConstantCheckLimAvg}
		Deciding whether a $\LimInfAvg$- or $\LimSupAvg$-automaton expresses a constant function, for a given constant or any constant, is \PSpaceC.
	\end{thm}
	\begin{proof}
		{\PSpaceH}ness is shown in \cref{cl:ConstantCheckLowerBound}.
		Consider a $\LimInfAvg$- or $\LimSupAvg$-automaton $\A$.
		We provide the upper bound as follows. First we construct in polynomial time a distance automaton $\D$, and then we reduce our statement to the limitedness problem of $\D$, which is decidable in \PSpace~\cite{Sim94}.
		
		By \cref{cl:ComputingTopValue}, one can first compute in polynomial time the top value of $\A$ denoted by $\top$.
		Thus, $\A$ expresses an arbitrary constant if and only if it expresses the constant function $\top$. 
		From $\A$, we construct the automaton $\A'$, by subtracting $\top$ from all transitions weights (by \cref{cl:ComputingTopValue}, $\top$ is guaranteed to be rational).
		By construction the top value of $\A'$ is $0$, i.e., $\A'(w)\leq 0$ for all $w$, and the question to answer is whether  $\A'$ expresses the constant function $0$, namely whether or not exists some word $w$ such that $\A'(w)<0$.
		
		Next, we construct from $\A'$, in which the nondeterminism is resolved by $\sup$ as usual, the opposite automaton $\A''$, in which the nondeterminism is resolved by $\inf$, by changing every transition weight $x$ to $-x$.
		If $\A'$ is a $\LimInfAvg$-automaton then $\A''$ is a $\LimSupAvg$-automaton, and vice versa.
		Observe that for every word $w$, we have $\A'(w) = - \A''(w)$.
		Now, we shall thus check if there exists a word $w$, such that $\A''(w)>0$.
		
		Since for every word $w$, we have that $\A''(w)\geq 0$, there cannot be a reachable cycle in $\A''$ whose average value is negative.
		Otherwise, some run would have achieved a negative value, and as the nondeterminism of $\A''$ is resolved with $\inf$, some word would have been mapped by $\A'$ to a negative value.
		Yet, there might be in $\A''$ transitions with negative weights.
		Thanks to Johnson's algorithm~\cite{Joh77} (see \cref{cl:Johnson} and the remark after it), we can construct from $\A''$ in polynomial time an automaton $\A'''$ that resolves the nondeterminism as $\A''$ and is equivalent to it, but has no negative transition weights.
		It is worth emphasizing that since the value of the automaton on a word is defined by the limit of the average values of forever growing prefixes, the bounded initial and final values that result from Johnson's algorithm have no influence.
		
		Finally, we construct from $\A'''$ the automaton $\B$ (of the same type), by changing every strictly positive transition weight to $1$.
		So, $\B$ has transitions weighted over $\{0, 1\}$.
		Observe that while $\A'''$ and $\B$ need not be equivalent, for every word $w$, we have $\A'''(w)>0$ if and only if $\B(w)>0$.
		This is because $x\cdot \B(w) \leq \A'''(w) \leq y\cdot \B(w)$, where $x$ and $y$ are the minimal and maximal strictly positive transition weights of $\A'''$, respectively.
		Further, we claim that $\B$ expresses the constant function $0$ if and only if the distance automaton $\D$, which is a copy of $\B$ where all states are accepting, is limited.
		
		If $\D$ is limited, then by \cref{cl:BoundedDistanceAutomata} there is a bound $b$, such that for every infinite word $w$, there exists an infinite run of $\D$ (and of $\B$) over $w$ whose summation of weights is bounded by $b$.
		Thus, the value of $\B$ (i.e., $\LimInfAvg$ or $\LimSupAvg$) for this run is $0$.
		
		If $\D$ is not limited, observe that the existence of an infinite word on which all runs of $\D$ are of unbounded value does not suffice to conclude.
		Indeed, the run that has weight $1$ only in positions $\{2^n \st n\in\NN\}$ has a limit-average of $0$.
		Nevertheless, we are able to provide a word $w$, such that the $\LimInfAvg$ and $\LimSupAvg$ values of every run of $\B$ over $w$ are strictly positive.
		
		By \cref{cl:UnBoundedDistanceAutomata}, there exists a finite nonempty word $u_1$, such that $\D(u_1)=1$ and the possible runs of $\D$ over $u$ lead to a set of states $S_1$, such that the distance automaton $\D^{S_1}$ (defined as $\D$ but where $S_1$ is the set of initial states) is of unlimited distance.
		We can then apply \cref{cl:UnBoundedDistanceAutomata} on $\D^{S_1}$, getting a finite nonempty word $u_2$, such that $\D^{S_1}(u_2)=1$, and the runs of $\D^{S_1}$ over $u_2$ lead to a set $S_2$, such that $\D^{S_2}$ is of unlimited distance, and so on.
		Since there are finitely many subsets of states of $\D$, we reach a set $S_\ell$, such that there exists $j<\ell$ with $S_j=S_\ell$. 
		We define the infinite word $w=u_1 \cdot u_2 \cdots u_j \cdot (u_{j{+}1} \cdot u_{j{+}2} \cdots u_\ell)^\omega$.	
		Let $m$ be the maximum length of $u_i$, for $i\in[1,\ell]$.
		Next, we show that the $\LimInfAvg$ and $\LimSupAvg$ values of every run of $\D$ (and thus the value of $\B$) over $w$ is at least $\frac{1}{m}$.
		
		Indeed, consider any infinite run $\rho$ of $\D$ over $w$.
		At position $|u_1|$, the summation of weights of $\rho$ is at least $1$, so the average is at least $\frac{1}{m}$.
		Since the run $\rho$ at this position is in some state $q\in S_1$ and $\D^{S_1}(u_2)=1$, the continuation until position $|u_1u_2|$ will go through at least another $1$-valued weight, having the average at position $|u_1u_2|$ is at least $\frac{1}{m}$.
		Then, for every position $k$ and natural number $i\in \NN$ such that $|u_1\cdots u_i| \leq k < |u_1\cdots u_{i+1}|$, we have $\frac{i-1}{i \cdot m} \leq \frac{i}{k} \leq \frac{i}{i\cdot m} = \frac{1}{m}$.
		Therefore, as $i$ goes to infinity, the running average of weights of $\rho$ converges to $ \frac{1}{m}$.
	\end{proof}
	
	\section{Safety of Quantitative Automata} \label{sec:safety}
	
	For studying the safety of automata, we build on the alternative characterizations of quantitative safety through threshold safety and continuity, as discussed in~\cref{sec:thresholdsafety,sec:continuity}.
	The characterizations for totally-ordered value domains hold in particular for properties expressed by quantitative automata.
	First, we extend the notion of safety from properties to value functions, allowing us to characterize families of safe quantitative automata.
	Finally, we provide algorithms to construct the safety closure of a given automaton $\A$ and to decide whether $\A$ is safe.

	\subsection{Safety of Value Functions} \label{sec:SafeVal}
	In this section, we focus on the value functions of quantitative automata, which operate on the value domain of real numbers.
	In particular, we carry the definitions of safety, co-safety, and discounting to value functions.
	This allows us to characterize safe (resp. co-safe, discounting) value functions as those for which all automata with this value function are safe (resp. co-safe, discounting).
	Moreover, we characterize discounting value functions as those that are safe and co-safe.
	
	Recall that we consider the value functions of quantitative automata to be bounded from below and above for every finite input domain $V \subset \QQ$.
	As the set $V^\omega$ can be taken as a Cantor space, just like $\Sigma^\omega$, we can carry the notions of safety, co-safety, and discounting from properties to value functions.
	
	\begin{defi}[Safety and co-safety of value functions]
		A value function $\Val : \QQ^\omega \to \RR$ is \emph{safe} when for every finite subset $V \subset \QQ$, infinite sequence $x \in V^\omega$, and value $v \in \RR$, if $\Val(x) < v$ then there exists a finite prefix $z \prefix x$ such that $\sup_{y \in V^\omega} \Val(zy) < v$.
		Similarly, a value function $\Val : \QQ^\omega \to \RR$ is \emph{co-safe} when for every finite subset $V \subset \QQ$, infinite sequence $x \in V^\omega$, and value $v \in \RR$, if $\Val(x) > v$ then there exists a finite prefix $z \prefix x$ such that $\inf_{y \in V^\omega} \Val(zy) > v$.
	\end{defi}
	
	\begin{defi}[Discounting value function]
		A value function $\Val : \QQ^\omega \to \RR$ is \emph{discounting} when for every finite subset $V \subset \QQ$ and every $\varepsilon > 0$ there exists $n \in \NN$ such that for every $x \in V^n$ and $y,y' \in V^\omega$ we have $|\Val(xy) - \Val(xy')| < \varepsilon$.
	\end{defi}
	
	We remark that by \cref{thm:safe:inf,thm:cosafeSup}, the value function $\Inf$ is safe and $\Sup$ is co-safe; moreover, $\DSum$ is discounting by definition.
	Now, we characterize the safety (resp. co-safety) of a given value function by the safety (resp. co-safety) of the automata family it defines.
	We emphasize that the proofs of the two statements are not dual.
	In particular, exhibiting a finite set of weights that falsifies the safety of a value function from a nonsafe automaton requires a compactness argument.
	
	\begin{thm}\label{cl:SafetyOfValAndOfAutomata} \label{cl:CoSafetyOfValAndOfAutomata}
		Consider a value function $\Val$.
		All $\Val$-automata are safe (resp. co-safe) iff $\Val$ is safe (resp. co-safe).
	\end{thm}
	\begin{proof}
		We show the case of safety and co-safety separately as they are not symmetric due to nondeterminism of automata.
		
		\subparagraph{Co-safety.}
		One direction is immediate, by constructing a deterministic automaton that expresses the value function itself: If $\Val$ is not co-safe then there exists some finite set $V \subset \QQ$ of weights with respect to which it is not co-safe.
		Consider the deterministic $\Val$-automaton over the alphabet $V$ with a single state and a self loop with weight $v\in V$ over every letter $v\in V$, that is, the letters coincide with the weights.
		Then, the automaton simply expresses $\Val$ and is therefore not co-safe.
		
		For the other direction, consider a co-safe value function $\Val$, a $\Val$-automaton $\A$ over an alphabet $\Sigma$ with a set of weights $V \subset \QQ$, a value $v \in \RR$, and a word $w$, such that $\A(w) > v$.
		We need to show that there exists a prefix $u \prefix w$ such that $\inf_{w' \in \Sigma^\omega} \A(uw') > v$.
		Let $\rho$ be some run of $\A$ on $w$ such that $\Val(\gamma(\rho)) >  v$. (Observe that such a run exists, since the value domain is totally ordered, as the supremum of runs that are not strictly bigger than $v$ is also not bigger than $v$.)
		
		Then, by the co-safety of $\Val$, there exists a prefix $\rho'$ of $\rho$, such that $\inf_{x' \in V^\omega} \Val(\gamma(\rho') x') >  v$. Let $u\prefix w$ be the prefix of $w$ of length $|\rho'|$.
		By the completeness of $\A$, for every word $w'' \in \Sigma^\omega$ there exists a run $\rho'\rho''$ over $uw''$, and by the above we have $\Val(\gamma(\rho'\rho'')) >  v$.
		Since $\A(uw'') \geq \Val(\gamma(\rho'\rho''))$, it follows that $\inf_{w' \in \Sigma^\omega} \A(uw') \geq  \inf_{x'\in V^\omega} \Val(\gamma(\rho')x') >  v$, as required.
		
		\subparagraph{Safety.}
		One direction is immediate: if the value function is not safe, we get a nonsafe automaton by constructing a deterministic automaton that expresses the value function itself, as detailed in the case of co-safety.
		
		As for the other direction, consider a nonsafe $\Val$-automaton $\A$ over an alphabet $\Sigma$ with a finite set $V \subset \QQ$ of weights. 
		Then, there exist a value $v \in \RR$ and a word $w$ with $\A(w) < v$, such that for every prefix $u \prefix w$, we have $\sup_{w' \in \Sigma^\omega} \A(uw') \geq v$.
		Let $v' \in (\A(w),v)$ be a value strictly between $\A(w)$ and $v$.
		For every prefix $u \prefix w$ of length $i>0$, let $w_i \in \Sigma^\omega$ be an infinite word and $r_i$ a run of $\A$ on $u w_i$, such that the value of $r_i$ is at least $v'$.
		Such a run exists since for all $u \prefix w$, the supremum of runs on $u w'$, where $w' \in \Sigma^\omega$, is larger than $v'$.
		
		Let $r'$ be a run of $\A$ on $w$, constructed in the spirit of K\"onig's lemma by inductively adding transitions that appear in infinitely many runs $r_i$.
		That is, the first transition $t_0$ on $w[0]$ in $r'$ is chosen such that $t_0$ is the first transition of $r_i$ for infinitely many $i\in\NN$. Then $t_1$ on $w[1]$, is chosen such that $t_0 \cdot t_1$ is the prefix of $r_i$ for infinitely many $i\in\NN$, and so on.
		Let $\rho$ be the sequence of weights induced by $r'$.
		Observe that $\Val(\rho) \leq \A(w) < v'$.
		Now, every prefix $\eta \prefix \rho$ of length $i$ is also a prefix of the sequence $\rho_i$ of weights induced by the run $r_i$, and by the above construction, we have $\Val(\rho_i)\geq v'$.
		Thus, while $\Val(\rho) < v'$, for every prefix $\eta \prefix \rho$, we have $\sup_{\rho' \in V^\omega} \Val(\eta \rho') \geq v'$, implying that $\Val$ is not safe.
	\end{proof}
	\noindent 
	Recall that a value function together with a finite set $V \subset \QQ$ of weights can be seen as a quantitative property over the finite alphabet $\Sigma = V$.
	Then, thanks to \cref{cl:SafeAndCosafeIffDiscountingProp}, we can characterize discounting value functions as those that are both safe and co-safe.
	
	\begin{cor} \label{cl:SafeAndCosafeIffDiscountingAut}
		A value function is discounting iff it is safe and co-safe.
	\end{cor}
	
	As a consequence of \cref{cl:CoSafetyOfValAndOfAutomata,cl:SafeAndCosafeIffDiscountingAut}, we obtain the following.
	
	\begin{cor} \label{cl:DiscountingOfValAndOfAutomata}
		All $\Val$-automata are discounting iff $\Val$ is discounting.
	\end{cor}
	
	\subsection{Deciding Safety of Quantitative Automata} \label{sec:SafeAutomata}
	We now switch our focus from generic value functions to families of quantitative automata defined by the common value functions $\Inf$, $\Sup$, $\LimInf$, $\LimSup$, $\LimInfAvg$, $\LimSupAvg$, and $\DSum$.
	As remarked in \cref{sec:SafeVal}, the value functions $\Inf$ and $\DSum$ are safe, thus all $\Inf$-automata and $\DSum$-automata express a safety property by \cref{cl:SafetyOfValAndOfAutomata}.
	Below, we focus on the remaining value functions of interest.
	
	Given a $\Val$-automaton $\A$ where $\Val$ is one of the nonsafe value functions above, we describe (i) a construction of an automaton that expresses the safety closure of $\A$, and (ii) an algorithm to decide whether $\A$ is safe.
	For these value functions, we can construct the safety closure as an $\Inf$-automaton.
	
	\begin{thm} \label{cl:SafetyClosure}
		Let $\Val \in \{\Sup, \LimInf, \LimSup, \LimInfAvg, \LimSupAvg\}$.\hspace{0.2em}
		Given a $\Val$-automaton $\A$, we can construct in \PTime an $\Inf$-automaton $\A'$ that expresses its safety closure.
		Moreover, if $\A$ is deterministic, then so is $\A'$.
	\end{thm}
	\begin{proof}
		Let $\A=(\Sigma,Q,\iota,\delta)$ be a $\Val$-automaton as above, where $\Val\neq\Sup$.
		We construct an $\Inf$-automaton $\A' = (\Sigma,Q,\iota,\delta')$ that expresses the safety closure of $\A$ by only changing  the weights of $\A$'s transitions, as follows.
		For every state $q\in Q$, we compute in \PTime, due to \Cref{cl:ComputingTopValue}, the top value $\top_{q}$ of the automaton $\A^q$.
		For every state $p \in Q$ and letter $\sigma \in \Sigma$, we define the transition function $\delta'(p, \sigma) = \{ (\top_{q}, q) \st \exists x \in \QQ : (x,q) \in \delta(p,\sigma)\}$.
		Notice that $\A$ and $\A'$ are identical except for their transition weights, therefore $\A'$ is deterministic if $\A$ is.
		
		Consider a run $\rho$ of $\A'$.
		Let $i \in \NN$ be the number of transitions before $\rho$ reaches its ultimate strongly connected component, i.e., the one $\rho$ stays indefinitely.
		By construction of $\A'$, the sequence $\gamma(\rho)$ of weights is nonincreasing, and for all $j \geq i$ we have that $\gamma(\rho[i]) = \gamma(\rho[j])$.
		Again, by construction, the value $\gamma(\rho[j])$ is the maximal value $\A$ can achieve after the first $j$ steps of $\rho$.
		Moreover, since $\gamma(\rho)$ is nonincreasing, it is the minimal value among the prefixes of $\gamma(\rho)$.
		In other words, $\gamma(\rho[i]) = \inf_{j \in \NN} \sup_{\rho' \in R_j} \Val(\gamma(\rho[..j] \rho'))$ where $R_j$ is the set of runs of $\A$ starting from the state reached after the finite run $\rho[..j]$.
		Notice that this defines exactly the value of the safety closure for the run $\rho$.
		Therefore, it is easy to see that $\A'(w) = \inf_{u \prec w} \sup_{w' \in \Sigma^\omega} \A(uw')$ for all $w \in \Sigma^\omega$.
		
		For $\Val = \Sup$, we use \Cref{cl:InfAndSuptoLim} to first translate $\A$ to a $\LimInf$- or $\LimSup$-automaton, which preserves determinism as needed.
	\end{proof}
	\noindent 
	For the prefix-independent value functions we study, the safety-closure automaton from the proof of \cref{cl:SafetyClosure} can be taken as a deterministic automaton with the same value function.
	
	\begin{thm} \label{cl:SafetyClosureValDet}
		Let $\Val \in \{\LimInf, \LimSup, \LimInfAvg, \LimSupAvg\}$.
		Given a $\Val$-automaton $\A$, we can construct in \PTime a $\Val$-automaton that expresses its safety closure and can be determinized in \ExpTime.
	\end{thm}
	\begin{proof}
		Let $\A$ be a $\Val$-automaton.
		We construct its safety closure $\A'$ as an $\Inf$-automaton in polynomial time, as in the proof of \cref{cl:SafetyClosure}.
		Observe that, by construction, every run $\rho$ of $\A'$ yields a nonincreasing weight sequence for which there exists $i \in \NN$ such that for all $j \geq i$ we have $\gamma(\rho[i]) = \gamma(\rho[j]) = \Inf(\gamma(\rho))$.
		Then, to construct a $\Val$-automaton that is equivalent to $\A'$, we simply copy $\A'$ and use the value function $\Val$ instead.
		Similarly, to obtain a deterministic $\Val$-automaton that is equivalent to $\A'$, we first determinize the $\Inf$-automaton $\A'$ in exponential time~\cite[Thm.~7]{DBLP:conf/vmcai/KupfermanL07}, and then the result can be equivalently considered as a $\Val$-automaton for the same reason as before.
	\end{proof}
\noindent 
	By contrast, this is not possible in general for $\Sup$-automata, as \cref{fig:Asup} witnesses.
	
		\begin{figure}[t]\centering
		\hspace*{-16pt}
		\noindent\begin{minipage}[c]{.48\linewidth}
			\scalebox{0.9}{
			\begin{tikzpicture}[bg={($\A$)}, node distance =2.2cm]
				\node[state, initial, label=center:$q_0$] (0) {};
				\node[state, left of = 0, label=center:$q_1$] (1) {};
				\node[state, right of = 0, label=center:$q_2$] (2) {};
				
				\path[transition]
				(0) edge[loop above] node[above] {$a:0$} (0)
				(0) edge[bend right=15] node[above] {$b:1$} (1)
				(0) edge[bend right=15] node[above] {$c:2$} (2)
				(1) edge[loop above] node[above] {$\Sigma:0$} (1)
				(2) edge[loop above] node[above] {$\Sigma:0$} (2)
				;	
			\end{tikzpicture}
			}
		\end{minipage}
		\noindent\begin{minipage}[c]{.48\linewidth}
			\scalebox{0.9}{
			\begin{tikzpicture}[bg={($\B$)}, node distance =2.2cm]
				\node[state, initial, label=center:$q_0$] (0) {};
				\node[state, left of = 0, label=center:$q_1$] (1) {};
				\node[state, right of = 0, label=center:$q_2$] (2) {};
				
				\path[transition]
				(0) edge[loop above] node[above] {$a:2$} (0)
				(0) edge[bend right=15] node[above] {$b:1$} (1)
				(0) edge[bend right=15] node[above] {$c:2$} (2)
				(1) edge[loop above] node[above] {$\Sigma:1$} (1)
				(2) edge[loop above] node[above] {$\Sigma:2$} (2)
				;	
			\end{tikzpicture}
			}
		\end{minipage}

		\caption{\label{fig:Asup}
			A $\Sup$-automaton $\A$ together with its safety closure $\B$ given as an $\Inf$-automaton, which cannot be expressed by a $\Sup$-automaton.}
	\end{figure}
	
	\begin{prop} \label{cl:SafetyClosureSup}
		Some $\Sup$-automaton admits no $\Sup$-automata that expresses its safety closure.
	\end{prop}
	\begin{proof}
		Consider the $\Sup$-automaton $\A$ given in \cref{fig:Asup}.
		We have $\safe{\A}(w) = 2$ if $w = a^\omega$ or the first $c$ in $w$ occurs before the first $b$ in $w$ (which may never occur), and $\safe{\A}(w) = 1$ otherwise.
		Suppose towards contradiction that there is a $\Sup$-automaton $\A'$ expressing $\safe{\A}$.
		Since $\A'$ has finitely many weights, it is $\sup$-closed, and $\A'(a^\omega) = 2$, there is a run $\rho$ of $\A'$ over $a^\omega$ in which the weight 2 occurs at least once, say at position $i$.
		Then, every valid continuation of the finite run $\rho[..i]$ over $\A'$ is mapped to at least 2.
		In particular, $\A'(a^i b^\omega) \geq 2$; however, $\safe{\A}(a^i b^\omega) = 1$.
	\end{proof}
	\noindent 
	We first prove the hardness of deciding safety by reduction from constant-function checks.
	
	\begin{lem}\label{cl:SafetyCheckLowerBound}
		Let $\Val \in \{\Sup, \LimInf, \LimSup, \LimInfAvg, \LimSupAvg\}$. 
		It is \PSpaceH to decide whether a $\Val$-automaton is safe.
	\end{lem}
	\begin{proof}
		We can reduce in \PTime the problem of whether a $\Val$-automaton $\A$ with the top value $\top$ expresses a constant function, which is \PSpaceH by \cref{cl:ConstantCheckLowerBound}, to the problem of whether a $\Val$-automaton $\A'$ is safe, by adding $\top$-weighted transitions over a fresh alphabet letter from all states of $\A$ to a new state $q_\top$, which has a $\top$-weighted self-loop over all alphabet letters. 
		
		Indeed, if $\A$ expresses the constant function $\top$, so does $\A'$ and it is therefore safe. Otherwise, $\A'$ is not safe, as a word $w$ over $\A$'s alphabet for which $\A(w)\neq \top$ also has a value smaller than $\top$ by $\A'$, while every prefix of it can be concatenated with a word that starts with the fresh letter, having the value $\top$.
	\end{proof}
	\noindent 
	For automata classes with \PSpace equivalence check, a matching upper bound is straightforward by comparing the given automaton and its safety-closure automaton.
	
	\begin{thm} \label{cl:SafetyCheckBasic}
		Deciding whether a $\Sup$-, $\LimInf$-, or $\LimSup$-automaton expresses a safety property is \PSpaceC.
	\end{thm}
	\begin{proof}
		{\PSpaceH}ness is shown in \cref{cl:SafetyCheckLowerBound}.
		For the upper bound, we construct in \PTime, due to \cref{cl:SafetyClosureValDet}, the safety-closure automaton $\A'$ of the given automaton $\A$, and then check in \PSpace if $\A = \A'$. Notice that equivalence-check is in \PSpace for these value functions in general~\cite[Thm.~4]{DBLP:journals/tocl/ChatterjeeDH10}.
	\end{proof}
	\noindent 
	On the other hand, even though equivalence of limit-average automata is undecidable \cite{DBLP:conf/csl/DegorreDGRT10,DBLP:conf/concur/ChatterjeeDEHR10,DBLP:journals/tcs/HunterPPR18}., we are able to provide a decision procedure using as a subroutine our algorithm to check whether a given limit-average automaton expresses a constant function (see \cref{cl:ConstantCheckLimAvg}).
	The key idea is to construct a limit-average automaton that expresses the constant function 0 iff the original automaton is safe.
	Our approach involves the determinization of the safety-closure automaton, resulting in an \ExpSpace complexity.
	Let us start with a lemma on checking the equivalence of limit-average automata.
	
	\begin{lem}\label{cl:LimAvgEquivalence}
		Let $\Val \in \{\LimInfAvg, \LimSupAvg\}$ and consider two $\Val$-automata $\A$ and $\B$.
		If $\B$ is deterministic and each of its runs yields an eventually-constant weight sequence, deciding whether $\A$ and $\B$ are equivalent is in \PSpace.
	\end{lem}
	\begin{proof}
		We construct $\C$ by taking the product between $\A$ and $\B$ where the weight of a transition in $\C$ is obtained by subtracting the weight of the corresponding transition in $\B$ from that in $\A$.
		We claim that $\A$ and $\B$ are equivalent iff $\C$ expresses the constant function 0.
		Indeed, consider a word $w \in \Sigma^\omega$.
		By definition, $\A(w) = \B(w)$ iff $\sup_{\rho_\A \in R^{\A}_w} \{\Val(\gamma(\rho_\A))\} - \Val(\gamma(\rho_\B)) = 0$ where $\rho_\B$ is the unique run of $\B$ on $w$.
		Equivalently $\sup_{\rho_\A \in R^{\A}_w} \{ \Val(\gamma(\rho_\A)) - \Val(\gamma(\rho_\B)) \} = 0$.
		We claim that $\Val(\gamma(\rho_\A)) - \Val(\gamma(\rho_\B)) = \Val(\gamma(\rho_\A) - \gamma(\rho_\B))$ where $\gamma(\rho_\A) - \gamma(\rho_\B)$ is the sequence obtained by taking the elementwise difference of the weight sequences produced by the runs $\rho_\A$ and $\rho_\B$.
		This claim does not hold for arbitrary sequences of weights, but it does hold if the sequence of weights $\gamma(\rho_B)$ is eventually constant and $\Val$ is prefix independent.
		As the weight sequence of $\rho_\B$ is eventually constant by our initial assumption and $\Val$ is prefix independent, we can subtract elementwise from the weight sequence of each run of $\A$ that of $\B$.
		Thus, we get $\sup_{\rho_\A \in R^{\A}_w} \{ \Val(\gamma(\rho_\A)) - \Val(\gamma(\rho_\B)) \} = 0$ iff $\sup_{\rho_\A \in R^{\A}_w} \{ \Val(\gamma(\rho_\A) - \gamma(\rho_\B)) \} = 0$.
		Observe that, by construction, each run of $\C$ produces a weight sequence that corresponds to this difference.
		Then, $\sup_{\rho_\A \in R^{\A}_w} \{ \Val(\gamma(\rho_\A) - \gamma(\rho_\B)) \} = 0$ iff $\sup_{\rho_{\C} \in R^{\C}_w} \{ \Val(\gamma(\rho_{\C})) \} = 0$ iff $\C(w) = 0$.
		Finally, to check the equivalence of $\A$ and $\B$, we can decide by \cref{cl:ConstantCheckLimAvg,cl:ComputingTopValue} if $\C(w) = 0$ for all $w \in \Sigma^\omega$.
	\end{proof}
	\noindent 
	Using the lemma above, we obtain an algorithm to check whether a given limit-average automaton is safe.
	
	\begin{thm} \label{cl:SafetyCheck}
		Deciding whether a $\LimInfAvg$- or $\LimSupAvg$-automaton expresses a safety property is in \ExpSpace.
	\end{thm}
	\begin{proof}
		Let $\Val \in \{\LimInfAvg, \LimSupAvg\}$ and let $\A$ be a $\Val$-automaton.
		We construct the safety-closure automaton of $\A$ whose weight sequences are eventually constant as in \cref{cl:SafetyClosure} and transform it into a deterministic $\Val$-automaton $\B$ as in the proof of \cref{cl:SafetyClosureValDet}.
		To check the safety of $\A$, we can decide by \cref{cl:LimAvgEquivalence} whether $\A$ and $\B$ are equivalent in \PSpace since $\B$ is deterministic and its weight sequences are eventually constant by construction. 
		Because the construction of $\B$ might cause up to an exponential size blow-up, the decision procedure for checking the safety of limit-average automata is in \ExpSpace.
	\end{proof}

	\section{Liveness of Quantitative Automata} \label{sec:liveness}
	
	In this section, we provide algorithms to check liveness of quantitative automata, and to decompose them into a safety automaton and a liveness automaton.
	We build on the alternative characterizations of quantitative liveness, as discussed in \cref{sec:notionsliveness}.
	In particular, our algorithms take advantage of the fact that liveness and top liveness coincide for $\sup$-closed properties (\cref{cl:LivenessCollapsingSupClosed}).

	\subsection{Deciding Liveness of Quantitative Automata} \label{sec:liveautomata}
	
	Let us start with the problem of checking whether a quantitative automaton is live.
	We first provide a hardness result by reduction from constant-function checks.
	
	\begin{lem}\label{cl:LivenessCheckLowerBound}
		Let $\Val \in \{\Inf, \Sup, \LimInf, \LimSup, \LimInfAvg, \LimSupAvg, \DSum\}$.
		Deciding whether a $\Val$-automaton $\A$ is live is \PSpaceH.
	\end{lem}
	\begin{proof}
		Let $\Val \in \{\Inf, \LimInf, \LimSup, \LimInfAvg, \LimSupAvg, \DSum\}$ be a value function.
		Consider a $\Val$-automaton $\A'$ that is constructed along the proofs of \cref{cl:ConstantCheckLowerBound}, in which we show that the constant-function check is \PSpaceH.
		Observe that $\A'$ either (i) expresses the constant function $\top$, and is therefore live; or (ii) has a value $\top$ on some word $w$ and a value $x<\top$ on some word $w'$, where there is a prefix $u$ of $w'$, such that for every infinite word $\hat{w}$, we have $\A'(u\hat{w})=x$, implying that $\A'$ is not live.
		Therefore, the {\PSpaceH}ness of the constant-function check extends to liveness-check.
		
		The proof for $\Val = \Sup$ goes by reduction from the constant-function check for $\Inf$-automata, which is \PSpaceH by \cref{cl:ConstantCheckLowerBound}.
		Given an $\Inf$-automaton $\A$ over an alphabet $\Sigma = \{a,b\}$, we construct in \PTime a $\Sup$-automaton $\A'$ such that $\A$ is constant iff $\A'$ is live.
		
		First, using \cref{cl:InfAndSuptoLim}, we transform $\A$ into an equivalent $\Inf$-automaton $\B=(\Sigma,Q_{\B},\iota,\delta_{\B})$ whose runs are nonincreasing.
		Let $\mathcal{S}_{\B} = \{S_1, \ldots, S_k\}$ be the set of strongly connected components of $\B$.
		Note that, by construction, each $S \in \mathcal{S}_{\B}$ (for which there is a transition whose target is in $S$) is associated with a weight $x$ such that all transitions whose target is in $S$ has weight $x$, which we denote by $\gamma_{\B}(S) = x$ with a slight abuse of notation.
		Notice that $\gamma_{\B}(S)$ is undefined when $S$ has no incoming transitions, which may happen if $S$ is a trivial strongly connected component containing the initial or an unreachable state.
		
		We now construct from $\B$ an $\Inf$-automaton $\C$.
		The automaton $\C$ is a copy of $\B$ over the alphabet $\Sigma_{\#} = \Sigma \cup \{\#\}$ with two additional states $Q_{\C} = Q_{\B} \uplus \{q_0, q_1\}$, modified transition weights, and some additional transitions.
		The transition function $\delta_{\C}$ is defined as follows:
		\begin{itemize}
			\item For every transition $(q, \sigma, x, p) \in \delta_{\B}$ with $x \geq \top_{\B}$, we have $(q, \sigma, 1, p) \in \delta_{\C}$.
			\item For every transition $(q, \sigma, x, p) \in \delta_{\B}$ with $x < \top_{\B}$, we have $(q, \sigma, 0, p) \in \delta_{\C}$.
			\item For every $\sigma \in \Sigma \cup \{\#\}$, we have $(q_1, \sigma, 1, q_1) \in \delta_{\C}$ and $(q_0, \sigma, 0, q_0) \in \delta_{\C}$.
			\item For every strongly connected component $S \in \mathcal{S}_{\B}$ with $\gamma_{\B}(S) \geq \top_{\B}$ and for every $q \in S$, we have $(q, \#, 1, q_1) \in \delta_{\C}$.
			\item For every strongly connected component $S \in \mathcal{S}_{\B}$ with $\gamma_{\B}(S) < \top_{\B}$ and for every $q \in S$, we have $(q, \#, 0, q_0) \in \delta_{\C}$.
		\end{itemize}
		\noindent 
		Notice that (i) we do not add transitions to $q_0$ or $q_1$ from strongly connected components for which the $\gamma_{\B}$ value is undefined, and (ii) by construction, the strongly connected components of $\C$ are given by the set $\mathcal{S}_{\C} = \{S_1, \ldots, S_k, T_0, T_1\}$ where, for $j \in \{0,1\}$, we have $T_j = \{q_j\}$.
		Moreover, for every $S \in \mathcal{S}_{\C}$, we have $\gamma_{\C}(S) = 1$ if $S = T_1$ or $S \in \mathcal{S}_{\B}$ with $\gamma_{\B}(T) \geq \top_{\B}$, and $\gamma_{\C}(S) = 0$ otherwise.
		
		We claim that $\A$ is constant iff $\C$ is constant.
		Since $\A$ and $\B$ are equivalent, we show that $\B$ is constant iff $\C$ is constant.
		
		Assume $\B$ is constant, i.e., $\B(w) = \top_{\B}$ for all $w \in \Sigma^\omega$.
		Let $w$ be a word with no occurrence of $\#$.
		There is a run of $\B$ over $w$ such that every strongly connected component $S \in \mathcal{S}_{\B}$ it visits satisfies $\gamma_{\B}(S) \geq \top_{\B}$.
		By construction, $\C$ has a run over $w$ following the same sequence of states, and thus the same strongly connected components, which satisfy $\gamma_{\B}(S) = 1$.
		Therefore, $\C(w) = 1$.
		Now, let $w$ be a word with an occurrence of $\#$, i.e., $w = u \# w'$ for some $u \in \Sigma^*$ and $w' \in \Sigma_{\#}^\omega$.
		Since $\B$ is constant and an $\Inf$-automaton, there is a finite run of $\B$ over $u$ that always stays in strongly connected components that are weighted at least $\top_{\B}$.
		Then, $\C$ has a finite run over $u$ staying only in 1-weighted components, reaching the 1-weighted bottom component $T_1$ after reading $u \#$, and thus $\C(w) = 1$.
		Therefore, $\C$ is also constant.
		
		Assume $\B$ is not constant.
		Then, there exists $w_1, w_2 \in \Sigma^\omega$ such that $\B(w_1) < \B(w_2) = \top_{\B}$.
		By similar arguments as above, we have that $\C(w_2) = 1$.
		Moreover, all runs of $\B$ over $w_1$ ultimately stay in a strongly connected component for which the $\gamma_{\B}$ value is strictly less than $\top_{\B}$.
		Again, similarly as above, each of these runs correspond to a run of $\C$ over $w_1$, and each corresponding run ultimately stays in a strongly connected component for which the $\gamma_{\C}$ value is 0, and thus $\C(w_1) = 0$.
		Therefore, $\C$ is also not constant.		
		
		Now, we construct from the $\Inf$-automaton $\C$ a $\Sup$-automaton $\A'$.
		The automaton $\A'$ is a copy of $\C$ with the only difference being the transition weights:
		for every $(q, \sigma, x, p) \in \delta_{\C}$, we have $(q, \sigma, x', p) \in \delta_{\A'}$ where $x'$ is the minimum over the values $\gamma_{\C}(S)$ such that the strongly connected component $S$ is reachable from the state $p$.
		In other words, the weight of a transition in $\A'$ is 0 if some run starting from the target state can achieve the value 0 in $\C$, and it is 1 otherwise.
				
		We claim that $\C$ expresses $\safe{\A'}$, which means $\C$ is constant iff $\A'$ is live, thanks to \cref{cl:ComputingTopValue,cl:LivenessCollapsingSupClosed}.
		First, observe that (i) $\mathcal{S}_{\C} = \mathcal{S}_{\A'}$, (ii) for every $S,S' \in \mathcal{S}_{\C}$, if $S'$ is reachable from $S$ and $\gamma_{\C}(S) = 0$, then $\gamma_{\C}(S') = \gamma_{\A'}(S') = 0$, and (iii) for every $S,S' \in \mathcal{S}_{\C}$, if $S'$ is reachable from $S$ and $\gamma_{\C}(S') = 1$, then $\gamma_{\C}(S) = 1$.
		
		Consider a word $w \in \Sigma_{\#}^\omega$ such that $\C(w) = 0$.
		We want to show that $\safe{\A'}(w) = 0$, i.e., there is a prefix $u \prefix w$ such that $\A'(uw') = 0$ for all $w' \in \Sigma_{\#}^\omega$.
		Since $\C(w) = 0$, every run of $\C$ over $w$ ultimately stays in a strongly connected component $S$ such that $\gamma_{\C}(S) = 0$.
		As $\A'$ only differs from $\C$ in transition weights, every run of $\A$ over $w$ follows the same states and the strongly connected components.
		Notice that whenever such a run visits a strongly connected component $T$ with $\gamma_{\C}(T) = 1$, we have $\gamma_{\A'}(T) = 0$ by construction (as the same run ultimately reaches a component $S$ with $\gamma_{\C}(S) = 0$).
		Moreover, due to observation (ii) above, every run of $\A'$ over $w$ ultimately stays in a strongly connected component $S$ such that $\gamma_{\A'}(S) = 0$.
		Then, by construction, there is a prefix $u \prefix w$ such that $\A'(uw') = 0$ for all $w' \in \Sigma_{\#}^\omega$.
		
		Consider a word $w \in \Sigma_{\#}^\omega$ such that $\C(w) = 1$.
		We want to show that $\safe{\A'}(w) = 1$, i.e., for every prefix $u \prefix w$ we have $\A'(uw') = 1$ for some $w' \in \Sigma_{\#}^\omega$.
		Since $\C(w) = 1$, some run of $\C$ over $w$ ultimately stays in a strongly connected component $S$ such that $\gamma_{\C}(S) = 1$.
		By construction of $\C$, the bottom strongly connected component $T_1$ is reachable from any such component $S$.
		Recall that $\A'$ only differs from $\C$ in transition weights.
		Then, every run $\rho$ of $\A'$ over $w$ follows the same states and the strongly connected components as $\C$, and thus the component $T_1$ is reachable from any component visited during $\rho$ by reading $\#$.
		Moreover, since $T_1$ is a bottom strongly connected component with $\gamma_{C}(T_1) = 1$, we have $\gamma_{\A'}(T_1) = 1$.
		Then, for every prefix $u \prefix w$ we have $\A'(u w') = 1$ for $w' = \#^\omega$.
	\end{proof}
\noindent 
	Recall that, thanks to \cref{cl:LivenessCollapsingSupClosed,cl:ComputingTopValue}, an automaton $\A$ expresses a liveness property iff $\safe{\A}$ expresses the constant function $\top$.
	For automata classes whose safety closure can be expressed as $\Inf$-automata, we provide a matching upper bound by simply checking the universality of the safety closure with respect to its top value.
	For $\DSum$-automata, whose universality problem is open, our solution is based on our constant-function-check algorithm (see \cref{cl:ConstantCheckForDSum}).
	
	\begin{thm} \label{cl:LivenessCheck}
		Deciding whether an $\Inf$-, $\Sup$-, $\LimInf$-, $\LimSup$-, $\LimInfAvg$-, $\LimSupAvg$- or $\DSum$-automaton expresses a liveness property is \PSpaceC.
	\end{thm}
	\begin{proof}
		{\PSpaceH}ness is shown in \cref{cl:LivenessCheckLowerBound}.
		Let $\A$ be a $\Val$-automaton and let $\top$ be its top value.
		Recall that liveness and top liveness coincide for $\sup$-closed properties by \cref{cl:LivenessCollapsingSupClosed}.
		As the considered value functions define $\sup$-closed properties, as proved in \cref{cl:ComputingTopValue}, we reduce the statement to checking whether $\safe{\A}$ expresses the constant function $\top$.
		
		For $\Val \in \{\Sup, \LimInf, \LimSup, \LimInfAvg, \LimSupAvg\}$, we first construct in \PTime an $\Inf$-automaton $\B$ expressing the safety closure of $\A$ thanks to \cref{cl:SafetyClosure}.
		Then, we decide in \PSpace whether $\B$ is equivalent to the constant function $\top$, thanks to \cref{cl:ConstantCheckBasic,cl:ComputingTopValue}
		For	$\Val=\DSum$, the safety closure of $\A$ is $\A$ itself, as $\DSum$ is a discounting value function due to \cref{cl:SafeAndCosafeIffDiscountingAut,cl:SafetyOfValAndOfAutomata}.
		Hence, we can decide in \PSpace whether $\A$ expresses the constant function $\top$, thanks to \cref{cl:ConstantCheckForDSum,cl:ComputingTopValue}.
	\end{proof}

	\subsection{Safety-Liveness Decompositions of Quantitative Automata} \label{sec:decompositions}
	
	We turn to safety-liveness decomposition, and start with the simple case of $\Inf$- and $\DSum$-automata, which are guaranteed to be safe. Their decomposition thus consists of only generating a liveness component, which can simply express a constant function that is at least as high as the maximal possible value of the original automaton $\A$. Assuming that the maximal transition weight of $\A$ is fixed, it can be done in constant time.
	
	Considering $\Sup$-automata, recall that their safety closure might not be expressible by $\Sup$-automata (\cref{cl:SafetyClosureSup}).
	Therefore, our decomposition of deterministic $\Sup$-automata takes the safety component as an $\Inf$-automaton.
	The key idea is to copy the state space of the original automaton and manipulate the transition weights depending on how they compare with the safety-closure automaton.
	
	\begin{thm} \label{cl:SafetyLivenessDecompositionSup}
		Given a deterministic $\Sup$-automaton $\A$, we can construct in \PTime a deterministic safety $\Inf$-automaton $\B$ and a deterministic liveness $\Sup$-automaton $\C$, such that $\A(w) = \min(\B(w), \C(w))$ for every infinite word $w\in\Sigma^\omega$. 
	\end{thm}
	\begin{proof}
		Given a deterministic $\Sup$-automaton, we can compute in \PTime, due to  \cref{cl:InfAndSuptoLim}, an equivalent deterministic $\Sup$-automaton $\A$ for which every run yields a nondecreasing weight sequence.
		We first provide the construction of the automata $\B$ and $\C$, then show that they decompose $\A$, and finally prove that $\B$ is safe and $\C$ is live.
		
		By \cref{cl:SafetyClosure}, we can construct in \PTime an $\Inf$-automaton $\B$ expressing the safety closure of $\A$, where every run of $\B$ yields a nonincreasing weight sequence.
		Observe that $\B$ is safe by construction, and that the structures of $\A$ and $\B$ only differ on the weights appearing on transitions, where each transition weight in $\B$ is the maximal value that $\A$ can achieve after taking this transition.
		In particular, $\B$ is deterministic because $\A$ is so.
		
		Then, we construct the deterministic $\Sup$-automaton $\C$ by modifying the weights of $\A$ as follows.
		For every transition, if the weight of the corresponding transitions in $\A$ and $\B$ are the same, then the weight in $\C$ is defined as the top value of $\A$, denoted by $\top$ here after.
		Otherwise, the weight in $\C$ is defined as the weight of the corresponding transition in $\A$.

		Next, we prove that $\A(w) = \min(\B(w), \C(w))$ for every word $w$.
		Let $\rho_\A, \rho_\B, \rho_\C$ be the respective runs of $\A$, $\B$, and $\C$ on $w$.
		There are the following two cases.
		\begin{itemize}
			\item 
			If the sequences of weights $\gamma(\rho_\A)$ and $\gamma(\rho_\B)$ never agree, i.e., for every $i \in \NN$ we have $\gamma(\rho_\A[i]) < \gamma(\rho_\B[i])$, then $\gamma(\rho_\C[i]) = \gamma(\rho_\A[i])$ for all $i \in \NN$ by the construction of $\C$.
			We thus get $\A(w) = \C(w) < \B(w)$, so $\A(w) = \min(\B(w) < \C(w))$, as required.
			\item
			Otherwise, the sequences of weights $\gamma(\rho_\A)$ and $\gamma(\rho_\B)$ agree on at least one position, i.e., there exists $i \in \NN$ such that $\gamma(\rho_\A[i]) = \gamma(\rho_\B[i])$.
			Since the run of $\A$ is guaranteed to yield nondecreasing weights and $\B$ is its safety closure, whose runs are nonincreasing, we have $\gamma(\rho_\A[j]) = \gamma(\rho_\B[j])$ for all $j \geq i$. Additionally, $\gamma(\rho_\C[i])=\top$ by the construction of $\C$.
			We thus get $\A(w) = B(w) < \C(w)$, so $\A(w) = \min(\B(w) < \C(w))$, as required.
		\end{itemize}
		\noindent 
		Finally, we show that $\C$ is live.
		By \cref{cl:LivenessCollapsingSupClosed}, it is sufficient to show that for every reachable state $q$ of $\C$, there exists a run starting from $q$ that visits a transition weighted by $\top$.
		Suppose towards contradiction that for some state $\hat{q}$, there is no such run.
		Recall that the state spaces and transitions of $\A$, $\B$, and $\C$ are the same.
		Moreover, observe that a transition weight in $\C$ is $\top$ if and only if the corresponding transitions in $\A$ and $\B$ have the same weight.
		
		If no transition with weight $\top$ is reachable from the state $\hat{q}$, then by the construction of $\C$, for every run $\rho_\A$ of $\A$ starting from $\hat{q}$ and the corresponding run $\rho_{\B}$ of $\B$, we have $\gamma(\rho_{\A}[i]) < \gamma(\rho_{\B}[i])$ for all $i \in \NN$.
		Recall that each transition weight in $\B$ is the maximal value $\A$ can achieve after taking this transition, and that for every finite word $u$ over which $\A$ reaches $\hat{q}$, we have  $\sup_{w'}\A(uw') = \B(uw')$.
		
		Hence, by the $\sup$-closedness of $\A$ and the fact that the sequences of weights in its runs are nondecreasing, for each prefix $r_{\A}$ of $\rho_{\A}$ and the corresponding prefix $r_{\B}$ of $\rho_{\B}$, there is an infinite continuation $\rho_{\A}'$ for $r_{\A}$ such that the corresponding infinite continuation $\rho'_{\B}$ for $r_{\B}$ gives  $\Sup(\gamma(r_{\A}\rho'_{\A})) = \Inf(\gamma(r_{\B} \rho'_{\B}))$.
		Note that this holds only if the two weight sequences have the same value after some finite prefix, in which case the weight of $\C$ is defined as $\top$.
		Hence, some run of $\C$ from $\hat{q}$ reaches a transition weighted $\top$, which yields a contradiction.
	\end{proof}
	\noindent 
	Using the same idea, but with a more involved reasoning, we show a safety-liveness decomposition for deterministic $\LimInf$- and $\LimSup$-automata.
	
	\begin{thm} \label{cl:SafetyLivenessDecompositionLiminf}
		Let $\Val\in\{\LimInf, \LimSup\}$.
		Given a deterministic $\Val$-automaton $\A$, we can construct in \PTime a deterministic safety $\Val$-automaton $\B$ and a deterministic liveness $\Val$-automaton $\C$, such that $\A(w) = \min(\B(w), \C(w))$ for every infinite word $w\in\Sigma^\omega$.
	\end{thm}
	\begin{proof}
		Consider a deterministic $\Val$-automaton $\A$.
		We construct $\B$ and $\C$ analogously to their construction in the proof of \cref{cl:SafetyLivenessDecompositionSup}, with the only difference that we use \cref{cl:SafetyClosureValDet} to construct $\B$ as a $\Val$-automaton rather than an $\Inf$-automaton.
		Once again, the structures of $\A$ and $\B$ only differ on the weights appearing on transitions, and $\B$ is deterministic because $\A$ is so.
		
		We first show that $\B$ and $\C$ decompose $\A$, and then prove that $\C$ is live. (Note that $\B$ is safe by construction.)
		
		Given an infinite word $w$, let $\rho_\A, \rho_\B, \rho_\C$ be the respective runs of $\A$, $\B$, and $\C$ on $w$.
		There are the following three cases.
		\begin{itemize}
			\item
			If the sequences of weights $\gamma(\rho_{\A})$ and $\gamma(\rho_{\B})$ agree only on finitely many positions, i.e., there exists $i \in \NN$ such that $\gamma(\rho_\A[j]) < \gamma(\rho_\B[j])$ for all $j \geq i$, then by the construction of $\C$, we have $\gamma(\rho_\C[j]) = \gamma(\rho_\A[j])$ for all $j \geq i$. Thus, $\A(w) = \C(w) < \B(w)$.
			\item
			If the sequences of weights $\gamma(\rho_\A)$ and $\gamma(\rho_{\B})$ disagree only on finitely many positions, i.e., there exists $i \in \NN$ such that $\gamma(\rho_\A[j]) = \gamma(\rho_\B[j])$ for all $j \geq i$, then by the construction of $\C$, we have  $\gamma(\rho_\C[j]) = \top$ for all $j \geq i$. Thus, $\A(w) = \B(w) \leq \C(w)$.
			\item
			Otherwise the sequences of weights $\gamma(\rho_\A)$ and $\gamma(\rho_{\B})$ both agree and disagree on infinitely many positions, i.e., for every $i \in \NN$ there exist $j, k \geq i$ such that $\gamma(\rho_A[j]) < \gamma(\rho_\B[j])$ and $\gamma(\rho_A[k]) = \gamma(\rho_\B[k])$.
			For $\Val=\LimInf$, we exhibit an infinite sequence of positions $\{x_i\}_{i\in\NN}$ such that $\gamma(\rho_A[x_i]) = \gamma(\rho_\C[x_i]) < \gamma(\rho_\B[x_i])$ for all $i\in\NN$.
			The first consequence is that $\A(w) < \B(w)$.
			The second consequence is that, by the construction of $\C$, if $\A(w) < \top$ then $\C(w) < \top$, which implies that $\A(w) = \C(w)$.
			For $\Val=\LimSup$, recall that every run of $\B$ yields a nonincreasing weight sequence.
			In particular, there exists $k\in\NN$ such that $\gamma(\rho_\B[k])=\gamma(\rho_\B(\ell))=\B(w)$ for all $\ell\geq k$.
			Then, we exhibit an infinite sequence of positions $\{y_i\}_{i\in\NN}$ such that $\gamma(\rho_A[y_i]) = \B(\gamma(\rho_\B[y_i])) = \B(w)$ and $\gamma(\rho_\C[y_i]) = \top$ for all $i\in\NN$.
			Consequently, $\C(w)=\top$ and $\A(w)=\B(w)$.
		\end{itemize}
		In either case, $\A(w) = \min(\B(w), \C(w))$.
		
		Next, we show that $\C$ is live using the same argument as in the proof of \cref{cl:SafetyLivenessDecompositionSup}:
		On the one hand, every word $w$ for which $\A(w) = \B(w)$ trivially satisfies the liveness condition as it implies $\C(w) = \top$.
		On the other hand, by \cref{cl:ComputingTopValue} every word $w$ for which $\A(w) < \B(w)$ is such that each finite prefix $u\prefix w$ admits a continuation $w'$ satisfying $\A(uw') = \B(uw')$.
		Hence, $\sup_{w'}\C(uw')=\top$ for all $u\prefix w$, implying the liveness condition.
	\end{proof}
\noindent 
	Finally, we provide a safety-liveness decomposition for nondeterministic automata with the prefix-independent value functions we consider.

	\begin{thm} \label{cl:SafetyLivenessDecompositionLimAvg}
		Let $\Val\in\{\LimSup, \LimInf, \LimInfAvg, \LimSupAvg\}$.
		Given a $\Val$-automaton $\A$, we can construct in \PTime a safety $\Val$-automaton $\B$ and a liveness $\Val$-automaton $\C$, such that $\A(w) = \min(\B(w), \C(w))$ for every infinite word $w\in\Sigma^\omega$.
	\end{thm}
		\begin{proof}
	Let $Q = \{q_1, \ldots, q_n\}$ be the set of states of $\A$, let $\Delta_{\A}$ be its transition relation, $\gamma_{\A}$ its weight function, and $X_{\A}$ its finite set of weights.
	We identify in \PTime the strongly connected components $Q_1 \uplus Q_2 \uplus \ldots \uplus Q_m$ of $\A$.
	For all $k \in \{1, \dots, m\}$, we compute in \PTime, thanks to \Cref{cl:ComputingTopValue}, the top value $\top_k$ of the automaton $\A^q$ for any $q\in Q_k$.
	Note that the choice of $q\in Q_k$ does not change $\top_k$ since the considered value function $\Val$ is prefix independent.
	Additionally, for all $k \in \{1, \dots, m\}$, we compute the highest value $\Theta_k$ achievable by some simple cycle $\pi_k$ within $Q_k$.
	To clarify, we emphasize that $\top_k \geq \Theta_k$ holds in general, and $\top_k > \Theta_k$ when all runs starting in $Q_k$ that achieve the top value $\top_k$ eventually leave the component $Q_k$.
	
	We explain briefly how $\Theta_k$ and $\pi_k$ are computed in \PTime.
	The value $\Theta_k$ is the top value of the automaton consisting of $Q_k$ and a sink absorbing all outgoing edges weighted with $\min X_{\A} - 1$.
	As discussed in the proof of~\cite[Thm.\ 3]{DBLP:journals/tocl/ChatterjeeDH10}, the top value of a $\Val$-automaton is attainable by a lasso run.
	Due to the properties of $\Val$, this lasso run can be transformed into a simple cycle, i.e., a cycle without inner cycles.
	Because $\Val$ is prefix independent, the path reaching the cycle of the lasso run can be removed to obtain a cycle run with the same value.
	Also, if the cycle $\rho =\rho_1\rho_2\rho_3$ has an inner cycle $\rho_2$, then $\rho$ can be shortened by keeping the cycle achieving the highest value between $\rho_2$ and $\rho_1\rho_3$.
	This proves that $\Theta_k$ is attainable by a simple-cycle run.

	Now, we briefly describe the computation of $\Theta_k$ and $\pi_k$.
	First, we consider $\Val\in\{\LimInf, \LimSup\}$.
	To compute $\Theta_k$, we first construct a $\Val$-automaton $\A_k$ that is a copy of the strongly connected component $Q_k$ extended to be total by adding a sink state with a self loop of weight $\min X_{\A} - 1$.
	Then, we compute the top value of $\A_k$, which is by definition $\Theta_k$.
	To compute $\pi_k$, we first construct a graph $G_k$ which is obtained from the underlying graph of $\A_k$ by removing all the edges corresponding to transitions of $\A_k$ whose weights are smaller than $\Theta_k$ if $\Val=\LimInf$ or greater than $\Theta_k$ if $\Val=\LimSup$. 	
	Then, we compute a cycle in $G_k$ using depth-first search and assign it to $\pi_k$.
	
	Second, we consider $\Val\in\{\LimInfAvg, \LimSupAvg\}$.
	To compute $\Theta_k$, we first construct $\A_k$ as above, take the underlying directed graph of $\A_k$, and multiply its edge weights by $-1$.
	Then, we use Karp's (dynamic programming) algorithm~\cite{DBLP:journals/dm/Karp78} to compute the minimum cycle mean in this directed graph, which gives us the value $-\Theta_k$.
	To compute $\pi_k$, it suffices to appropriately maintain the backtracking pointers in Karp's algorithm~\cite{DBLP:journals/ipl/ChaturvediM17}.
	Recall that the top value of an automaton can be computed in \PTime thanks to \Cref{cl:ComputingTopValue}, and note that the constructions described above are also in \PTime.
	
	We define the set of states of $\C$ as $P=\{p_1, p'_1, \ldots, p_n, p'_n, p_\bot\}$, in particular $|P| = 2|Q|+1$.
	In the following, we define the transition relation $\Delta_{\C}$ of $\C$.
	The states $\{p_i \st 1 \leq i \leq n\}$ are used to copy $\A$, i.e., $(p_i, \sigma, p_j) \in \Delta_{\C}$ if and only if $(q_i, \sigma, q_j) \in \Delta_{\A}$.
	Additionally, for all $k\in\{1, \ldots, m\}$, if $\top_k = \Theta_k$ then for all transitions of the simple cycle $\pi_k$ of the form $(q_i, \sigma, q_j)\in\Delta_{\A}$, we have $(p'_i, \sigma, p'_j)\in \Delta_{\C}$ and $(p_i, \sigma, p'_j)\in\Delta_{\C}$.
	Finally, for all $p'_i$ and $\sigma$, we have $(p'_i, \sigma, p_\bot)\in\Delta_{\C}$ and $(p_\bot, \sigma, p_\bot)\in\Delta_{\C}$.
	Now, we define the weight function $\gamma_{\C}$ of $\C$.
	For all transitions of the form $t = (p_i, \sigma, p_j)\in\Delta_{\C}$, we have $\gamma_{\C}(t)=\gamma_{\A}(q_i, \sigma, q_j)$.
	For all transitions of the from $t = (p, \sigma, p')$ with $p\in P\setminus\{p_{\bot}\}$ and $p'\in\{p'_i\st 1 \leq i \leq n\}$, we have $\gamma_{\C}(t)=\top_{\A}$.
	Finally, $\gamma_{\C}(p_\bot, \sigma, p_\bot) = \min X_{\A}$ for all $\sigma$.
	An example is given in \cref{fig:avgDecomp}.

	Next, we prove that $\C$ is live.
	The key argument is that, for each component $Q_k$ for which $\top_k=\Theta_k$, the automaton $\C$ provides a continuation leading to achieve the highest weight of $\A$.
	Recall that liveness and top liveness coincide for $\sup$-closed properties by \cref{cl:LivenessCollapsingSupClosed}.
	As the considered value function $\Val$ defines $\sup$-closed properties, as proved in \cref{cl:ComputingTopValue}, the liveness of $\C$ reduces to checking whether $\safe{\C}$ expresses the constant function $\top_{\A}$.
	In fact, by construction, all finite runs ending in $P\setminus\{p_{\bot}\}$ admit a continuation leading to achieve $\top_{\A}$.
	Additionally, for all finite runs ending in $p_\bot$, there is another run over the same word that follows the states of $\A$.
	Hence, the safety closure of $\C$ maps every words to $\top_{\A}$, implying the liveness of $\C$.
	
	By \cref{cl:SafetyClosureValDet}, we can construct in \PTime a $\Val$-automaton $\B$ expressing the safety closure of $\A$.
	We prove that the automata $\B$ and $\C$ yield a safety-liveness decomposition of $\A$.
	For all $w\in\Sigma^\omega$, if there is a run of $\A$ over $w$ of the form $\pi \pi_k^\omega$ for some finite run $\pi$ in $\A$, then $\top_k =  \B(w) =\A(w)  \leq \C(w) = \top_{\A}$, otherwise $\A(w)=\C(w)$.
	Since $\A(w) \leq \B(w)$ by construction, we have $\A(w) = \min(\B(w), \C(w))$, for all $w\in\Sigma^\omega$.
	
	Finally, let us note that the liveness component $\C$ constructed here may differ from the liveness component $\varPsi$ of the decomposition in \cref{thm:decomp}.
	To construct $\C$ efficiently, we only take into account one simple cycle $\pi_k$ that achieves the value $\Theta_k$ within each strongly connected component $S_k$.
	However, there may be many cycles within $S_k$ achieving $\Theta_k$, which would need to be taken into account to express $\varPsi$.
\end{proof}
	\noindent 
	Nondeterministic $\Sup$-automata can be handled as $\LimInf$- or $\LimSup$-automata (\cref{cl:InfAndSuptoLim}) and decomposed accordingly.
	For deterministic automata, the decomposition in \cref{cl:SafetyLivenessDecompositionLimAvg} yields a deterministic safety component, but its liveness component may be nondeterminizable.
	Whether deterministic $\LimInfAvg$- and $\LimSupAvg$-automata can be decomposed into deterministic automata remains open.
	
	\begin{figure}
		\begin{minipage}{.49\linewidth}
			\scalebox{.9}{
			\begin{tikzpicture}[bg={($\A$)}, node distance=2cm]
				\node[state,label=center:$q_0$] (q0) {};
				\draw[transition](q0.90)++(90:13pt) -- (q0);
				\node[state, left of = q0, label=center:$q_1$] (q1) {};
				\node[state, right of = q0, label=center:$q_2$] (q2) {};
				\node[state, right of = q2, label=center:$q_3$] (q3) {};
				
				\path[transition]
				(q0) edge node[above] {$a$} (q1)
				(q0) edge node[above] {$b$} (q2)
				(q1) edge[loop above] node[above] {$\Sigma:2$} (q1)
				(q2) edge[loop above] node[above] {$a:1$} (q2)
				(q2) edge node[below] {$b:0$} (q3)
				(q3) edge[bend right = 20] node[above] {$a:3$} (q2)
				(q3) edge[loop above] node[above] {$\Sigma:0$} (q3)
				;	
			\end{tikzpicture}
			}
			\bigskip\\
			\scalebox{.9}{
			\begin{tikzpicture}[bg={($\B$)}, node distance=2cm]
				\node[state,label=center:$q_0$] (q0) {};
				\draw[transition](q0.90)++(90:13pt) -- (q0);
				\node[state, left of = q0, label=center:$q_1$] (q1) {};
				\node[state, right of = q0, label=center:$q$] (q) {};
				
				\path[transition]
				(q0) edge node[above] {$a$} (q1)
				(q0) edge node[above] {$b$} (q2)
				(q1) edge[loop above] node[above] {$\Sigma:2$} (q1)
				(q) edge[loop above] node[above] {$\Sigma:3/2$} (q)
				;
			\end{tikzpicture}
			}
		\end{minipage}
		\begin{minipage}{.49\linewidth}\centering
			\scalebox{.9}{
			\begin{tikzpicture}[bg={($\C$)}, node distance=2cm]
				\node[state,label=center:$q_0$] (q0) {};
				\draw[transition](q0.90)++(90:13pt) -- (q0);
				\node[state, left of = q0, label=center:$q_1$] (q1) {};
				\node[state, right of = q0, label=center:$q_2$] (q2) {};
				\node[state, right of = q2, label=center:$q_3$] (q3) {};
				
				\path[transition]
				(q0) edge node[above] {$a$} (q1)
				(q0) edge node[above] {$b$} (q2)
				(q1) edge[loop above] node[above] {$\Sigma:2$} (q1)
				(q2) edge[loop above] node[above] {$a:1$} (q2)
				(q2) edge node[below] {$b:0$} (q3)
				(q3) edge[bend right = 20] node[above] {$a:3$} (q2)
				(q3) edge[loop above] node[above] {$\Sigma:0$} (q3)
				;

				\node[state, yshift=-1.5cm, label=center:$q'_1$] (qq1) at (q1) {};
				\node[state, yshift=-1.5cm, label=center:$q'_2$] (qq2) at (q2) {};
				\node[state, yshift=-1.5cm, label=center:$q'_3$] (qq3)at (q3) {};
				\node[state, yshift=-3cm, label=center:$q_\bot$] (qbot) at (q0){};
				
				\path[transition]
				(q1) edge  node[left] {$a$} (qq1)
				(q2) edge[pos=.1]  node[below] {$b$} (qq3)
				(q3) edge[pos=.18]  node[below] {$a$} (qq2)
				(qq1) edge[loop below] node[below] {$a:3$} (qq1)
				(qq2) edge node[above] {$b:3$} (qq3)
				(qq3) edge[bend left = 20] node[below] {$a:3$} (qq2)
				(qq1) edge node[above] {$\Sigma$} (qbot)
				(qq2) edge node[above] {$\Sigma$} (qbot)
				(qq3) edge[bend left=40] node[above] {$\Sigma$} (qbot)
				(qbot) edge[loop above] node[above] {$\Sigma:0$} (qbot)
				;	
			\end{tikzpicture}
			}
		\end{minipage}
	\caption{A nondeterministic $\LimInfAvg$-automaton $\A$ and its safety-liveness decomposition into $\LimInfAvg$-automata $\B$ and $\C$, as presented in the proof of \cref{cl:SafetyLivenessDecompositionLimAvg}.}\label{fig:avgDecomp}
	\end{figure}

	\section{Conclusions} \label{sec:Conclusions}
	We presented a generalization of safety and liveness that lifts the safety-progress hierarchy to the quantitative setting of~\cite{DBLP:journals/tocl/ChatterjeeDH10} while preserving major desirable features of the boolean setting such as the safety-liveness decomposition and connections to topology.
	Then, we instantiated our framework with the specific classes of quantitative properties expressed by automata.
	
	Monitorability identifies a boundary separating properties that can be verified or falsified from a finite number of observations, from those that cannot.
	Safety-liveness and co-safety-co-liveness decompositions allow us separate, for an individual property, monitorable parts from nonmonitorable parts.
	The larger the monitorable parts of the given property, the stronger the decomposition.
	We provided the strongest known safety-liveness decomposition, which consists of a pointwise minimum between a safe part defined by a quantitative safety closure, and a live part which corrects for the difference.
	
	Moreover, we studied the quantitative safety-liveness dichotomy for properties expressed by $\Inf$-, $\Sup$-, $\LimInf$-, $\LimSup$-, $\LimInfAvg$-, $\LimSupAvg$-, and $\DSum$-automata.
	To this end, and solved the constant-function problem for these classes of automata.
	We presented automata-theoretic constructions for the safety closure of these automata and decision procedures for checking their safety and liveness.  
	We proved that the value function $\Inf$ yields a class of safe automata and $\DSum$ both safe and co-safe.
	For all common automata classes, we provided a decomposition into a safe and a live component.
	We emphasize that the safety component of our decomposition algorithm is the safety closure, and thus the best safe approximation of a given automaton.
	We note that most of these algorithms have been recently implemented in a tool~\cite{DBLP:conf/isola/ChalupaHMS24,chalupa2025automatinganalysisquantitativeautomata}.
	
	We focused on quantitative automata~\cite{DBLP:journals/tocl/ChatterjeeDH10} because their totally-ordered value domain and their $\sup$-closedness make quantitative safety and liveness behave in particularly natural ways; 
	a corresponding investigation of weighted automata~\cite{DBLP:journals/iandc/Schutzenberger61b} remains to be done. 
	We left open the complexity gap in the safety check of limit-average automata, and the study of co-safety and co-liveness for nondeterministic quantitative automata, which is not symmetric to safety and liveness due to the nonsymmetry in resolving nondeterminism by the supremum value of all possible runs.

	\bibliographystyle{alphaurl}
	\bibliography{bib}
	
\end{document}